\documentclass[fleqn,usenatbib,referee]{mnras}

\usepackage[T1]{fontenc}
\usepackage{natbib}
\usepackage{color}
\newcommand{\cm}{\,{\rm cm}^{-3} }
\newcommand{\msun}{\thinspace M_\odot} 
\newcommand{\vect}[1]{\mbox{\boldmath$#1$}}
\usepackage{bm}
\usepackage{cite}

\newcommand{\so}[1]{{\color[rgb]{0., 0., 0.}{#1}}}
\newcommand{\red}[1]{{\color[rgb]{0., 0., 0.}{#1}}}

\usepackage{graphicx}	
\usepackage{amsmath}	
\usepackage{amssymb}	

\title[Dependence of Hall Coefficient]{Dependence of Hall Coefficient on Grain Size and Cosmic Ray Rate and Implication for Circumstellar Disk Formation}

\author[S. Koga et al.]{
Shunta Koga,$^{1}$\thanks{E-mail: s.koga.098@s.kyushu-u.ac.jp} Yusuke Tsukamoto,$^{2}$ Satoshi Okuzumi$^{3}$ and Masahiro N. Machida$^{1}$
\\
$^{1}$Department of Earth and Planetary Sciences, Faculty of Sciences, Kyushu University, Fukuoka 819-0395, Japan\\
$^{2}$Department of Earth and Space Science, Graduate Schools of Science and Engineering, Kagoshima University, Kagoshima, Japan\\
$^{3}$Department of Earth and Planetary Sciences, Tokyo Institute of Technology, Tokyo, Japan
}

\begin{document}
\maketitle

\begin{abstract}
The Hall effect plays a significant role in star formation because it induces rotation in the infalling envelope, which in turn affects the formation and evolution of the circumstellar disk. 
The importance of the Hall effect varies with the Hall coefficient, and this coefficient is determined by the fractional abundances of charged species. These abundance values are primarily based on the size and quantity of dust grains as well as the cosmic ray intensity, which respectively absorb and create charged species. 
Thus, the Hall coefficient varies with both the properties of dust grains and the cosmic ray rate (or ionization source).
In this study, we explore the dependence of the Hall coefficient on the grain size and cosmic ray ionization rate using a simplified chemical network model.
Following this, using an analytic model, we estimate the typical size of a circumstellar disk induced solely by the Hall effect. 
The results show that the disk grows during the main accretion phase to a size of $\sim 3-100$\ au, with the actual size depending on the parameters.
These findings suggest that the Hall effect greatly affects circumstellar disk formation, especially in the case that the dust grains have a typical size of $\sim 0.025\, \mu {\rm m} - 0.075\, \mu {\rm m}$. 
\end{abstract}

\begin{keywords}
stars: formation --stars: magnetic field -- ISM: clouds -- cosmic rays--  dust, extinction
\end{keywords}



\section{Introduction}
\label{sec:intro}
\ \ It is important to elucidate the process by which stars are formed, because stars are the fundamental constituents of the universe and because star formation is associated with the origin of planets. 
Observations indicate that the magnetic energy in prestellar clouds is comparable to the gravitational energy \citep{2010ApJ...725..466C}. 
This large quantity of magnetic energy (that is, a strong Lorentz force)  affects the star formation process.
For example, magnetic fields play a significant role in determining the angular momentum distribution in  star-forming cores.   
In the star formation process, the magnetic fields suppress  disk formation.
Some researchers pointed out that  no disk appears in the early star formation phase, because the disk angular momentum  is excessively transferred by the  magnetic effect (the so-called `magnetic braking catastrophe' problem) \citep{2008ApJ...681.1356M}. 
There is another issue to the effect of magnetic fields during the star formation process (the so-called magnetic flux problem).
The magnetic flux of a star-forming core  is approximately five orders of magnitude greater than that of a protostar \citep[e.g.][]{1984FCPh....9..139N}.
Because magnetic flux is a conserved quantity, the magnetic field should dissipate during the course of the star formation.

The magnetic field is coupled with charged particles, and the amount of charged particles is considerably less in star-forming clouds due to their low temperature and high density.
However, during the early gas collapse phase, which is associated with a number density of $n\lesssim 10^{10}\cm$, the magnetic field is closely coupled to the neutral gas as a result of efficient momentum exchange between charged particles and neutral species.
Conversely, as the cloud density increases, dust grains absorb charged particles and the momentum exchange becomes inefficient \citep{1980PASJ...32..405U}. 
Therefore, the magnetic field cannot couple with neutral species and the magnetic field dissipates to a greater extent \citep{2002ApJ...573..199N}. 
The magnetic dissipation helps  disk formation, because the magnetic braking, which suppresses the formation and evolution of the disk,  is alleviated by the dissipation.  
Thus, the magnetic field and its dissipation are related to major issues (magnetic braking catastrophe and magnetic flux problems)  of the star formation process.

Both the degree of ionization (that is, the abundance of charged particles) and the extent of magnetic dissipation during star formation had been investigated in detail using one-zone calculations by Nakano and his collaborators for approximately three decades \citep{1990MNRAS.243..103U}. 
The methodology of this group is well established and is frequently applied to estimate the abundance of various chemicals as well as magnetic dissipation during the star formation process \citep[e.g.][]{2017PASJ...69...95T,2018MNRAS.476.2063W}.

Beginning in the 1990s and subsequent to the introduction of the one-zone calculations, which basically estimate the chemical abundance of each charged species, multi-dimensional 
magnetohydrodynamics (MHD) simulations of collapsing star-forming clouds were reported. In this process, calculations are performed until a protostar begins to form from the prestellar cloud core stage. 
One pioneering work by \citet{1998ApJ...502L.163T,2000ApJ...528L..41T,2002ApJ...575..306T} investigated cloud evolution up to the point of protostar formation using two-dimensional ideal MHD simulations. 
There are three non-ideal MHD effects: Ohmic dissipation, ambipolar diffusion and the Hall effect.
\citet{2006ApJ...647L.151M,2007ApJ...670.1198M} first calculated protostar formation using non-ideal MHD simulations, although only Ohmic dissipation was considered. \so{\citet{2009ApJ...706L..46D} performed three dimensional non-ideal MHD simulations including only ambipolar diffusion and  showed  the early formation of  outflow and disk.}
Subsequently, \citet{2015MNRAS.452..278T} and \citet{2015ApJ...801..117T} included ambipolar diffusion in addition to Ohmic dissipation. 

However, until recently, the Hall effect has been ignored in multi-dimensional core collapse simulations because it is believed to be not directly related to the magnetic flux problem. 
It should also be noted that, although both Ohmic dissipation and ambipolar diffusion substantially reduce the magnetic flux in the star-forming core, the Hall effect only changes the magnetic field direction to generate toroidal magnetic fields from global poloidal fields \citep{1999MNRAS.303..239W}. 

Recently, the Hall effect has been considered with regard to the formation and evolution of circumstellar disks \citep{2011ApJ...733...54K,2015ApJ...810L..26T,2017PASJ...69...95T,2016MNRAS.457.1037W}. 
\citet{2011ApJ...733...54K} have reported that, without rotation, a toroidal magnetic field component that is decoupled from the neutral gases is generated by the Hall effect. 
In the case that coupling between the magnetic field (or charged particles) and neutral species is recovered, the toroidal field (or twisted magnetic field) imparts a rotation to the infalling gas as a result of magnetic tension.  
In summary, even when the initial prestellar cloud has no rotation, rotation can be induced by the Hall effect.
\so{It should be noted that when a rotation motion is induced by the Hall effect, the magnetic field produces an counter-rotation in a star forming core due to the angular momentum conservation law, in which the angular momentum of the anti-rotation is transferred to the outer envelope from the cloud centre by the Alfv\'en wave. In other words, both clockwise and anti-clockwise rotations coexist in the initially non-rotating cloud, as reported by past studies  \citep{2011ApJ...733...54K,2015ApJ...810L..26T,2016MNRAS.457.1037W,2017PASJ...69...95T,2018arXiv180808731M}. }
Thus, it is important to consider the Hall effect when examining  the angular momentum problem and  the formation and evolution of the circumstellar disk during the star formation process. 
Note that we should not ignore the Hall effect even when considering the dissipation of magnetic fields (i.e. the magnetic flux problem).
Since the Hall effect changes the configuration of the magnetic field,  it  would  affect the dissipation process of the magnetic field \citep{2017ApJ...836...46B}.

\citet{1999MNRAS.303..239W} also pointed out that the Hall effect (or Hall conductivity) strongly depends on the chemical abundance, which in turn is determined by the properties of the dust in star the formation process and the ionization (that is, cosmic ray) rate. 
Thus, it is important to estimate the effect of the Hall conductivity (or Hall coefficient) in conjunction with various grain sizes and cosmic ray rates and, for this purpose, some previous studies have examined grain size ranges and various cosmic ray rates.
\citet{2018arXiv180400848H} investigated the grain sizes in galaxy halos using the extent of cosmic extinction of distant background quasars, and found that the most likely size range is 0.01-0.3 $\rm{\mu m}$. 
In addition, based on theoretical and observational studies, cosmic ray rates have been estimated to be $2 \times 10^{-16} \ \rm{s^{-1}}$\citep{2009A&A...501..619P} and $3.4 \times 10^{-18} \ \rm{s^{-1}}$\citep{1994MNRAS.269..641F} in a diffuse core and in the Bok globule B335, respectively. 
Thus, it is necessary to consider a wide range of possible grain sizes and cosmic ray rates.

However, it is quite difficult to perform multi-dimensional simulations of cloud collapse including non-ideal MHD effects and incorporating different dust properties and cosmic ray rates because of the limited CPU resources.
\so{For example, the calculation time of the simulation including the Hall effect is about 10 times longer than that of the simulation not including the Hall effect, because \red{the whistler waves, which are the right-handed waves induced by the Hall term, should be resolved and} implicit methods can not be applied to the Hall effect \citep{2015ApJ...810L..26T}.}
In addition, calculations regarding the formation and evolution of circumstellar disks that resolve the protostar using multi-dimensional non-ideal MHD simulations have yet to be realized. 
It is important to note that erroneous results tend to be produced when introducing a sink method to model the long-term evolution of the circumstellar disk \citep{2014MNRAS.438.2278M}.
Also, performing calculations to resolve both a protostar and circumstellar disk having large resistivities (or small conductivities) requires exceedingly short time steps during simulations \citep{2018arXiv180108193V}, and so such calculations cannot be executed. 
Multi-dimensional calculations up to the point of several years after protostar formation are possible without employing a sink \citep[e.g.,][]{2015ApJ...801..117T,2016A&A...587A..32M,2017PASJ...69...95T,2018MNRAS.475.1859W,2018ApJ...868...22T}, but it is important to keep in mind that a sink can introduce numerical artifacts and these should be carefully considered, as discussed in \citet{1998ApJ...508L..95B}, \citet{2010ApJ...724.1006M}, \citet{2011MNRAS.413.2767M} \& \citet{2014MNRAS.438.2278M}. 
In addition to the large computational cost and sink problems, there is a further serious problem in non-ideal MHD simulations including the Hall term.  
 \citet{2018arXiv180808731M}  pointed out that the angular momentum is not conserved after the first core formation in their Hall MHD simulations.
Therefore,  the disk formation simulation including the Hall term is not an easy task.

In the present study, we analytically estimated the influence of the Hall effect on the formation and evolution of a circumstellar disk.
In particular, we focused on the dependence of the Hall coefficient on the grain size (distribution) and the cosmic ray rate. 
We initially calculated the fractional abundance of charged species by solving chemical networks in conjunction with the grain size (distribution) and cosmic ray strength as input parameters, and derived the Hall coefficient in a realistic parameter space. 
Then, using the Hall coefficient, we estimated the disk size. On this basis, we discuss the importance of the Hall effect with regard to the disk evolution.

The remainder of the present paper is organized as follows. 
In \S\ref{sec:hallcoefficient}, we describe the method used to calculate the Hall coefficient, while in \S\ref{sec:chemicalresult} we describe the chemical abundance and Hall coefficient results obtained in association with various grain sizes (distributions) and cosmic ray strengths. In \S\ref{sec:diskmethod}, we describe the disk model obtained by considering the Hall effect and the resulting disk sizes based on the parameters employed. In \S\ref{sec:caveats}, we discuss the time scale for the Hall effect and the effect of the magnetic field strength. Finally, a summary is presented in \S\ref{sec:summary}.

\section{Chemical Reactions and Hall Coefficient}
\label{sec:hallcoefficient}

\subsection{The Hall Effect and Hall Drift Velocity}
\label{subsec:hallvelocity}
The induction equation including the Hall term is written as 
\begin{equation}
\frac{\partial \bm{B}}{\partial t}=\nabla \times \left(\bm{v} \times \bm{B} \right) - \nabla \times \left(\eta_{\rm H} \left(\nabla \times \bm{B} \right) \times \hat{\bm{B}} \right),
\label{induction}
\end{equation}
where $\bm{v}$, $\bm{B}$,  $\bm{\hat{B}}$ and  $\eta_{\rm H}$ are the fluid velocity, the magnetic field, the unit vector for the magnetic flux density and the Hall coefficient, respectively.
Ohmic dissipation and ambipolar diffusion terms are not included in this section because this study focuses on the Hall effect. 
We define the Hall drift velocity as 
\begin{equation}
\bm{v}_{\rm Hall} \equiv -\eta_H \frac{\nabla \times \bm{B}}{|\bm{B}|}.
\label{vhall}
\end{equation}
Substituting equation~(\ref{vhall}) into equation~(\ref{induction}), the induction equation can be rewritten as 
\begin{equation}
\frac{\partial \bm{B}}{\partial t}=\nabla \times \left[ (\bm{v}+\bm{v}_{\rm Hall}) \times \bm{B} \right].
\label{induction2}
\end{equation}
Equation~(\ref{induction2}) indicates that the magnetic field can be amplified by the Hall velocity even when the gas fluid has no velocity (i.e., $\bm{v}=0$). 
In addition, the Hall velocity is rewritten as
\begin{equation}
\bm{v}_{\rm Hall}  = -\eta_{\rm H} \frac{c \bm{J}}{4 \pi |\bm{B}|},
\label{vhall2}
\end{equation}
and is also parallel to the current ($\vect{v}_{\rm Hall} \parallel \vect{J}$).
The equations (\ref{induction2}) and (\ref{vhall2}) indicate that magnetic field drifts to the direction parallel to $\bm{J}$ with the velocity of $\bm{v}_{\rm Hall}$. The drifted magnetic field (or toroidal field) induces gas rotation due to the magnetic tension force.

Assuming only uniform vertical magnetic fields in the initial state, the toroidal component of the magnetic field is continuously amplified by the Hall effect until $\vect{v}=-\vect{v}_{\rm Hall}$ is realized. 
Thus, the Lorentz force changes the gas velocity unless the condition $\vect{v}+\vect{v}_{\rm Hall}=0$ is fulfilled, and the generation of the toroidal field gradually decreases as the gas velocity, $\vect{v}$, approaches $-\vect{v}_{\rm Hall}$.
Finally, further generation of the toroidal field is completely suppressed and the gas has a velocity of $\vect{v}=-\vect{v}_{\rm Hall}$, such that the right hand side of equation (\ref{induction2}) becomes zero. 
The timescale for the gas velocity to reach $\vect{v}=-\vect{v}_{\rm Hall}$ is discussed in \S\ref{subsec:halltimescale}. 
 
The gas fluid should have a rotational velocity of $\vect{v}=-\vect{v}_{\rm Hall}$ without the initial cloud rotation.
That is, even in the absence of rotation of the initial cloud, a rotational motion will result from the Hall effect \so{\citep{2011ApJ...733...54K,2016PASA...33...10T,2018arXiv180808731M}.}

\subsection{The Hall Coefficient}
\label{methodetaH}
Investigating the influence of the Hall effect requires calculation of the Hall coefficient, $\eta_{\rm H}$, as described in Section \S\ref{subsec:hallvelocity}. 
The Hall coefficient is written as 
\begin{equation}
\eta_H=\frac{c^2 \sigma_H}{4 \pi \sigma_\bot^2},
\label{eq:etah} 
\end{equation}
where $c$ is the speed of light and $\sigma_\bot$ is described as 
\begin{equation}
\sigma_\bot=\sqrt{\sigma_p^2+\sigma_H^2},
\label{eq:sigmav}
\end{equation}
in which the Hall $\sigma_H$ and Petersen $\sigma_P$ conductivities are defined  as  
\begin{equation}
\sigma_H=-\frac{c}{B}{\displaystyle \sum_\nu} \frac{\beta_\nu^2}{1+\beta_\nu^2} Q_\nu n_\nu,
\label{eq:sigmah}
\end{equation}
and
\begin{equation}
\sigma_P=\frac{c}{B}{\displaystyle \sum_\nu} \frac{\beta_\nu}{1+\beta_\nu^2} |Q_\nu| n_\nu. 
\end{equation}
Here, $B$ is the magnetic field strength, $\nu$ indicates the physical quantity of each chemical species (for details, see Section \S\ref{subsec:chemicalnetwork}), and $Q_\nu$, $n_\nu$ and $\beta_\nu$ are the charge number, number density and Hall parameter for each charged species, respectively.
The Hall parameter, $\beta_\nu$, is defined as
\begin{equation}
\beta_\nu=\tau_\nu \frac{|Q_\nu|  B}{m_\nu c},  
\label{eq:betanu}
\end{equation}
where $\tau_\nu$ and $m_\nu$ are the stopping time (see below) and mass of each species, respectively.
The number density, $n_\nu$, of each chemical species, $\nu$, is calculated using the chemical reaction networks described in Section \S\ref{subsec:chemicalnetwork}.

Based on the abundance of each species, $\nu$, relative to that of atomic hydrogen (hereafter termed the fractional abundance), $x_\nu$ ($\equiv n_\nu/n_{\rm H}$), the Hall and Pedersen conductivities can be rewritten as 
\begin{equation}
\sigma_H=-\frac{c \ n_{\rm H}}{B}{\displaystyle \sum_\nu} \frac{Q_\nu x_\nu \beta_\nu^2}{1+\beta_\nu^2},
\label{eq:sigmahall}
\end{equation}
and
\begin{equation}
\sigma_P=\frac{c \ n_{\rm H}}{B}{\displaystyle \sum_\nu} \frac{|Q_\nu| x_\nu \beta_\nu}{1+\beta_\nu^2}, 
\label{eq:sigmap}
\end{equation}
where $n_{\rm H}$ is the number density of hydrogen atoms and $x_\nu$ is calculated according to  the procedure described in Section \S\ref{subsec:chemicalnetwork}.

To calculate the Hall coefficient $\eta_H$ or Hall parameter $\beta_{\nu}$, 
one needs to know the stopping time, $\tau_\nu$, in equation (\ref{eq:betanu}).
In a weakly ionized gas, the stopping times for electrons, ions and dust grains are determined 
by collisions with neutral gas particles.
The stopping times for electrons and ions are given by 
\begin{equation}
\tau_{e} = \frac{m_e + m_n}{\langle \sigma \nu \rangle_{e n} \rho_n}, 
\end{equation}
and
\begin{equation}
\tau_{i}=  \frac{m_i + m_n}{\langle \sigma \nu \rangle_{i n} \rho_n}, 
\end{equation}
respectively. Here, $m_n$, $m_e$ and $m_i$ are the masses of neutrals, electrons and 
ions, respectively, $\rho_n$ is the number density of the neutrals, and $\langle \sigma \nu \rangle_{en}$ and $\langle \sigma \nu \rangle_{in}$ are the momentum-transfer rate coefficients 
for electron--neutral and ion--neutral collisions.
According to \citet{1983ApJ...264..485D}, the scattering cross-sections between a neutral and an electron, $\langle \sigma \nu \rangle_{e n}$, and between a neutral and an ion, $\langle \sigma \nu \rangle_{i n}$ are respectively described by 
\begin{equation}
\langle \sigma \nu \rangle_{e n}=1.0 \times 10^{-15} {\rm cm^2}\sqrt{\frac{8 k_B T}{\pi m_e}} ,
\label{eq:electroncollsion}
\end{equation}
and
\begin{equation}
\langle \sigma \nu \rangle_{i n}=1.8 \times 10^{-9} \rm{cm^3 s^{-1}},
\label{eq:ioncollsion}
\end{equation}
where $k_B$ is the Boltzmann constant and $T$ is the gas temperature. 
In equations (\ref{eq:electroncollsion}) and (\ref{eq:ioncollsion}), the suffix $n$ indicates neutral atoms and molecules ($\rm{H,H_2,He}$).
The stopping time for dust grains can be written as 
\begin{equation}
\tau_{{\rm dust}}=\frac{\rho_{\rm d} \, a }{ \rho_n \left[8 k_B T/(\pi m_n) \right]^{1/2}},
\label{eq:collisiondust}
\end{equation}
where $a$ and $\rho_d$ are the radius and total density of the dust grains (i.e., the total mass of the dust per $\rm{cm^{3}}$), respectively.
Equation (\ref{eq:collisiondust}) is the stopping time based on the Epstein drag law. 
Using the process described in \S \ref{subsec:chemicalnetwork}, we calculated $x_{\nu}$ for the density range of $10^4\cm < n_{\rm H} < 10^{14}\cm$, within which the sizes of dust grains are much smaller than the mean free path of the neutrals, meaning that equation (\ref{eq:collisiondust}) can be utilized.
The grain size (distribution) $a$ is discussed in the following section. 

\subsection{Grain Size Distribution}
\label{sec:dustsize}
Dust grains play a significant role in determining the Hall coefficient, $\eta_{\rm Hall}$, because they greatly affect the chemical abundance that in turn governs $\eta_{\rm Hall}$.
Thus, calculation of the Hall coefficient requires the dust grain density and size (distribution) to be ascertained.
Although the true size distribution of dust grains is uncertain, the sizes can be estimated as described in \S\ref{sec:intro}, and so the present work includes the grain size as a parameter. 
In this subsection, we describe the two types of grain size distributions utilized in this study: MRN \citep{1977ApJ...217..425M} and single-sized distributions. 
The MRN grain size distribution can be written as
\begin{equation}              
dn_{\rm{d,tot}}(a)=C a^{-q} da,
\label{eq:mrn}
\end{equation}
where $n_{\rm d, tot}$ and $a$ are the dust grain number density and radius, respectively, $q$ is typically given a value of 3.5 \citep{1977ApJ...217..425M}and $C$ is a normalization factor determined by the total dust grain mass. 
Herein, we define the dust-to-gas mass ratio $f_{\rm dg} = \rho_{\rm d}/\rho_{\rm g}$, where $\rho_{\rm g}$ is the gas density.
Although $f_{\rm dg}=0.01$ is adopted in the present calculations as the fiducial value, the value was varied over the range of \so{$0.005 \le f_{\rm dg} \le 0.0341$} so as to investigate the dependence of the amount of dust grains on the Hall coefficient. 
In the case on an MRN distribution, the dust density can be calculated as
\begin{equation}
\begin{split}
\rho_{\rm d}&=\frac{4}{3} \pi \rho_s \int_{a_{\rm min}}^{a_{\rm max}} a^3\frac{dn_{\rm{d,tot}}}{da}da \\
&=\frac{4}{3} \pi \rho_s C \int_{a_{\rm min}}^{a_{\rm max}} a^3 a^{-q} da \\
&=\frac{4}{3} \pi \rho_s C \frac{1}{4-q} \left(a_{\rm max}^{4-q}-a_{\rm min}^{4-q} \right),
\end{split}
\label{eq:dustdensity}
\end{equation}
where $\rho_s$, $a_{\rm max}$ and $a_{\rm min}$ are the density of a dust grain and the maximum and minimum grain radius, respectively. 
Using equation~(\ref{eq:dustdensity}), the normalization factor $C$ is derived as 
\begin{equation}
C=\frac{3 (4-q) \, f_{\rm dg} \rho_{\rm g}}{4 \pi \rho_s \left(a_{\rm max}^{4-q}-a_{\rm min}^{4-q} \right)}.
\end{equation}
In this work, the dust grain density was defined as $\rho_s=2\,\rm{g\, cm^{-3}}$ and the MRN dust distribution was discretized into ten logarithmically-spaced bins.
The number of partitions was $s_{\rm{max}}$ (having a value of 10 in the present case), where {\it s} (with values from 1 -10) indicates the bin number.  
\so{We define the typical (or averaged) dust grain radius included in the $s$-th bin by dividing the size range ($a_{\rm min} \le a < a_{\rm max}$) on the log scale and the typical radius in the $s$-th bin can be written as} 
\begin{equation}
a_s = a_{\rm min} \left(\frac{a_{\rm max}}{a_{\rm min}} \right)^{\frac{2s-1}{\rm 2 s_{\rm{max}}}}.   
\label{eq:sbinsize}
\end{equation}
In addition, we calculated the typical (or averaged) dust grain cross-section \so{$\sigma_s$} in the $s$-th bin as 
\begin{equation}
\begin{split}
\so{\sigma_s} &=\frac{\int_{a_{s-0.5}}^{a_{s+0.5}} \pi a^2 \frac{dn_{\rm{d,tot}}}{da}da}{\int_{a_{s-0.5}}^{a_{s+0.5}} \frac{dn_{\rm{d,tot}}}{da}da} \\
&=\frac{\int_{a_{s-0.5}}^{a_{s+0.5}} \pi a^2 C a^{-q} da}{\int_{a_{s-0.5}}^{a_{s+0.5}} C a^{-q} da} \\
&=\pi \frac{1-q}{3-q} \frac{a_{s+0.5}^{3-q}-a_{s-0.5}^{3-q}}{a_{s+0.5}^{1-q}-a_{s-0.5}^{1-q}},
\end{split} 
\end{equation}
\so{in order to define the collisional rate between dust grains and charged particles (see Appendix Appendix \S\ref{sec:reactioneq})} where $a_{s+0.5}$ and $a_{s-0.5}$ are the maximum and minimum dust grain radius values in the $s$-th bin, respectively.
The number density of the dust grains in the $s$-th bin is also calculated as 
\begin{equation}
\label{eq:initialdustabunMRN}
\begin{split}
n_{{\rm d},s} &= \int_{a_{s-0.5}}^{a_{s+0.5}} \frac{dn_{\rm{d,tot}}}{da}da \\
&=\frac{C}{1-q} \left(a_{s+0.5}^{1-q}-a_{s-0.5}^{1-q}  \right),
\end{split}
\end{equation}
using $n_{{\rm d},s}$ as the initial abundance of dust grains in each bin.

In addition to the MRN distribution calculations, we also constructed single-sized dust grain models. 
To make the model easier to understand and analogous to the MRN distribution, we defined the dust radius for single-sized grains as $a_{\rm single}$ as $a_{\rm single} = a_{\rm max}=a_{\rm min}$ and set $s_{\rm{max}}$=1. 
Based on this treatment, the collision cross-section, $\sigma_{\rm s}$, dust number density, $n_{\rm d}$, and total dust density, $\rho_{\rm d}$, were, respectively, described as 
\begin{equation}
\sigma_{\rm s}=\pi \so{a_{\rm single}}^2, 
\end{equation}
\begin{equation}
\label{eq:initialdustabunsingle}
n_{\rm d} = \frac{3 \rho_{\rm g}}{4 \pi \so{a_{\rm single}}^3 \rho_s} f_{\rm dg},
\end{equation}
and
\begin{equation}
\rho_{\rm d} = f_{\rm dg} \rho_{\rm g} = 1.4 f_{\rm dg} m_n n_{\rm H}.
\end{equation}
The factor of 1.4 is derived from the assumption of the neutral gas component $n_{\rm He} = 0.2 \ n_{\rm H_2}$ in neutral gas.
Substituting $\rho_{\rm d}$ and $a_s$ into equation~(\ref{eq:collisiondust}) allows the collisional timescale between neutrals and dust grains, $\tau_{\nu, {\rm dust}}$, to be calculated.
We employed $n_{\rm d}$ as the initial abundance of dust grains in the single-sized dust models while, in the MRN distribution model, we calculated $\tau_{\nu, {\rm dust}}$ for each bin (that is, the $s=1-10$\,th bins).

In this study, we considered four MRN distribution models having different values of $a_{\rm min}$ and $a_{\rm max}$ along with
six single-sized grain models. 
The minimum and maximum radii in each MRN dust grain model (s1a, s1b, s1c and s2) are provided in Table~\ref{tab:dust}, along with the grain radii for the single-sized models (s3 -- s8). 
The dust-to-gas ratio and total dust grains surface area for each model are presented in the sixth and seventh 
columns of the table, respectively. 
  
\begin{table}
\caption{Dust grain model parameters.}
  \begin{center}
    \begin{tabular}{l}
          \begin{tabular}{c|c|c|c|c|c|c}
          \hline
         Model & $a_{\rm min}$ [$\rm{\mu m}$] & $a_{\rm single}$ [$\rm{\mu m}$] & $a_{\rm max}$ [$\rm{\mu m}$] & $s_{\rm{max}}$ &  $f_{\rm dg}$ & $\sigma_{\rm tot}/n_{\rm{H}}$ [$\rm{cm^5}$]  \\ \hline \hline
            s1a & 0.005 & & 0.25 & 10 & 0.01 & $2.83052 \times 10^{-21}$ \\ \hline
            s1b & 0.005 & & 0.25 & 10 & 0.005 & $1.41526 \times 10^{-21}$ \\ \hline
            s1c & 0.005 & & 0.25 & 10 & 0.02 & $5.66105 \times 10^{-21}$ \\ \hline
            s2 & 0.0181 & & 0.9049 & 10 & 0.0341 & $2.66649 \times 10^{-21}$ \\ \hline
            s3 & & 0.3 & & 1 & 0.01 & $5.83531 \times 10^{-22}$ \\ \hline
            s4 & & 0.1 &  & 1 & 0.01 & $1.74486 \times 10^{-21}$ \\ \hline
            s5 & & 0.075 & & 1 & 0.01 & $2.33644 \times 10^{-21}$ \\ \hline
            s6 & & 0.05 & & 1 & 0.01 & $3.5035 \times 10^{-21}$ \\ \hline
            s7 & & 0.025 & & 1 & 0.01 & $7.00006 \times 10^{-21}$ \\ \hline
            s8 & & 0.01 & & 1 & 0.01 & $1.74486 \times 10^{-20}$ \\ \hline
          \end{tabular}
    \end{tabular}
     \label{tab:dust}
  \end{center}
\end{table}

\subsection{Chemical Reactions and Networks} 
\label{subsec:chemicalnetwork}
Calculation of the Hall coefficient also required determination of the fractional abundance of each charged species, $x_\nu$ (see equations (\ref{eq:sigmahall}) and (\ref{eq:sigmap})).  
In this subsection, we describe the method by which the chemical abundance of each charged particle was determined, based on a series of papers by \citet{1990MNRAS.243..103U}. 
In these calculations, the relative abundance of each neutral species (H$_2$, He, CO, O$_2$, O and Mg) relative to the number of hydrogen atoms was fixed as shown in Table~\ref{tab:neu}.
\begin{table}
\caption{
Abundance of each neutral species relative to the number of hydrogen atoms \citep{2000ApJ...543..486S}. 
$\delta$ indicates the fraction of each element remaining in the gas phase.
The fractional abundances of molecules, $\delta {\rm m}$, and of metal ions, $\delta {\rm M}$, in the interstellar gas phase are $\delta {\rm m}=0.2$ and 
$\delta {\rm M}= 0.02$, respectively, while the abundance of oxygen molecules in the gas phase is $\delta {\rm O}_2=0.7$ \citep{1974ApJ...193L..35M}.
}       
\begin{tabular}{cc}
\hline
Chemical Species & Relative Abundances  \\ \hline
            $\rm{H_2}$ & x$\rm{H_2} = 0.5$  \\
            He & xHe = $9.75 \times 10^{-2}$  \\
            CO & xCO = $3.62 \times 10^{-4} \ \delta {\rm m}$  \\
            $\rm{O_2} $ & x$\rm{O_2}$ = $0.5 \ \delta {\rm O}_2 \ (8.53-3.62) \times 10^{-4} \delta {\rm m}$  \\
            O &  $\rm{xO} =  (1-\delta {\rm O}_2) (8.53-3.62) \times 10^{-4} \delta {\rm m}$  \\
            Mg & xMg = $ 7.97 \times 10^{-5} \delta {\rm M}$  \\
            \hline
          \end{tabular}
\label{tab:neu}
\end{table}
As described in \S\ref{methodetaH}, $\rm{{H_3}^+,m^+,Mg^+,He^+,C^+ and \ H^+}$ were considered as the charged species in addition to the charged dust grains (see below), where m$^+$ represents molecular ions other than ${\rm H_3^+}$.
Because the most abundant charged molecular ion species is HCO$^+$, we identified m$^+$ with HCO$^+$ \citep{2016A&A...592A..18M}. 
A total of 25 chemical reactions was considered.

The four types of charged particle reactions incorporated in the calculations were
\begin{enumerate}
\renewcommand{\labelenumi}{(\arabic{enumi})}
\item $\rm{j \to i+e^{-}}$ \ \ (ionization of neutral particles),
\item $\rm{i + e^{-} \to j}$ \ \ (recombination of positive ions and electrons),
\item $\rm{i + k \to j+ \ast}$ (recombination of charged ions), and
\item $\rm{G({\it q})+ i/e^- \ ({\rm ion \ or \ electron}) \to G({\it q} \pm 1)}$ (absorption of charged particles onto dust grains following collisions).
\end{enumerate}
Here, i indicates a charged species, j and k represent different neutral species, * is the product of a charged species and G($q$) is dust having charge $q$.
In the following discussions, for convenience, we assign a number to each charged particle (ion or electron), as shown in Table~\ref{tab:num}.

In the case of reaction (1), cosmic rays represent the ionization source.
Here, the parameter $\zeta$ is applied, equal to the total ionization rate of a hydrogen molecule, and we set the reaction rate of $\rm{H_2}$ and He as shown in Table 2 in \citet{1990MNRAS.243..103U}.
The cosmic ray rates employed in this study are provided in Table~\ref{tab:cr}, with $\zeta=10^{-17}$\,s$^{-1}$ as the fiducial value.

The chemical reactions and the associated reaction rates in points (2) and (3), above, are taken from the UMIST database \citep{2013A&A...550A..36M} and are summarized in Table~\ref{tab:rea} in Appendix \S\ref{sec:reactinrate}.
\begin{table*}
\begin{center}
\caption{Chemical species considered in this study and assigned numbers.}
\renewcommand{\arraystretch}{1.2}
\begin{tabular}{|c|c|}
\hline
species & assigned number i \\
\hline
 e$^-$ & 0 \\
 H$_3^+$ & 1\\
 m$^+$ & 2 \\
 Mg$^+$ & 3 \\
 He$^+$ & 4 \\
 C$^+$ & 5 \\
 H$^+$ & 6 \\
\hline
\end{tabular}
\label{tab:num}
\end{center}
\end{table*} 

Since dust grains absorb charged particles, they play a significant role in determining the fractional abundances of such particles.
In addition, because the dust grains themselves are charged, they also contribute to the Hall coefficient.  
In Appendix \S\ref{secdust}, we discuss reaction (4), which is based on collisions between dust grains and charged particles with subsequent absorption of the charged particles on the grain surfaces. 

\begin{table*}
\caption{Cosmic ray rates employed in this study.}
    \begin{tabular}{l}
          \begin{tabular}{c|c|c|c|c|c|c}
	  \hline
            & & &  $\zeta$ \ [ s$^{-1}$ ] & \\ 
	  \hline
            & 10$^{-18}$ &10$^{-17.5}$ &10$^{-17}$ &10$^{-16.5}$ &10$^{-16}$  \\
          \hline
          \end{tabular}
    \end{tabular}
\label{tab:cr}
\end{table*}

At this point, we explain the calculations used to derive the abundances of charged particles and dust grains. 
For the sake of convenience, these calculations are based on the abundance of charged particles and dust grains relative to the number of hydrogen atoms, and so the abundances of each charged species and dust grain is described as $x_i=n_i/n_{\rm H}$ and $x_{\rm d}=n_{\rm d}/n_{\rm H}$.
In this process, the maximum and minimum charges of the dust grains at each gas density are determined \citep[for details, see][]{2009ApJ...698.1122O}, and the time-derivative equation for each species is provided in Appendix \S\ref{sec:reactioneq}.
The fractional abundance of each charged particle and dust grain as a function of the gas density was calculated according to the procedure:
\begin{enumerate}
\renewcommand{\labelenumi}{(\alph{enumi})}
\item beginning from the initial chemical abundance values in Table~\ref{tab:neu}, the equations given in Appendix \S\ref{sec:reactioneq} were implicitly solved for a fixed gas density \so{and gas temperature (see \S\ref{sec:magtemp})} until an equilibrium state was realized, and the initial dust grains abundance values were pre-calculated for each dust distribution model (using equation (\ref{eq:initialdustabunMRN}) for MRN models and equation (\ref{eq:initialdustabunsingle}) for single-sized models), following which
\item the gas density was increased by 0.1 dex and step (a) was repeated.
\end{enumerate}
Using this procedure, the fractional abundance of each species was derived over the range of $10^4\cm < n_{\rm H} < 10^{14}\cm$.
It should be noted that we confirmed that the elapsed time required to obtain an equilibrium state was much shorter than the freefall time scale $t_{\rm ff}$ (=$(3\pi/(32\,G\,\rho))^{1/2}$) in the density range considered in this study. 
\so{We compare  the elapsed timescale and freefall timescale and show that equilibrium time is much shorter than freefall time in Appendix \S\ref{sec;quilibriumtime}.}  
Finally, using the abundance of each species and the equations described in \S\ref{methodetaH}, we derived the Hall coefficient, $\eta_{\rm H}$, for each gas density.
Employing ten different dust grain models (Table~\ref{tab:dust}) and five different cosmic ray rates (Table~\ref{tab:cr}), we calculated the chemical abundance values and the Hall coefficients for 50 models in total. 

\subsection{Gas Temperature and Magnetic Field Strength}
\label{sec:magtemp}
When solving the chemical network for a given gas density, it is necessary to assume the gas temperature.
According to \citet{1969MNRAS.145..271L}, \citet{2000ApJ...531..350M} and \citet{2013ApJ...763....6T}, the gas temperature can be obtained using the barotropic equation of state
\begin{equation}
T = 10 \left[1+\gamma \left(\dfrac{n_{\rm H}}{10^{11}\cm}\right)^{\gamma-1} \right] \rm K,
\label{eqn:temperature}
\end{equation}
where $n_{\rm H}$ and $\gamma$ (=5/3) are the number density and specific heat ratio of hydrogen.
Because the Hall coefficient depends on the magnetic field strength, $B$, as discussed in Section \S\ref{sec:hallcoefficient}, we also need to assume the magnetic field strength.
In the present work, the plasma beta, $\beta$, was employed as an index of the magnetic field, defined as 
\begin{equation}
\beta  \equiv  \frac{P_{\rm th}}{P_{\rm mag}}, 
\end{equation}
where the thermal $P_{\rm th}$ and magnetic $P_{\rm mag}$ pressures are respectively written as 
\begin{equation}
P_{\rm th} =  \frac{\rho_{\rm g} \, k_B T}{\mu m_{\rm H}}
\label{eq:pth}
\end{equation}
and
\begin{equation}
P_{\rm mag} = {\frac{B^2}{8 \pi}}.
\label{eq:pmag}
\end{equation}
Here, $\rho_{\rm g}$, $k_B$, $\mu$ and $m_{\rm H}$ are the fluid density, Boltzmann constant, mean molecular weight and proton mass, respectively.  
In this study, a constant mean molecular weight of $\mu=2.4$ was adopted because the work focused solely on the low-temperature region of the collapsing cloud (see Section \S\ref{subsec:Hallvelocity}).
Based on equations~(\ref{eq:pth}) and (\ref{eq:pmag}), the magnetic field strength can be written as 
\begin{equation}
B = \sqrt{\dfrac{8\pi \rho_{\rm g} k_{\rm B} T }{\mu m_{\rm H} \beta} }.
\label{eq:magstrength}
\end{equation} 
The gas density is related to the number density according to the relationship $\rho_{\rm g} = 1.4 n_{\rm H} m_{\rm H}$.
Thus, the gas temperature is a function of the (number) density (eq.~\ref{eqn:temperature}) and the magnetic field strength is a function of both the number density, $n_{\rm H}$, and the plasma beta, $\beta$. 
Because the plasma beta is not uniquely determined, we referred to recent publications regarding non-ideal MHD simulations \citep[e.g.][]{2018MNRAS.476.2063W}, and on this basis adopted $\beta=100$ as the fiducial value.
Note that the effect of varying $\beta$ on the Hall coefficient is discussed in Section \S\ref{subsec:magneticfielddependence}.

\begin{figure}
	\includegraphics[width=7.5cm]{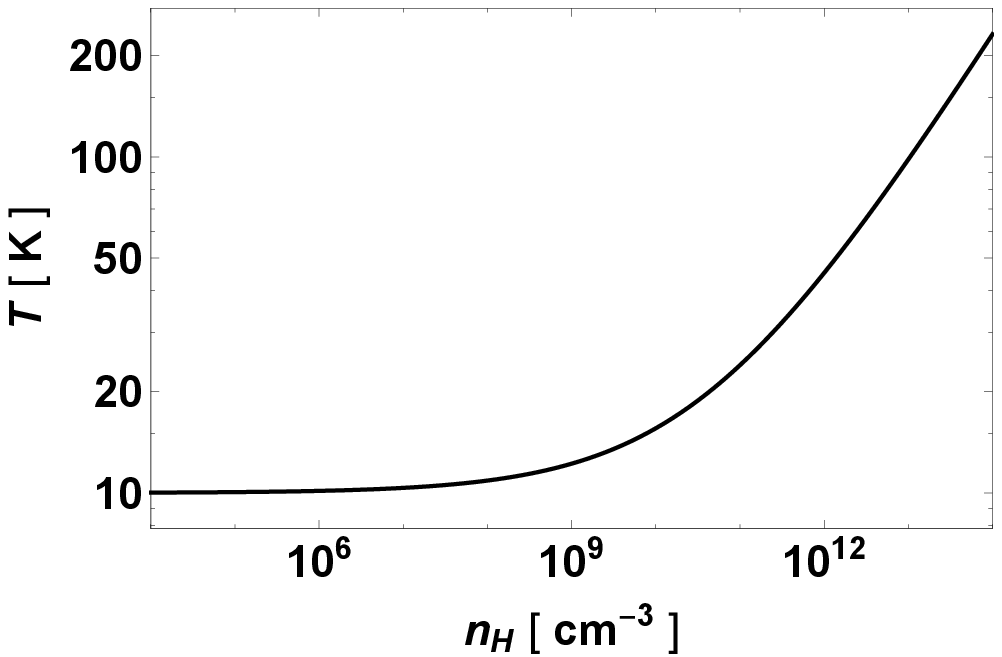}
	\includegraphics[width=7.5cm]{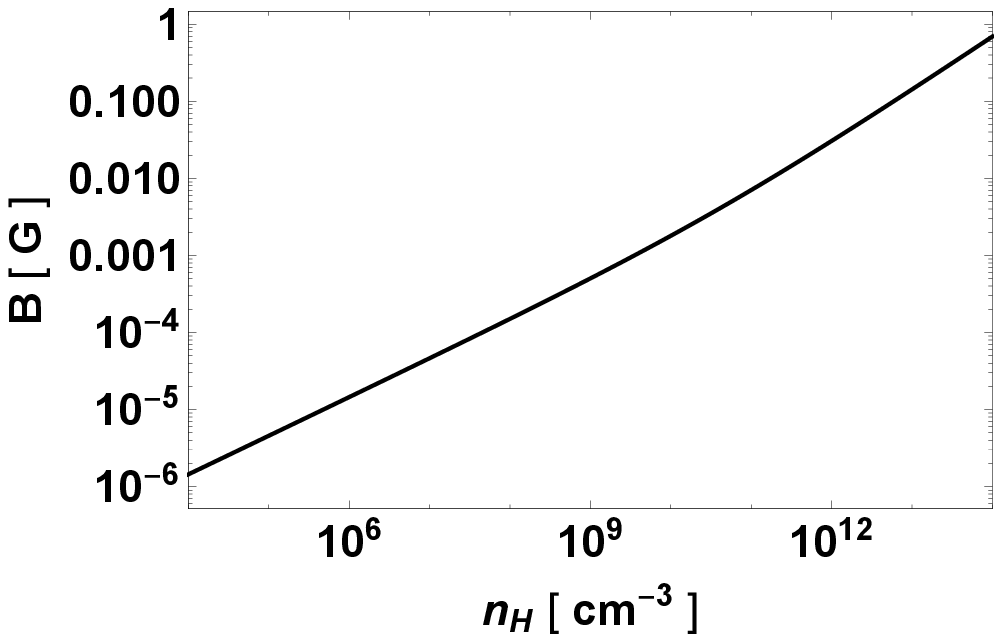}
    \caption{Gas temperature (left) and magnetic field strength (right) as functions of number density.}
    \label{fig:tempmag}
\end{figure}

Fig~\ref{fig:tempmag} plots the gas temperatures (left) derived from equation~(\ref{eqn:temperature}) and the magnetic field strengths (right) derived from equation~(\ref{eq:magstrength}) with $\beta=100$ against the number density. 
These plots are in good agreement with recently published one- and multi-dimensional simulations \citep{2015MNRAS.452..278T,2015ApJ...801..117T}. 

\section{Results of Chemical Reaction Calculations}
\label{sec:chemicalresult}
\begin{figure*}
\begin{tabular}{ccc}
\begin{minipage}{0.36\hsize}
\includegraphics[width=\columnwidth]{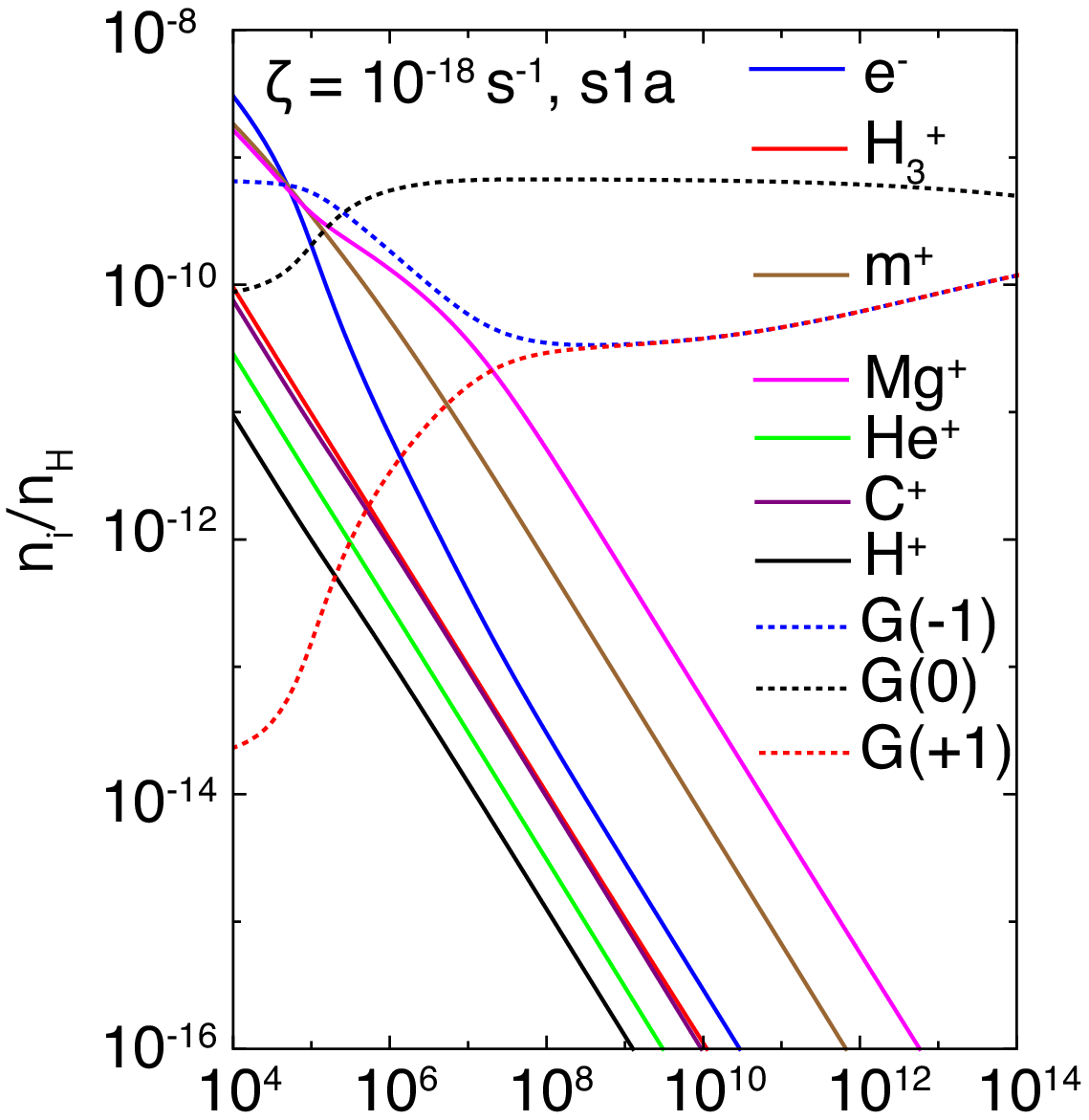}
\end{minipage}
\hspace{-5pt}
\begin{minipage}{0.33\hsize}
\includegraphics[width=\columnwidth]{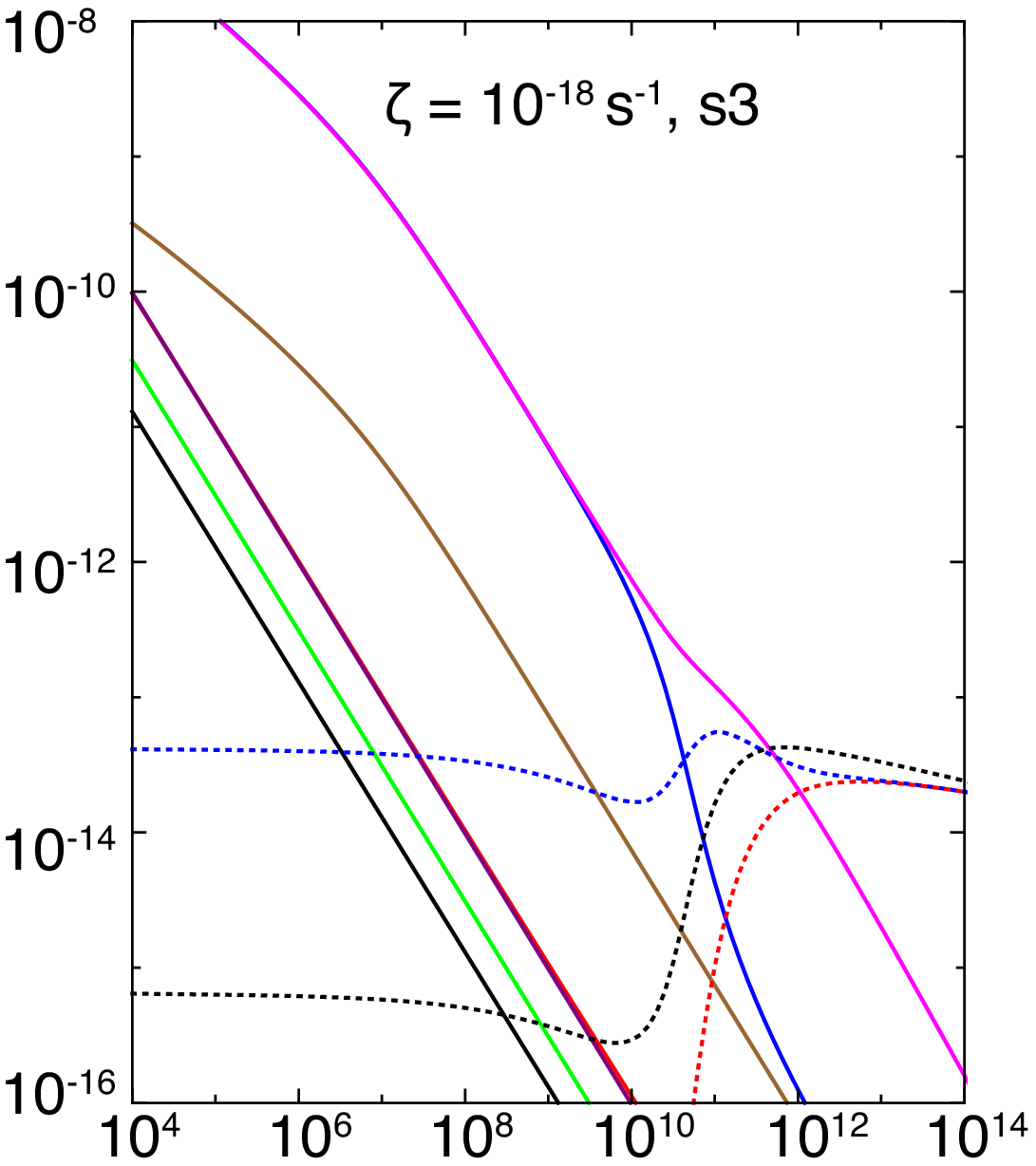}
\end{minipage}
\begin{minipage}{0.33\hsize}
\includegraphics[width=\columnwidth]{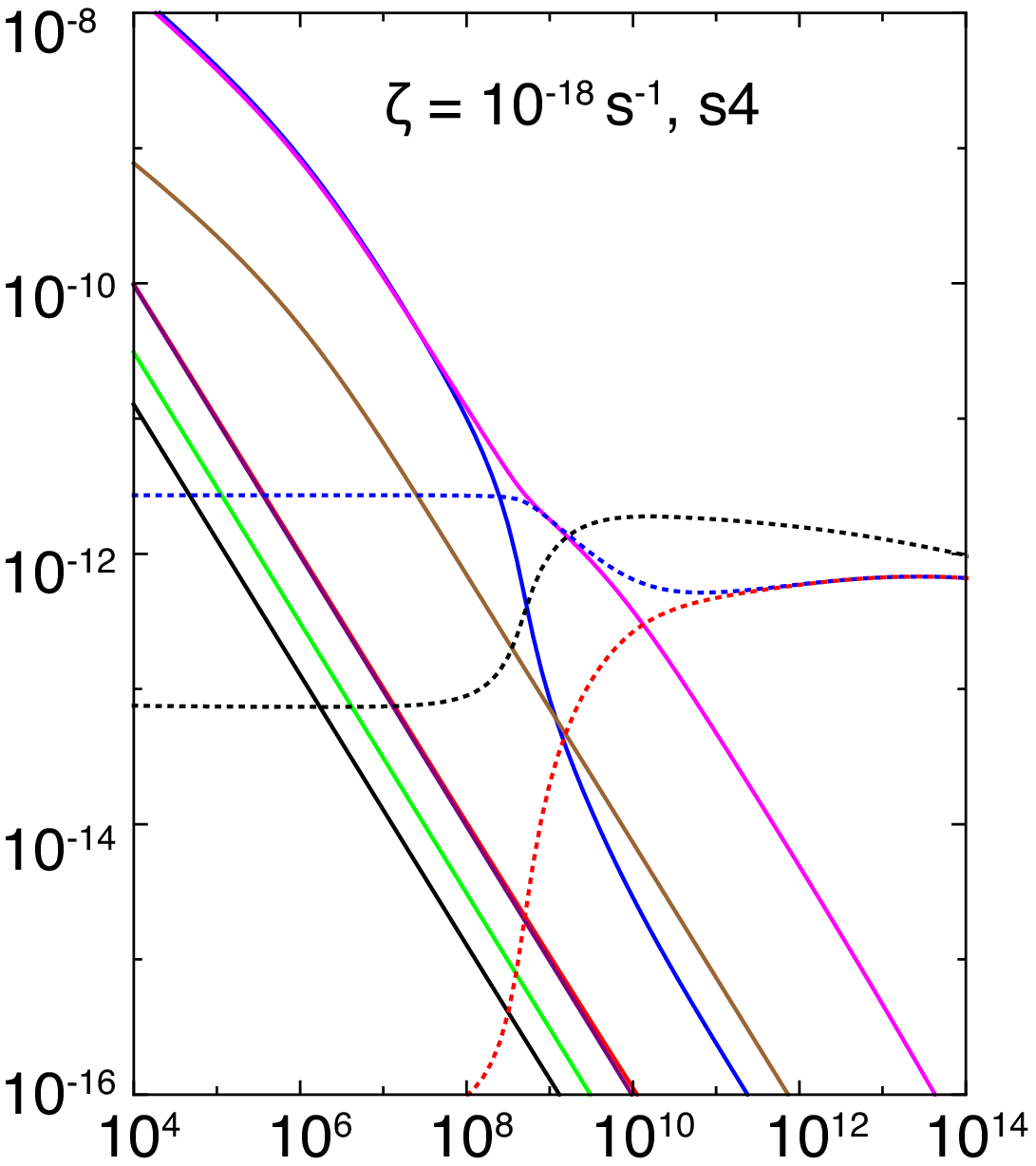}
\end{minipage}
\hspace{-5pt}
\\
\\
\begin{minipage}{0.36\hsize}
\includegraphics[width=\columnwidth]{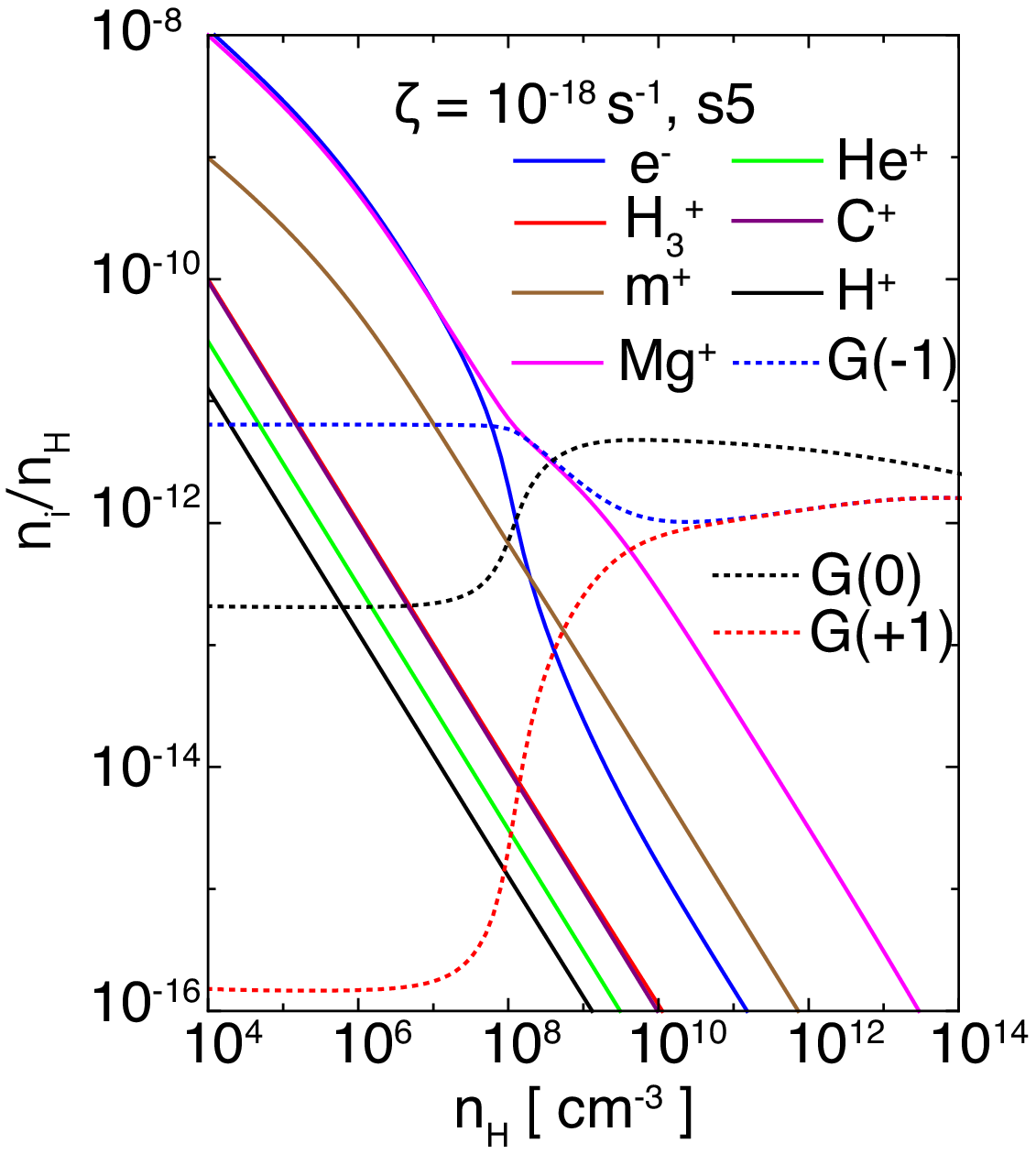}
\end{minipage}
\hspace{-5pt}
\begin{minipage}{0.33\hsize}
\includegraphics[width=\columnwidth]{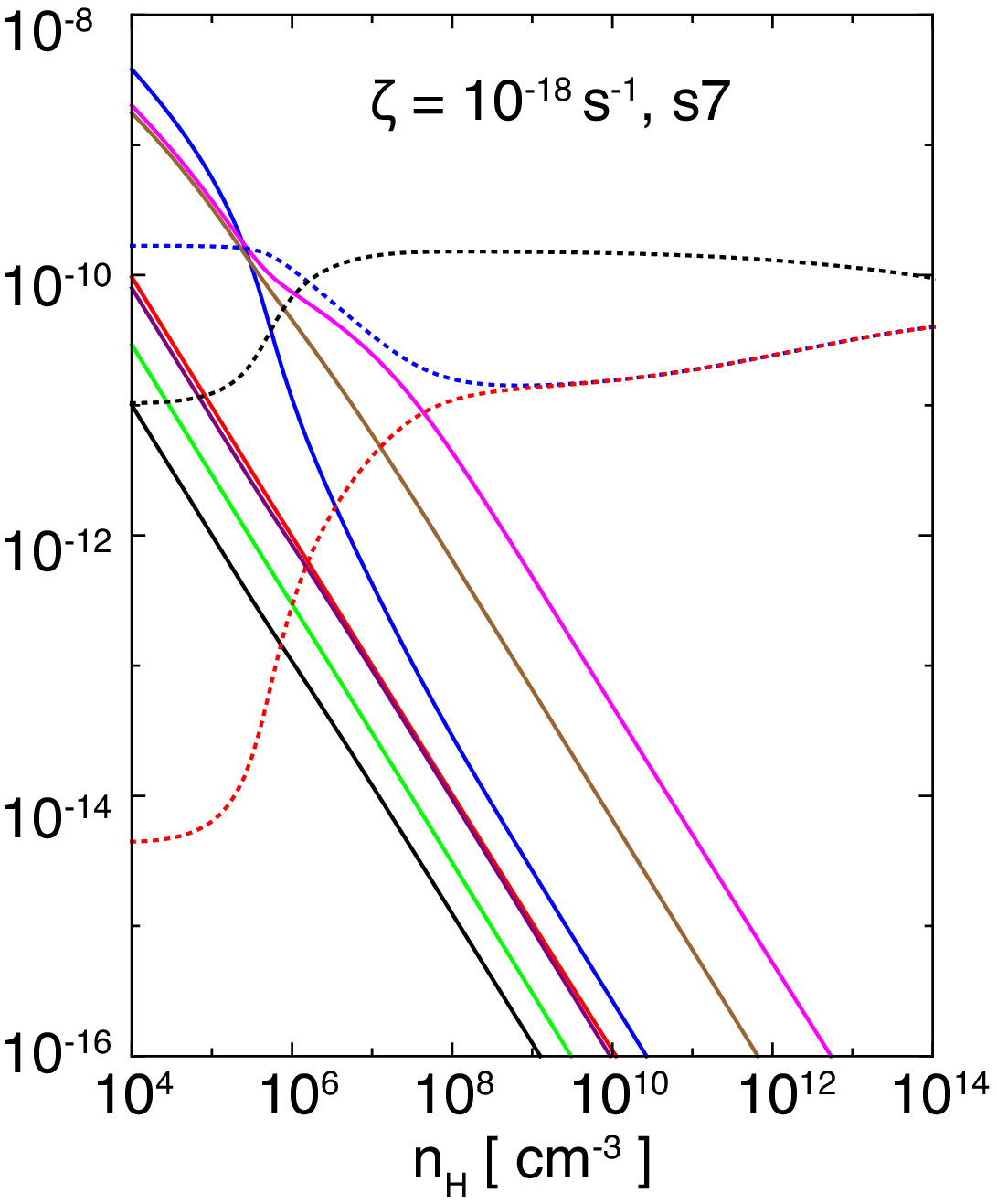}
\end{minipage}
\hspace{-5pt}
\begin{minipage}{0.33\hsize}
\includegraphics[width=\columnwidth]{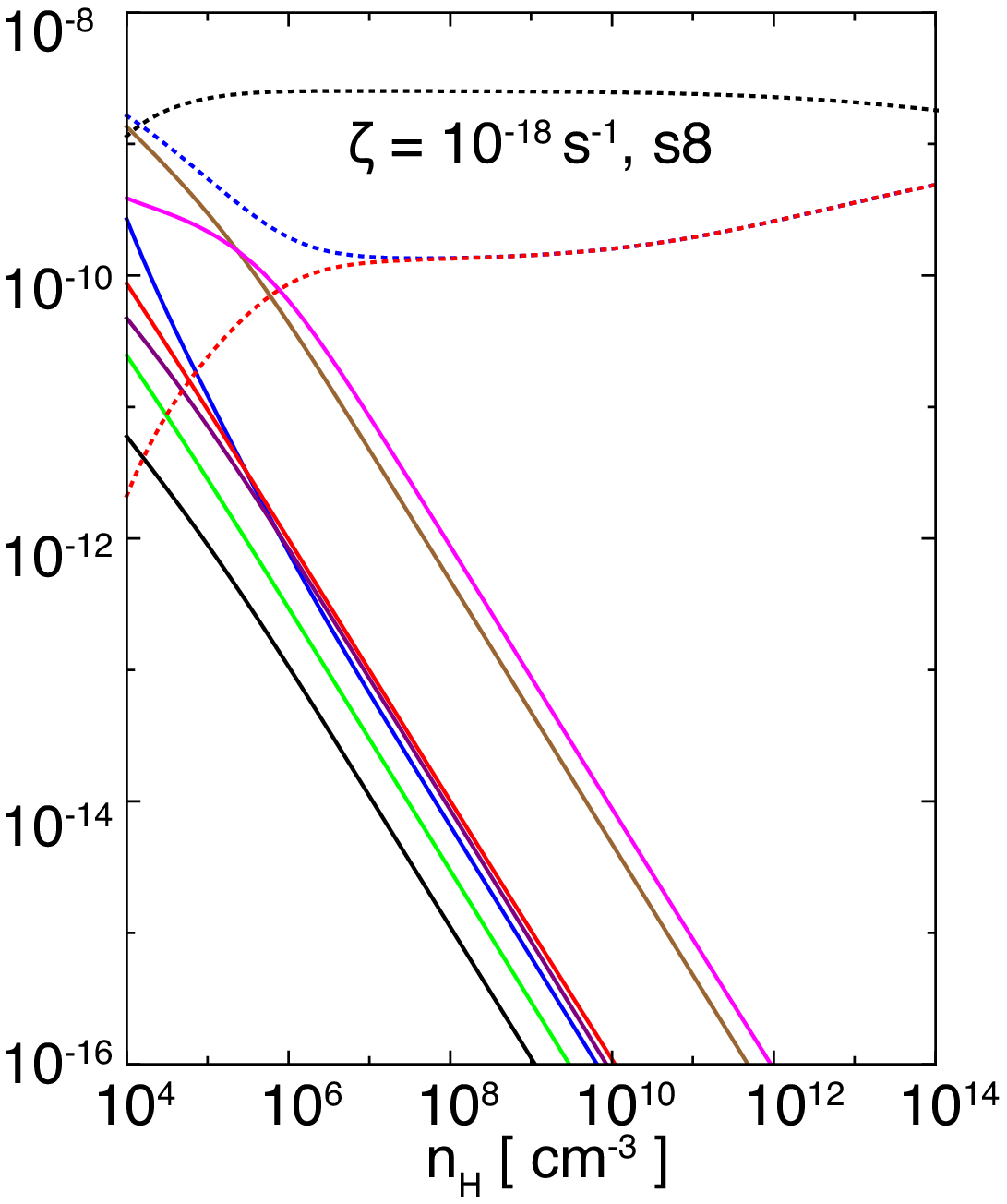}
\end{minipage}
\end{tabular}
\caption{Chemical abundances of charged species, neutral species and charged dust grains ($Z=\pm$1) for different dust models (s1a, s3, s4, s5, s7 and s8), obtained using a constant ionization rate of $\zeta=10^{-18}$\,s$^{-1}$, as functions of the number density.}
\label{fig:ca18}
\end{figure*}

\begin{figure*}
\begin{tabular}{ccc}
\begin{minipage}{0.36\hsize}
\includegraphics[width=\columnwidth]{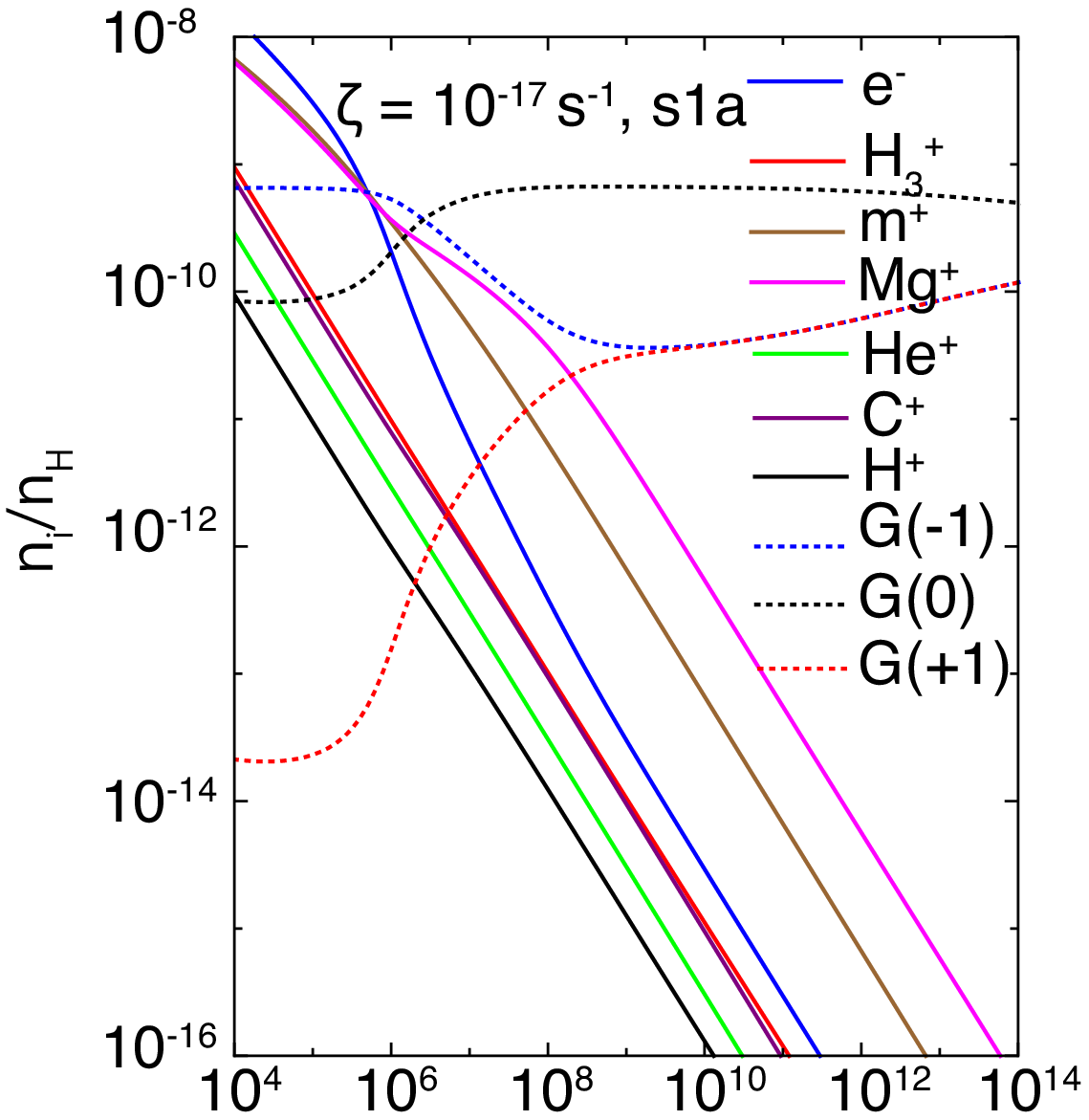}
\end{minipage}
\hspace{-5pt}
\begin{minipage}{0.33\hsize}
\includegraphics[width=\columnwidth]{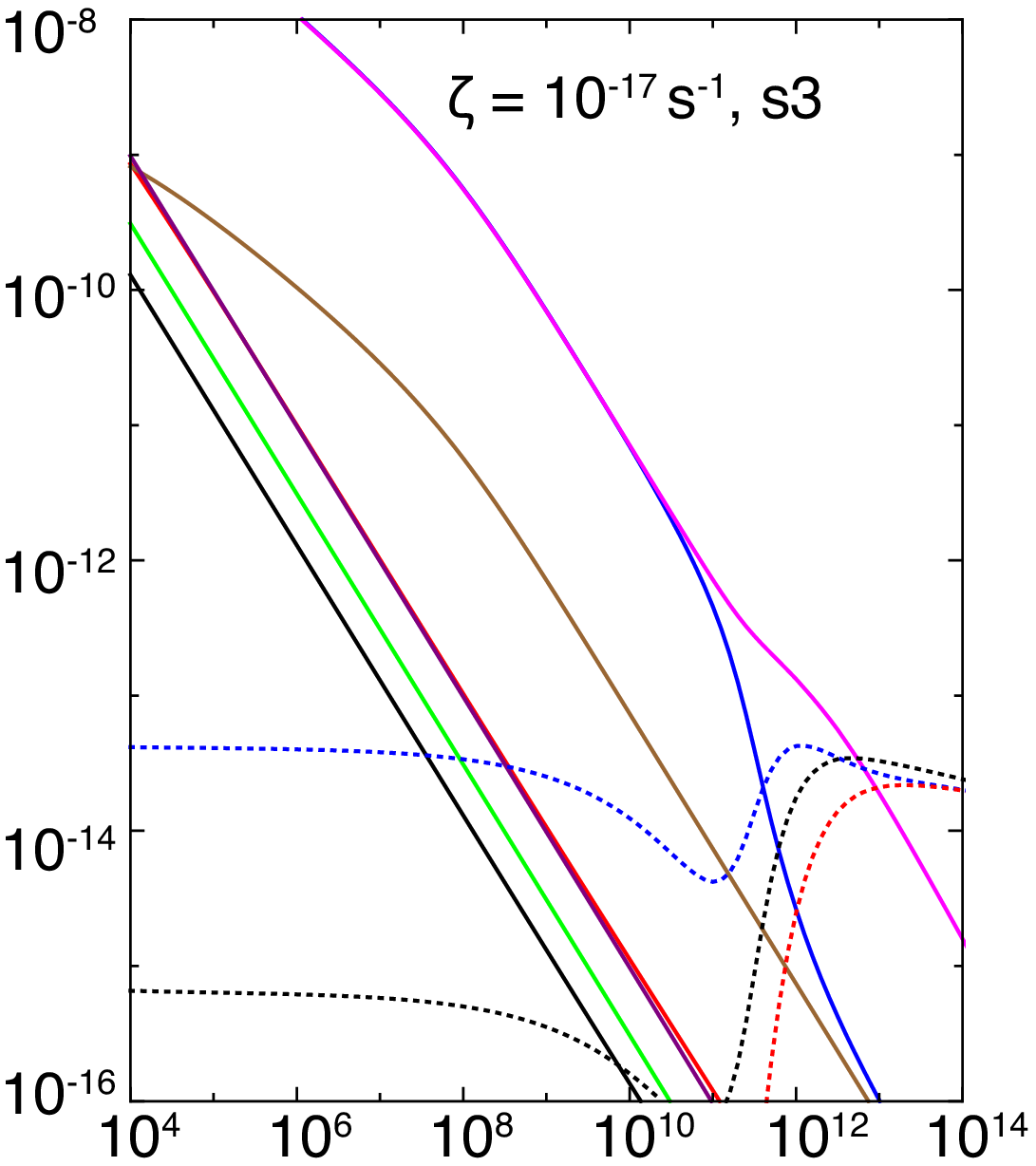}
\end{minipage}
\begin{minipage}{0.33\hsize}
\includegraphics[width=\columnwidth]{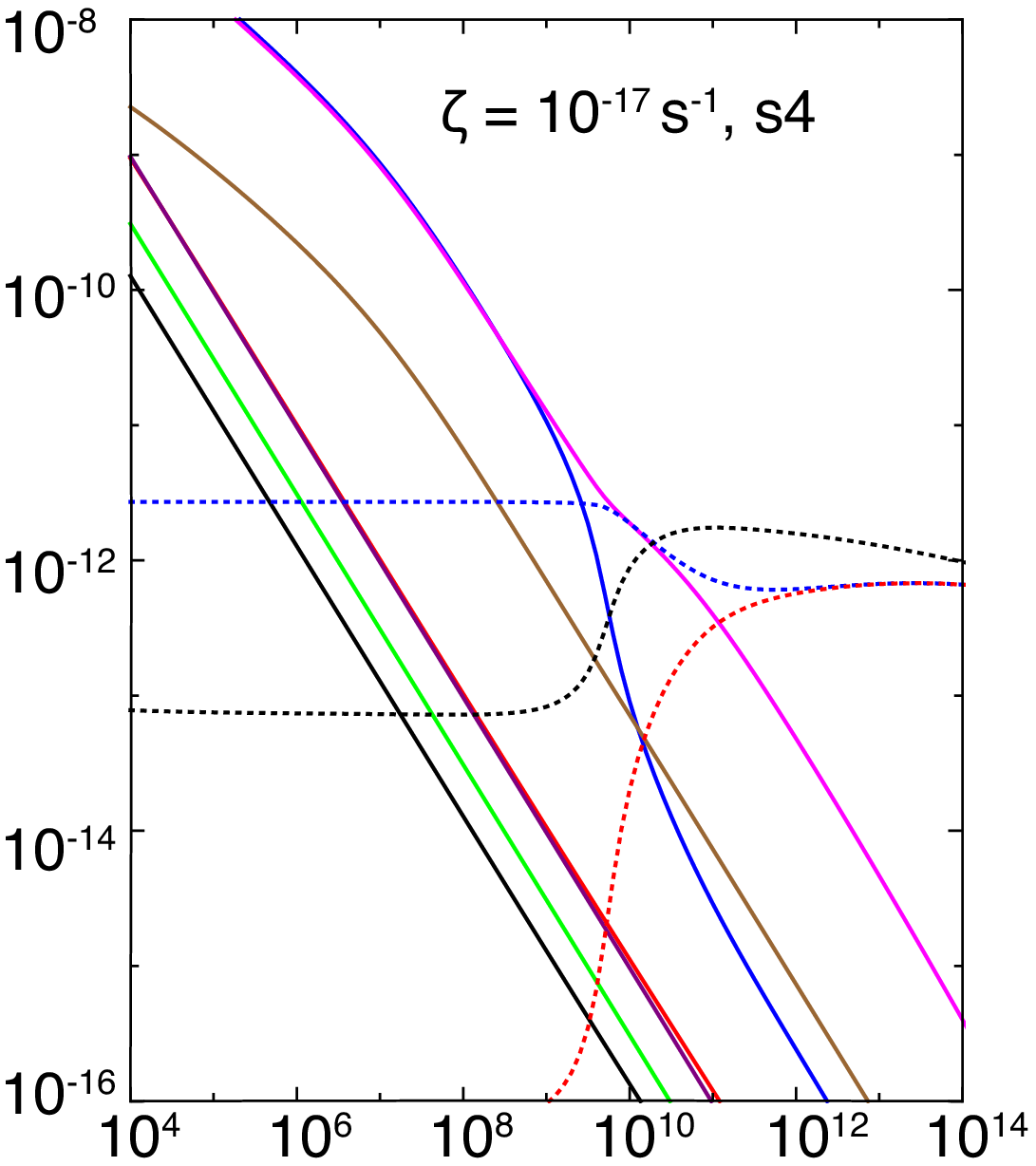}
\end{minipage}
\hspace{-5pt}
\\
\\
\begin{minipage}{0.36\hsize}
\includegraphics[width=\columnwidth]{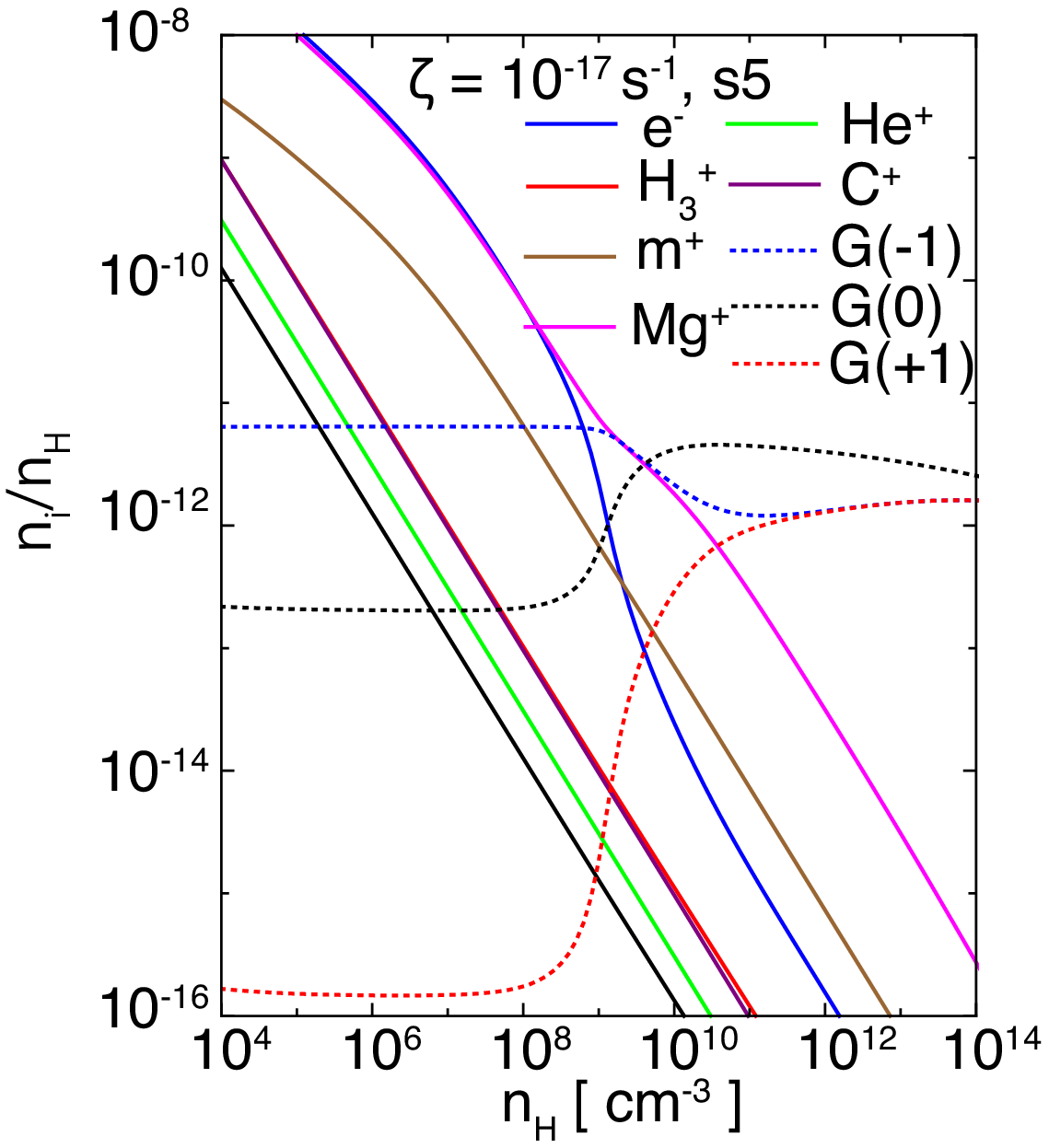}
\end{minipage}
\hspace{-5pt}
\begin{minipage}{0.33\hsize}
\includegraphics[width=\columnwidth]{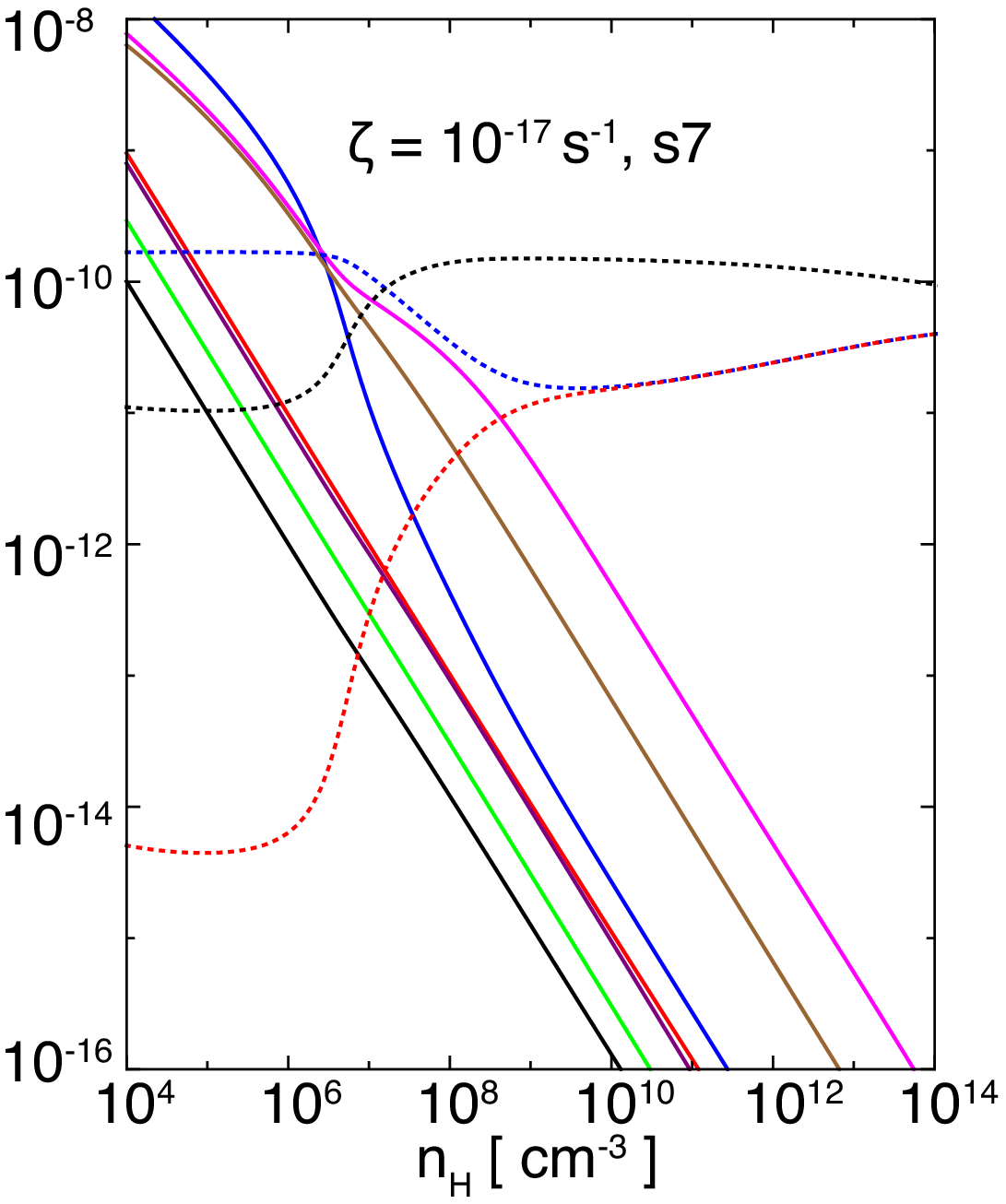}
\end{minipage}
\hspace{-5pt}
\begin{minipage}{0.33\hsize}
\includegraphics[width=\columnwidth]{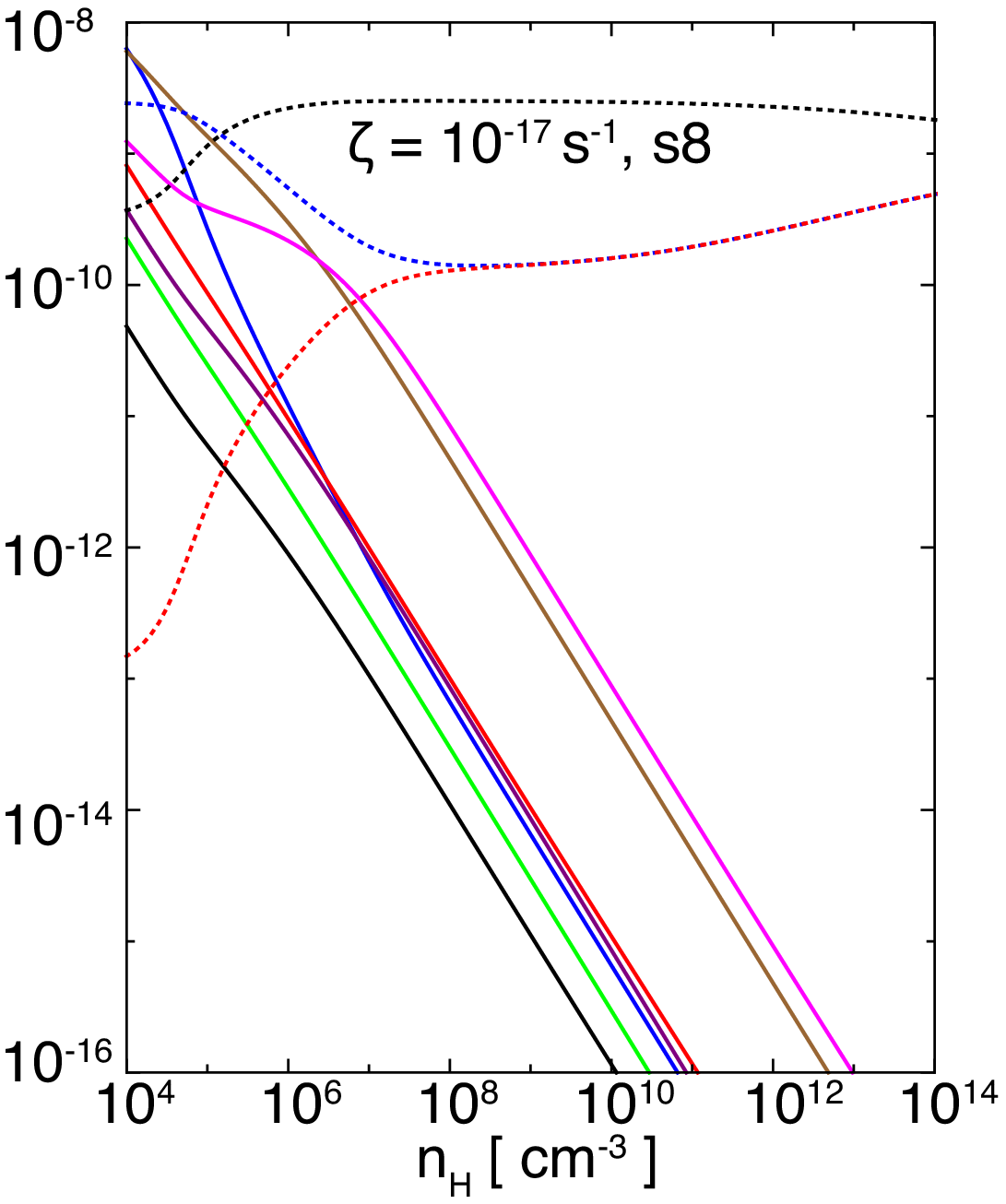}
\end{minipage}
\end{tabular}\caption{The same data as in Fig.~\ref{fig:ca18} but for $\zeta=10^{-17} \rm{s^{-1}}$ using models (s1a, s3, s4, s5, s7 and s8).}
\label{fig:ca17}
\end{figure*}

\begin{figure*}
\begin{tabular}{ccc}
\begin{minipage}{0.36\hsize}
\includegraphics[width=\columnwidth]{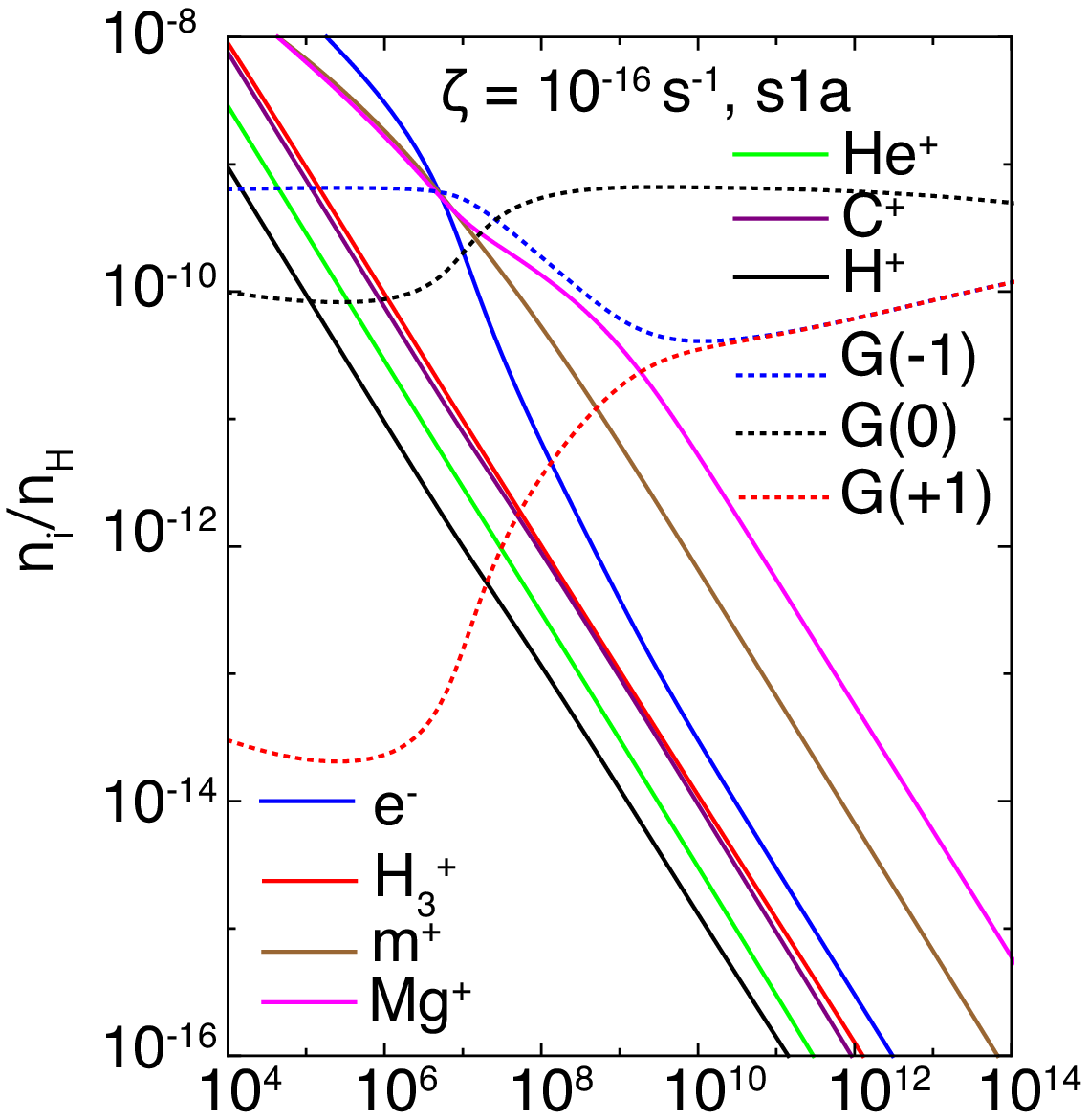}
\end{minipage}
\hspace{-5pt}
\begin{minipage}{0.33\hsize}
\includegraphics[width=\columnwidth]{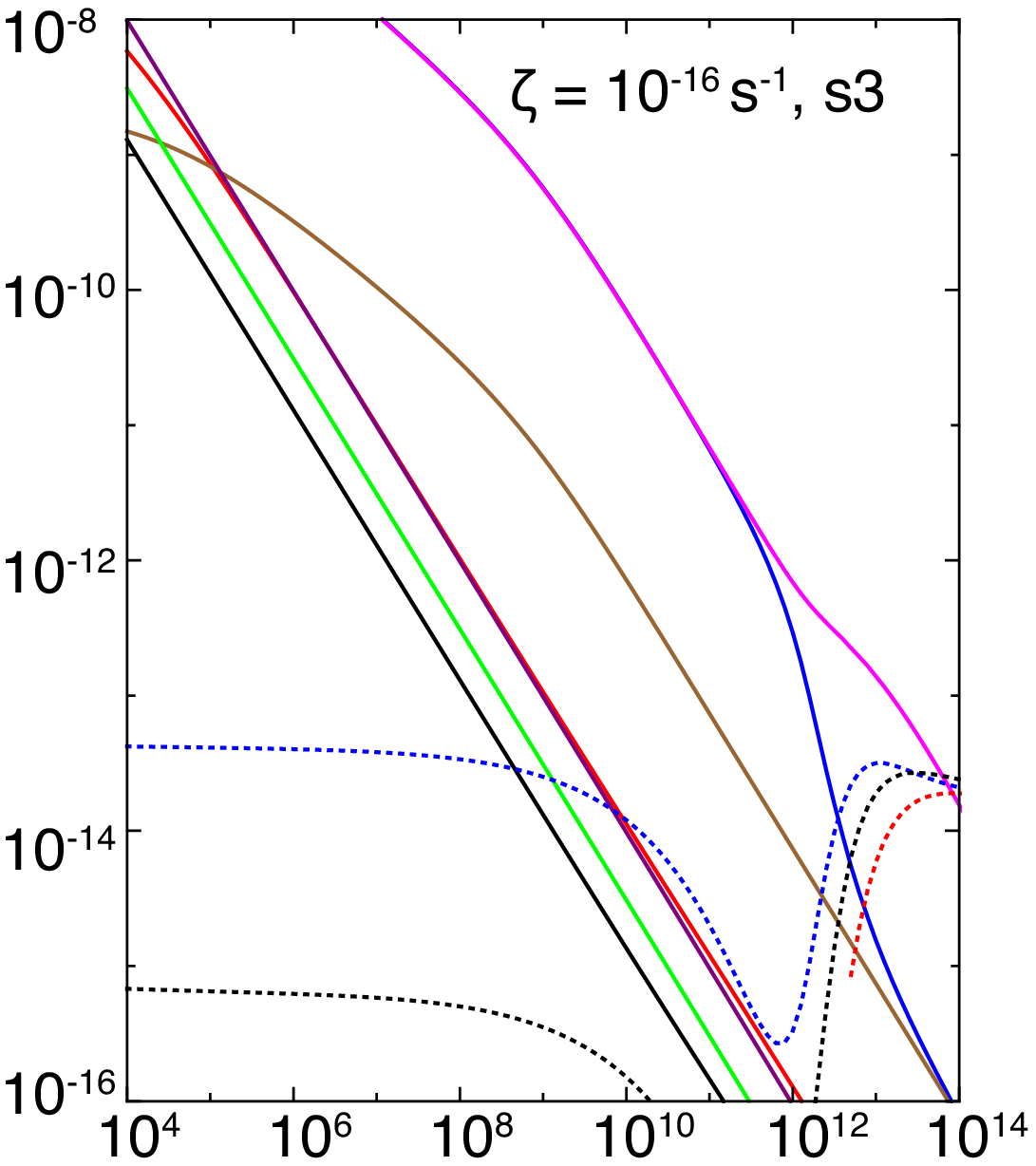}
\end{minipage}
\begin{minipage}{0.33\hsize}
\includegraphics[width=\columnwidth]{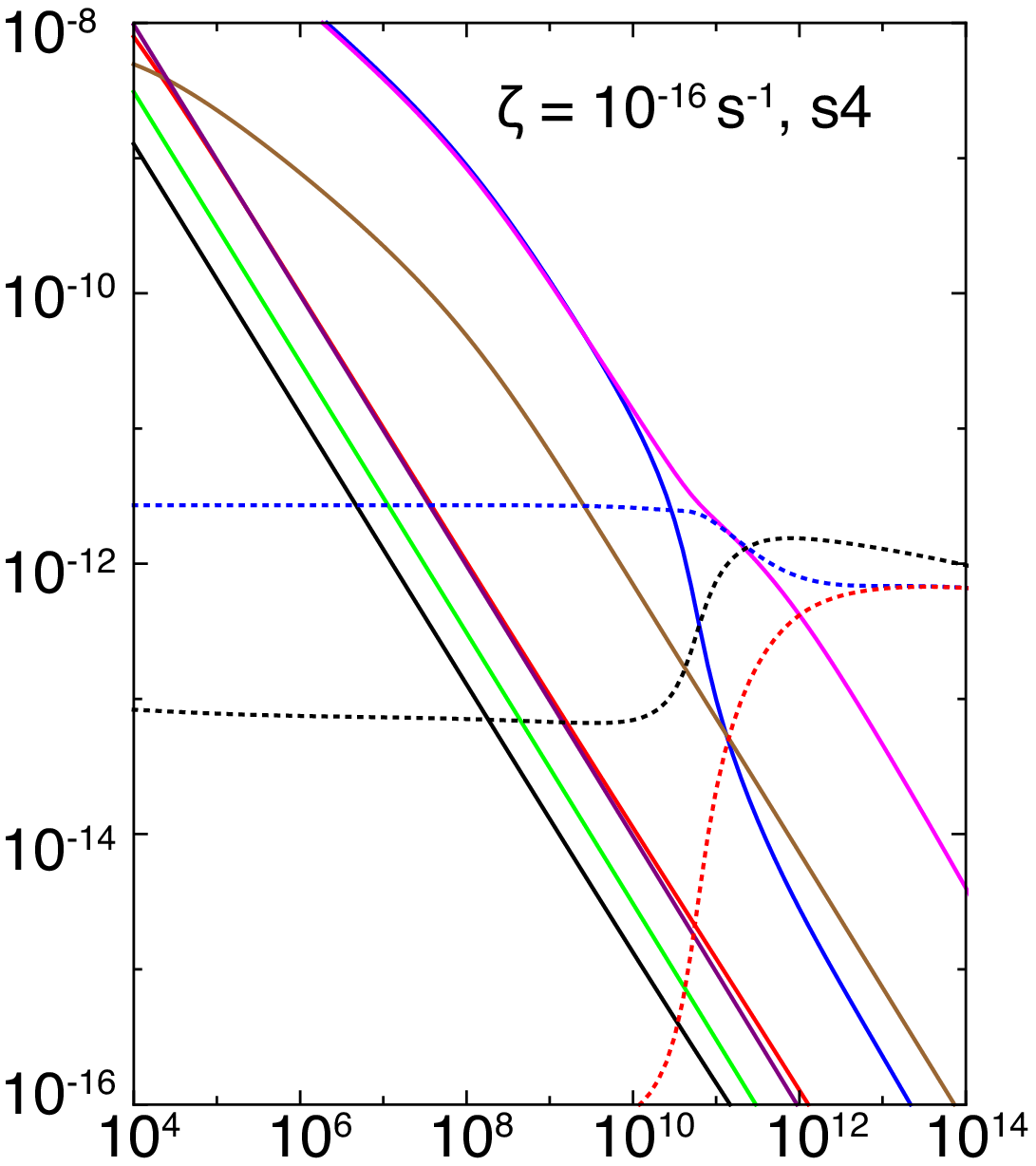}
\end{minipage}
\hspace{-5pt}
\\
\\
\begin{minipage}{0.36\hsize}
\includegraphics[width=\columnwidth]{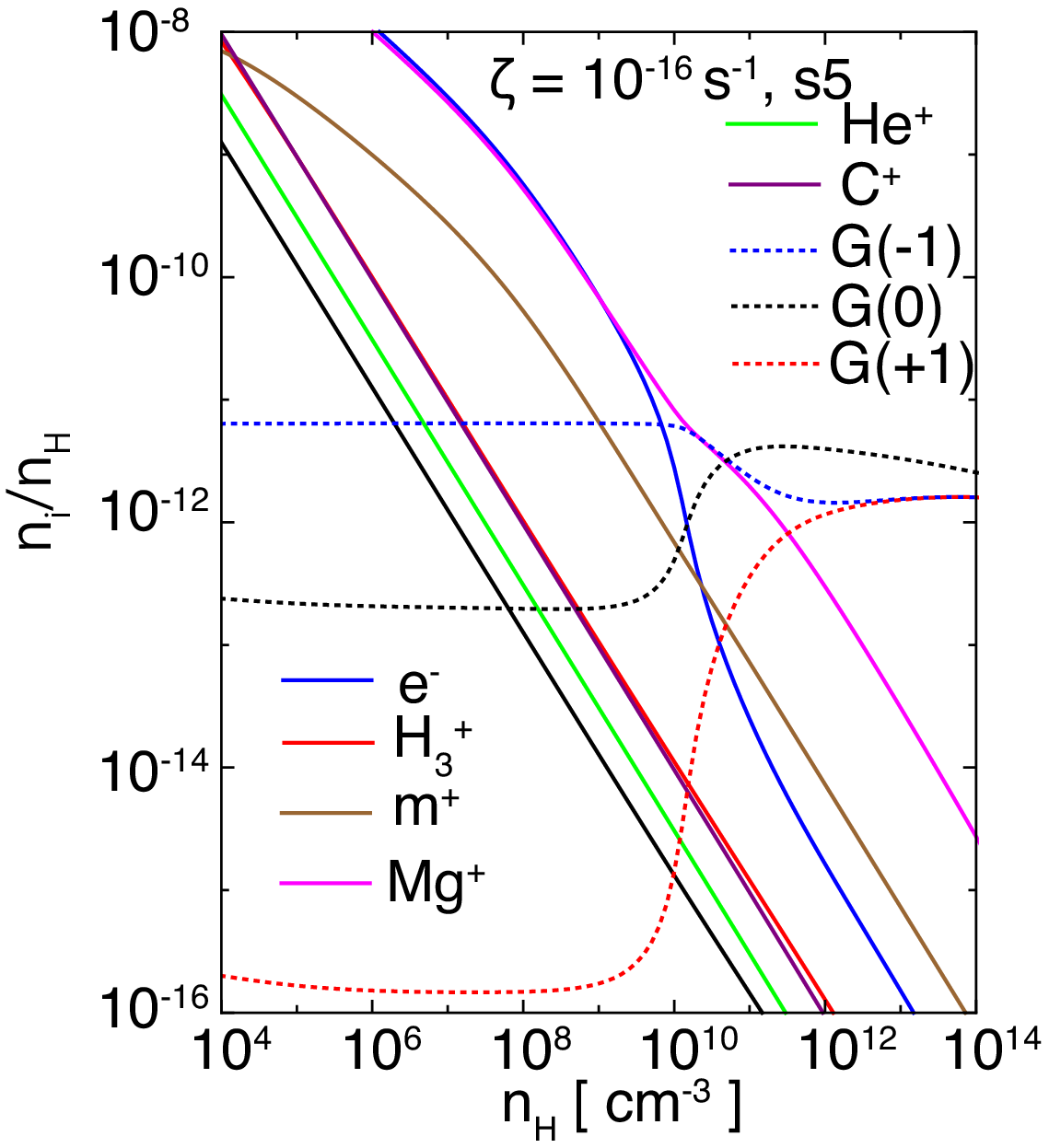}
\end{minipage}
\hspace{-5pt}
\begin{minipage}{0.33\hsize}
\includegraphics[width=\columnwidth]{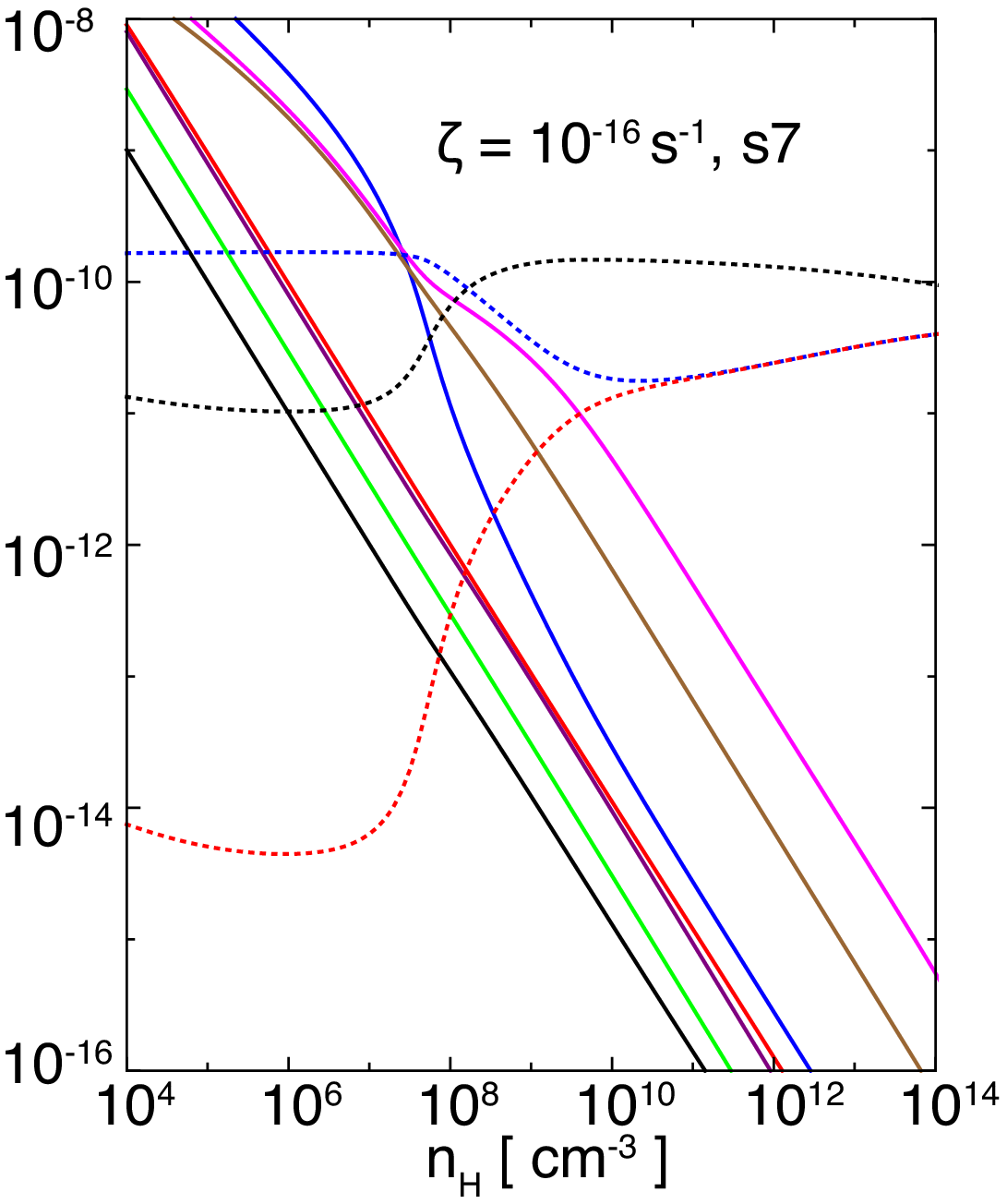}
\end{minipage}
\hspace{-5pt}
\begin{minipage}{0.33\hsize}
\includegraphics[width=\columnwidth]{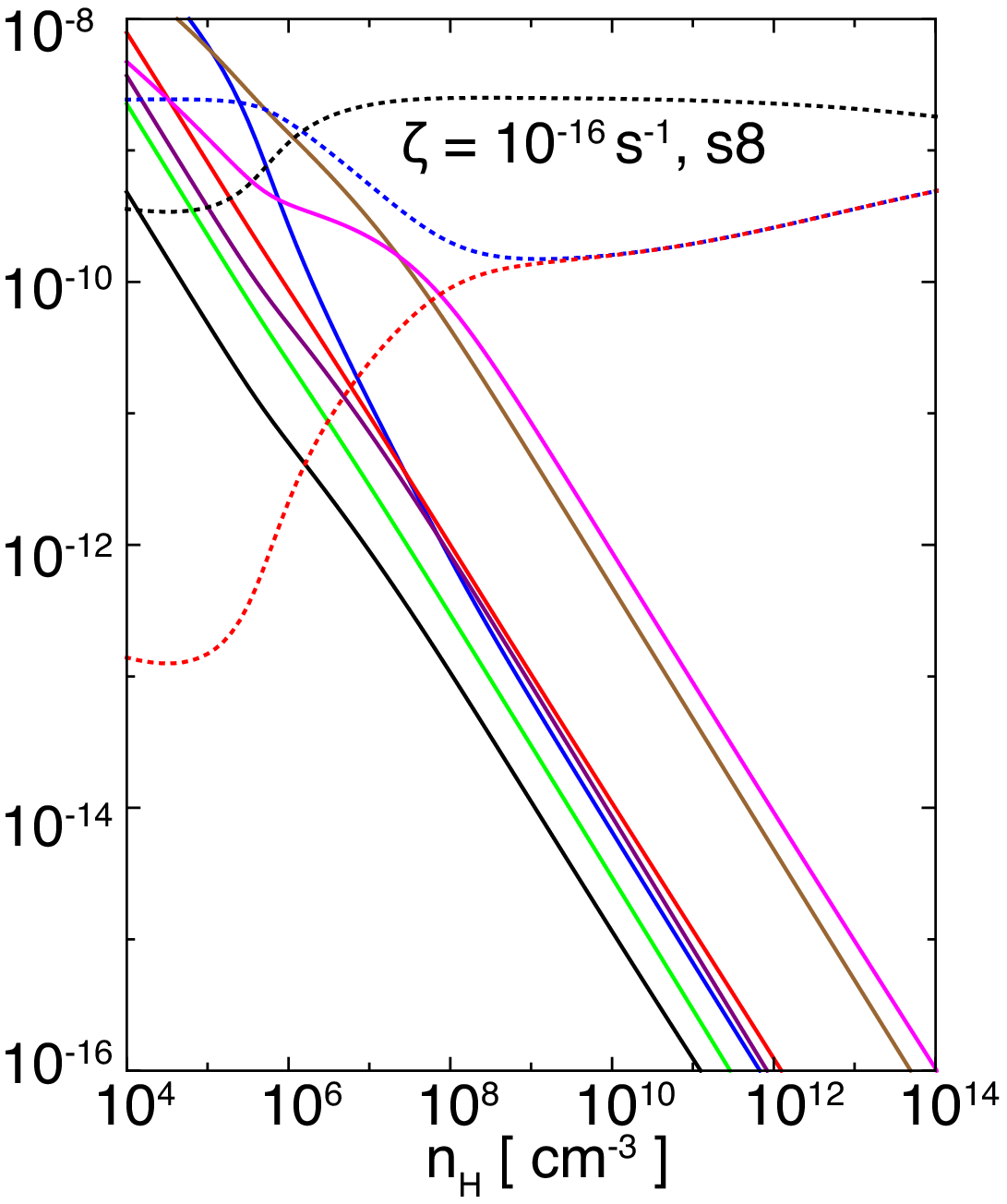}
\end{minipage}
\end{tabular}
\caption{The same data as in Fig.~\ref{fig:ca18} but for $\zeta=10^{-16} \rm{s^{-1}}$ using models (s1a, s3, s4, s5, s7 and s8).}
\label{fig:ca16}
\end{figure*}

Chemical reactions were calculated according to the method outlined in \S\ref{sec:hallcoefficient}.
To begin with, we present the chemical abundance results obtained using different dust models in conjunction with $\zeta=10^{-18}$\,s$^{-1}$ (Fig.~\ref{fig:ca18}), $10^{-17}$\,s$^{-1}$ (Fig.~\ref{fig:ca17}) and $10^{-16}$\,s$^{-1}$ (Fig.~\ref{fig:ca16}).
These data indicate that the chemical abundances are significantly affected by both the dust distribution and cosmic ray rate.
As the grain size becomes smaller, the adsorption of charge onto the grains becomes more effective, such that the number of ions and electrons decreases, as shown in Fig.~\ref{fig:ca17}. 
\so{In addition, we checked the relation between the number of grains and $f_{\rm dg}$ comparing models s1a, s1b and s1c, and confirmed that the fractional abundances of grains increase as $f_{\rm dg}$ decreases.}
\so{Figs. \ref{fig:ca18} - \ref{fig:ca16} also show that the fractional abundances of dust grains in single sized models s3 - s8 are roughly proportional to $a^{-3}$ because the total mass of grains is fixed.}
Models s1a and s7 in Fig.~\ref{fig:ca17} give similar results because the average size in MRN model s1a is 0.035$\rm{\mu m}$ (as calculated using equation (\ref{eq:sbinsize})), which is close to the grain size assumed in model s7 (0.025\,$\mu$m).
In contrast, models s3 (a$_{\rm single}=0.3 \mu$m) and s8 (a$_{\rm single}=0.01 \mu$m) provide noticeably different results, 
with the abundances of Mg$^+$ and e$^{-}$ at $n_{\rm H}=10^8\cm$ from model s3 being two and four orders of magnitudes higher, respectively, 
than those from model s8.
These large differences in the fractional abundances of charged species greatly affects the Hall coefficient. 
In addition, a comparison of Fig.~\ref{fig:ca18}, Fig.~\ref{fig:ca17} and Fig.~\ref{fig:ca16} shows that the cosmic ray ionization rate significantly affects the fractional abundances of charged particles.
To a first approximation, the fractional abundances of charged particles scale linearly with the ionization rate.

\begin{figure*}
	\includegraphics[width=5.3cm]{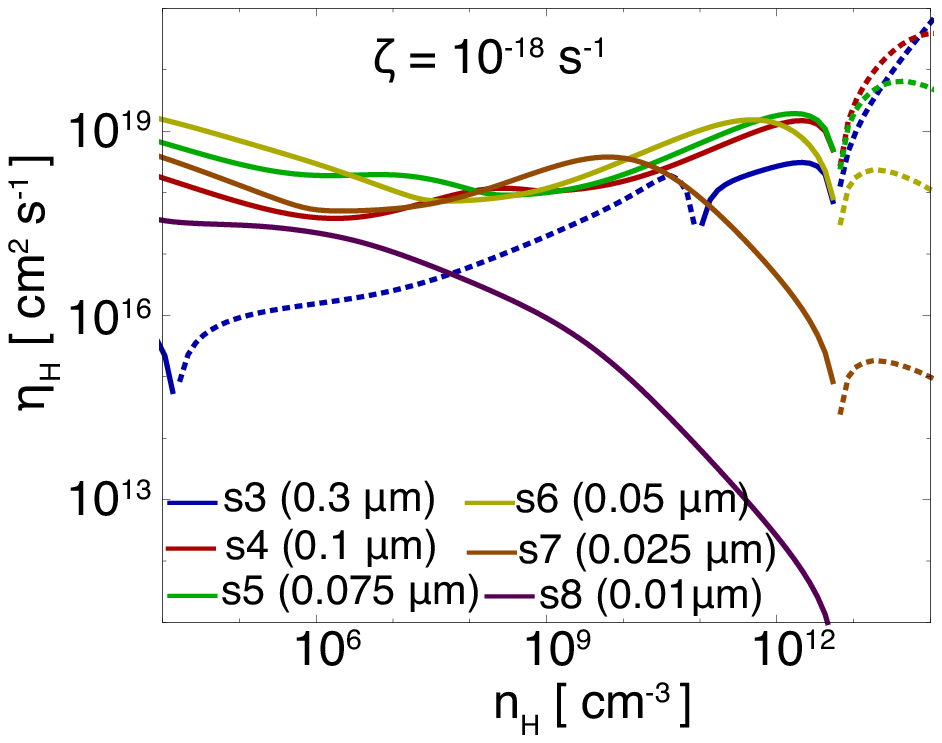}
	\includegraphics[width=5.3cm]{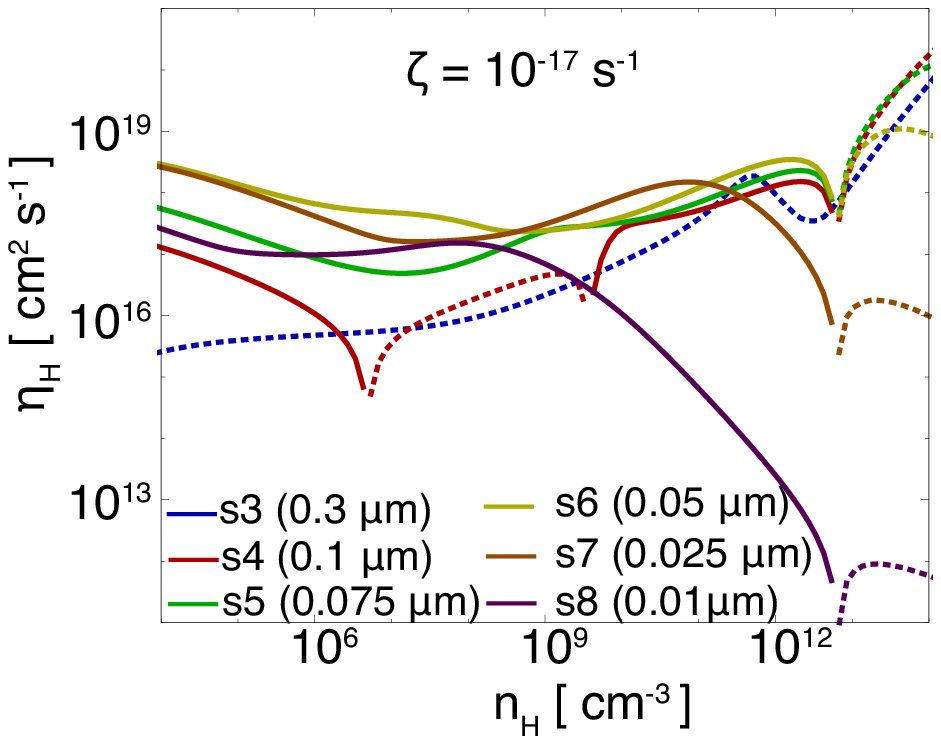}
    \includegraphics[width=5.3cm]{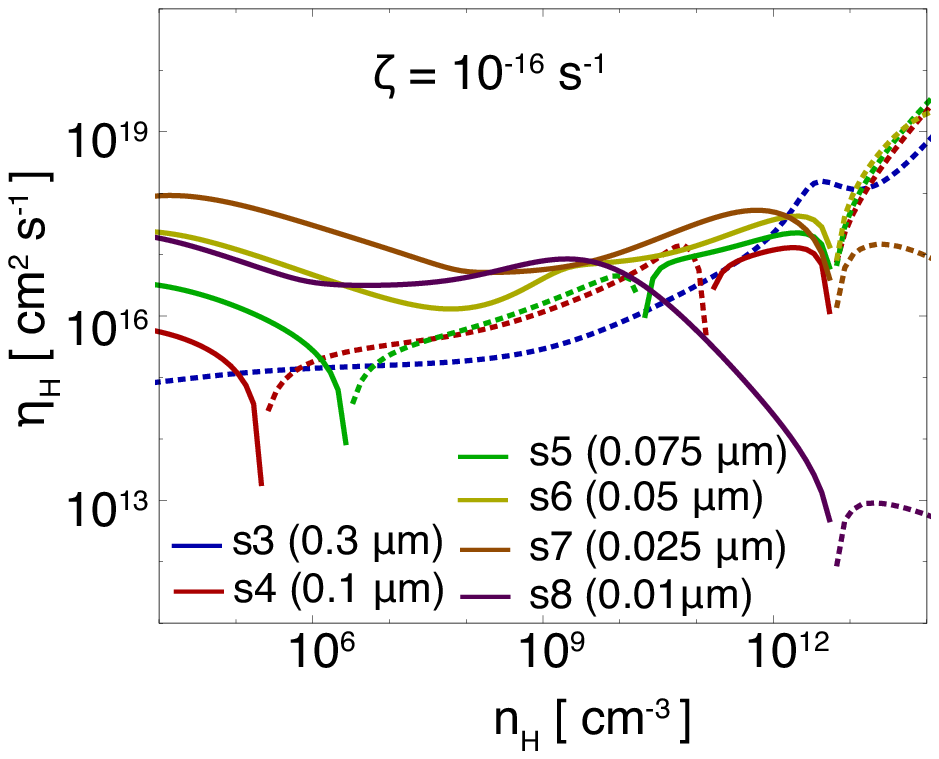} 
\caption{
Hall coefficients for single size dust models at $\zeta=10^{-18}$\,s$^{-1}$ (left), $10^{-17}$\,s$^{-1}$ (center) and $10^{-16}$\,s$^{-1}$ (right). The results from models s3, s4, s5, s6, s7 and s8 are plotted in blue, red, green, yellow, orange and purple. For each line, the positive and negative values of $\eta_{\rm H}$ are indicated by broken and solid lines, respectively.}
\label{fig:etahn}
\end{figure*}

\begin{figure*}
\includegraphics[width=5.3cm]{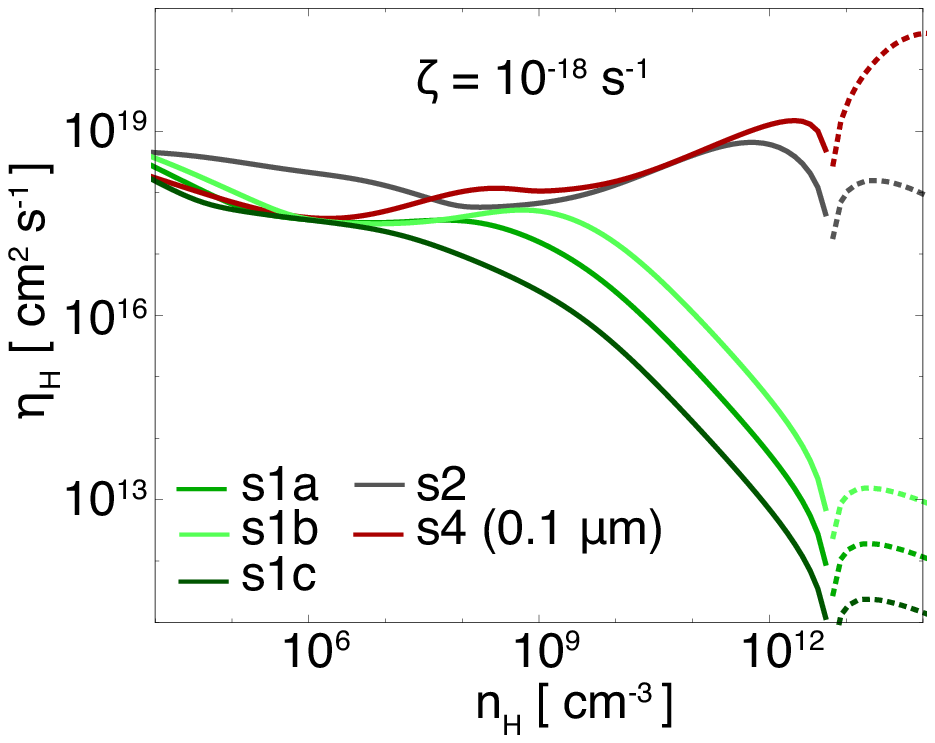}
\includegraphics[width=5.3cm]{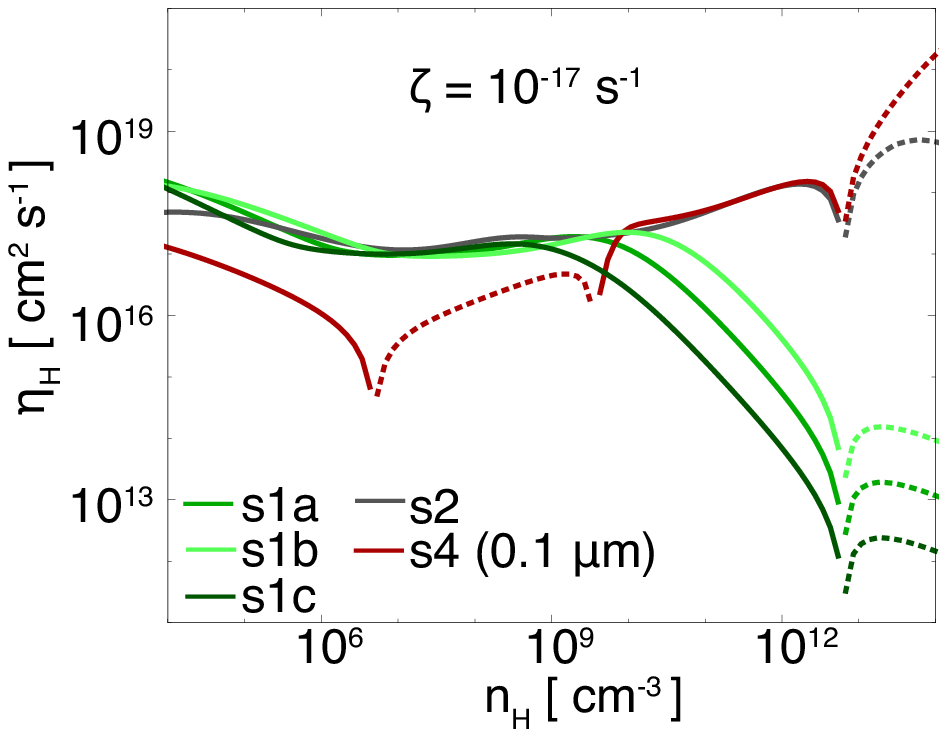}
\includegraphics[width=5.3cm]{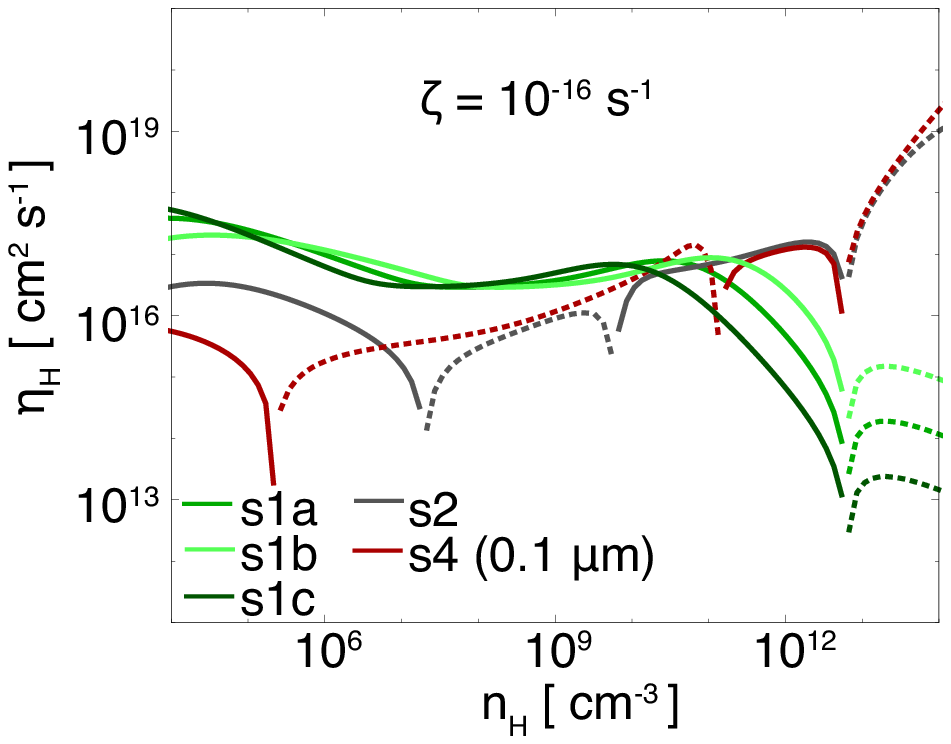}
\caption{
Hall coefficients for various MRN dust distributions. Models s1a, s1b, s1c, s2 and s4 are plotted in green, light green, dark green, gray and red, respectively. For each line, positive and negative values of $\eta_{\rm H}$ are indicated by solid and broken lines, respectively.
}
\label{fig:etahnmrn}
\end{figure*}

Fig.~\ref{fig:etahn} and Fig.~\ref{fig:etahnmrn} show the Hall coefficients
calculated for the different dust models using equations (\ref{eq:etah})-(\ref{eq:collisiondust}).
These plots indicate that the absolute value of the Hall coefficient is modified by both the grain size and grain size distribution, as well as by the cosmic ray strength. 
Furthermore, as seen in Fig.~\ref{fig:etahn} and Fig.~\ref{fig:etahnmrn}, the sign of the Hall coefficient can change depending on the grain size or grain size distribution. 
As an example, in the center panel of Fig.~\ref{fig:etahn} ($\zeta=10^{-17}$\,s$^{-1}$), the Hall coefficient is positive over the entire range of $10^4\cm \lesssim n_{\rm H} \lesssim 10^{14}\cm$ for model s3, while it is positive only in the range of $n_{\rm H} > 10^{13}\cm$ for model s5. 
Thus, even a change in the grain size by a factor of four can cause a flip in the sign of the Hall coefficient at $n_{\rm H} \sim 10^{13}\cm$. 
\citet{2018MNRAS.478.2723Z} also showed that large-sized grains (or removal of small-sized grains; $\lesssim 0.1 \rm{\mu m}$) noticeably decrease the Hall coefficient, which agrees well with our results (see the center panel of Fig.~\ref{fig:etahn}).

\section{Application: Disk Size Induced by the Hall Effect}
\label{sec:diskmethod}
As described in Section \S\ref{sec:intro}, the purpose of this study was to estimate the disk size considering solely the Hall effect.
To correctly determine the disk size, non-ideal MHD simulations were required \citep[e.g.][]{2015ApJ...810L..26T}.
However, during the star formation process, a small disk formed around a protostar gradually evolves. 
Thus, at minimum, the spatial scale of the protostar should be resolved \citep{2018arXiv180108193V} when examining the formation and evolution of the circumstellar disk. 
Despite this, the timescale becomes extremely short when resolving the protostar, and so it is impossible to simulate disk evolution and determine the disk size with adequate spatial resolution. 
As an example, a very recent simulation managed to calculate the disk evolution over a time span of several months following protostar formation \citep[e.g.,][]{2015ApJ...801..117T,2016A&A...587A..32M,2017PASJ...69...95T}.
In contrast, observations have indicated that the main accretion phase, during which the disk evolves via mass accretion, lasts for $\sim10^5$\,yr \citep{1994ApJ...420..837A,2008ApJ...684.1240E}. 
In addition, the Hall coefficient, which is related to the angular momentum associated with transport in the infalling gas and disk evolution, is greatly affected by specific dust properties and the environmental effects of the ionization source (or the cosmic ray strength). 
Thus, to determine the disk size, it is necessary to calculate the disk evolution over a span of at least $\sim10^5$\,yr, while resolving the protostar using the grain size and cosmic ray strength as parameters. 
Since such calculations are difficult to execute even using the fastest supercomputers presently available, we analytically estimated the disk size by referring to the simulation results. 

Because the Hall coefficient is determined by the abundance of charged particles, the charged particle properties and chemical networks were discussed in the previous sections.
Below, in order to simplify our assumptions, we estimate and discuss the influence of the Hall effect on the evolution and formation of the circumstellar disk. 
This discussion begins with an explanation of the process used to estimate the disk size induced by the Hall effect.  

\begin{figure*}
\includegraphics[width=10cm]{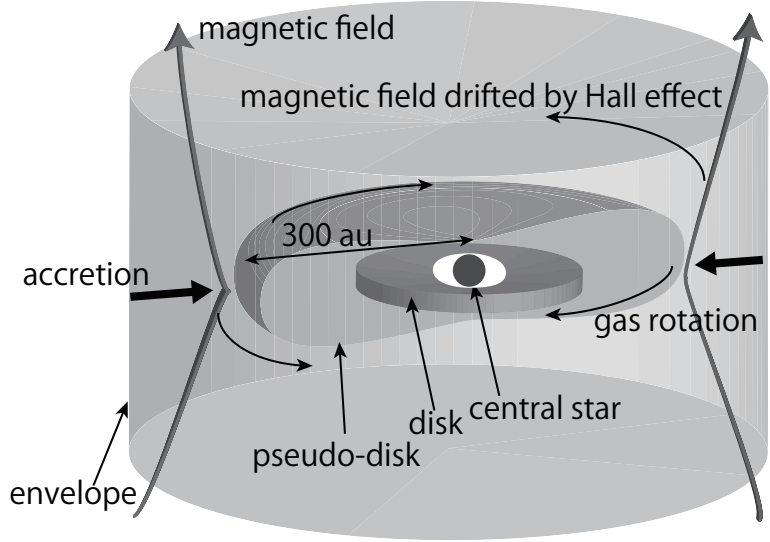}
\caption{Schematic of the envelope, pseudo-disk and disk.}
\label{fig:model}
\end{figure*}

The (specific) angular momentum of the infalling gas resulting from the Hall effect must first be evaluated, as this determines the disk size. 
Note that the angular momentum is redistributed in the disk and a part of the angular momentum is also ejected by protostellar outflow, meaning that the estimates provided herein actually refer to the upper limit of the disk size. 
Even so, it is useful to assess the effects of both the dust properties and cosmic ray strength on the Hall coefficient and resultant disk size.
Our approach to estimating the disk size is illustrated schematically in Fig.~\ref{fig:model}. 
To begin, we assume a non-rotating prestellar cloud. 
Although this assumption is not actually correct, it is a useful approach to estimating the angular momentum resulting solely from the Hall effect.
Next, we calculate the chemical networks and derive the Hall coefficient as a function of cloud density, which can be converted to a length scale (or Jeans length). 
Subsequently, the angular momentum of the infalling gas is derived by defining $r_{\rm Hall}$ and assuming that the infalling gas instantaneously acquires the angular momentum at $r_{\rm Hall}$. 
In this study, we use the term $r_{\rm Hall}$ for the Hall radius or Hall point and employ a value of $r_{\rm Hall}=300$\,au, which is similar to the pseudo disk size \citep{2017PASJ...69...95T}.
A Jeans length of $r_{\rm Hall}=300$\,au corresponds to a number density of $n_{\rm Hall} = 10^8 {\rm cm^{-3}}$.

Even without rotation, the magnetic field is twisted by the Hall effect such that a toroidal component is produced. 
Subsequently, after coupling between the neutral gas and magnetic field (or charged particles) is recovered, the magnetic tension force caused by the toroidal component of the magnetic field imparts a rotational motion. This is the reason why the non-rotating gas acquires an angular momentum as a result of the Hall effect.

\subsection{Hall Velocity}
\label{subsec:Hallvelocity}
We assume that the gas velocity passing through the Hall radius ($r=r_{\rm Hall}$) converges to the Hall drift velocity.
Thus, the azimuthal component of the gas velocity can be written as
\begin{eqnarray}
v_\phi &\sim& -\bm{v}_{\rm Hall,\phi},  \nonumber \\
&=& \eta_H \frac{(\nabla \times \bm{B})_\phi}{|B_z|}. \label{eq:hallvelphi}
\end{eqnarray}
\so{The same definition of the Hall drift velocity is also used in \citet{2016PASA...33...10T}}.
The convergence of the gas rotation velocity is explained in Section \S\ref{subsec:hallvelocity}, and the validity of the gas velocity assumption is demonstrated in Fig. 3 of \citet{2011ApJ...733...54K}, which confirms that the azimuthal gas velocity converges to the Hall drift velocity. 
The timescale of the $\bm{v}_\phi$ change is discussed in Section \S\ref{subsec:halltimescale}.

Using a cylindrical coordinate system, the Hall-induced rotational velocity can be written as 
\begin{equation}
v_{\phi} = \frac{\eta_H}{B_z}  \left( \frac{\partial B_{r,s}}{\partial z}-\frac{\partial B_z}{\partial r}  \right),
\label{eq:vhallder}
\end{equation}
where $B_{r,s}$ is the magnetic field strength in the radial direction at the surface of the pseudo-disk. 
These calculations allow the approximation $|{\partial B_{r,s}}/{\partial z}| \sim B_{r,s}/H_{\rm env}$, where $H_{\rm env}$ is the scale height of the infalling envelope.
In addition, the radial derivative term is assumed to be zero (i.e., $\frac{\partial B_{r,s}}{\partial z} \gg \frac{\partial B_z}{\partial r}$; \citealt{1994MNRAS.267..235L}).
Thus, equation~(\ref{eq:vhallder}) can in turn be approximated by 
\begin{equation}
v_{\phi} = \frac{\eta_H}{|B_z|} \frac{B_{r,s}}{H_{\rm env}}.
\label{eq:vhall3}
\end{equation}

In these calculations, $B_{r,s}$ is rewritten in terms of the mass-to-flux ratio. The magnetic flux threading the core is given by
\begin{equation}
\Phi_{\rm tot} =2 \pi r^2 B_{r,s},  
\label{eq:magneticflux}
\end{equation}
based on the monopole approximation employed in \citet{1998ApJ...504..247C} and \citet{2012MNRAS.422..261B}.

We also define the mass-to-flux ratio normalized by the critical value as 
\begin{equation}
\mu_{\rm tot} \equiv  \left( \frac{M_{\rm tot}}{\Phi_{\rm tot}} \right) / \left( \frac{M}{\Phi} \right)_{\rm crit},
\end{equation}
where $(M/\Phi)_{\rm crit}=(0.53/3\pi) (5/G)^{1/2}$ is the critical mass-to-flux ratio \citep{1976ApJ...210..326M,2004RvMP...76..125M}
and $M_{\rm tot}$ is the total core mass, as discussed below (see equation (\ref{eq:mtot})).
Using these equations, equation~(\ref{eq:vhall3}) can be rewritten as 
\begin{eqnarray}
v_{\phi} &=& \frac{\eta_{\rm H}}{|B_z|} \frac{B_{r,s}}{H_{\rm env}} \frac{(M/\Phi)_{\rm tot}}{\mu_{\rm tot} (M/\Phi)_{\rm crit}}  \nonumber \\
&=& \frac{\eta_{\rm H}}{|B_z|} \frac{1}{\mu_{\rm tot} (M/\Phi)_{\rm crit}} \frac{M_{\rm tot}}{2 \pi r^2 H_{\rm env}}. 
\label{eq:vhallmu}
\end{eqnarray}
Here, $\mu_{\rm tot}=1$ is adopted based on the simulation results of \citet{2011MNRAS.413.2767M}. 
The vertical direction of the magnetic field is given by equation~(\ref{eq:magstrength}), using $B=B_z$. 

According to \citet{1998ApJ...493..342S}, the scale height of the envelope can be approximated by  
\begin{equation}
H_{\rm env}=\frac{\sqrt{2} c_s}{\sqrt{\pi G \rho_{\rm env}}} \sim \frac{c_s}{\sqrt{G \rho_{\rm env}}},
\label{eq:scaleheightH}
\end{equation}
where $\rho_{\rm env}=3.8\times10^{-16}$\,g$\cm$ ($n_{\rm env}=10^8\cm$) is adopted on the basis of \citet{2017PASJ...69...95T}.
The speed of sound is determined employing equation~(\ref{eqn:temperature}).
Using this approach, the scale height of the envelope is estimated to be $H_{\rm env}=360$\ au, in good agreement with recently-reported simulation results \citep[e.g.,][]{2017PASJ...69...95T}.  

\subsection{Disk Model and Angular Momentum Estimated using the Hall Effect}
\label{subsec:diskmodel}
Recent simulations of disk formation have indicated that the circumstellar disk in the early star formation phase is massive \citep{2010ApJ...718L..58I,2015MNRAS.452..278T,2015ApJ...801..117T,2018arXiv180108193V}.  
Thus, Toomre's  $Q$-parameter, which is given by $Q=c_s \Omega_{\rm kep}/(\pi G \Sigma_{\rm disk}$), is expected to be small. 
Using the $Q$-parameter, we can represent the disk surface density as 
\begin{equation}
\Sigma_{\rm disk}=\frac{c_s \Omega_K}{\pi G Q},
\label{eq:sigma}
\end{equation}
where the value of $Q$ is assumed to be constant over the entire disk. 
According to \citet{1970PThPh..44.1580K} and \citet{1997ApJ...490..368C}, the disk temperature can be determined using the equation 
\begin{equation}
T_{\rm disk} = 150 \left( \frac{r}{1~\rm{au}}  \right)^{- \frac{3}{7}} \ \ \ {\rm K}. 
\end{equation}
Substituting the temperature, $T_{\rm disk}$, and Keplerian angular velocity, $\Omega_{\rm k}$, the dimensional surface density of 
the disk can be described as 
\begin{equation}
\Sigma_{\rm disk}= c_{\rm sd} \Sigma_0 \left( \frac{Q}{1} \right)^{-1} \left(\frac{M_{\rm star}}{0.1 M_\odot} \right)^{\frac{1}{2}} \left(\frac{r}{100 \rm{au}} \right)^{-\frac{12}{7}} \rm{g}\, {\rm cm}^{-2},
\label{eq:sigmanon}
\end{equation}
where $M_{\rm star}$ is the protostellar mass, $\Sigma_0$ is a constant estimated to have a value of $\Sigma_0=8.2$\,g\,cm$^{-2}$ based on the above equations, and $c_{\rm sd}$ is a control parameter used to adjust the disk surface density.
Using equation~(\ref{eq:sigmanon}), the disk mass is estimated as 
\begin{equation}
M_{\rm disk}=\int_0^r \Sigma_{\rm disk}\, 2 \pi r dr.
\end{equation}
In addition, we define the total mass (which includes the disk and protostellar masses) as
\begin{equation}
M_{\rm tot}=M_{\rm star}+M_{\rm disk}.
\label{eq:mtot}
\end{equation}
The total mass, $M_{\rm tot}$, should equal the overall mass of the infalling gas passing through the Hall radius. 
Note that we ignore mass ejection via protostellar outflow in this study. 
The infalling gas passing through the Hall radius, $r_{\rm Hall}$, is assumed to acquire an angular momentum via the Hall effect, and this momentum can be estimated as 
\begin{eqnarray}
J_{\rm Hall} &\sim& M_{\rm tot} j_{\rm Hall} \nonumber \\  
 &=& M_{\rm tot} r_{\rm Hall} v_{\phi}, 
\end{eqnarray}
where $r=r_{\rm Hall}$ and equation~(\ref{eq:vhallmu}) are employed to obtain $v_{\phi}$.
Thus, the Hall angular momentum, $J_{\rm Hall}$, can be written as $J_{\rm Hall} \sim M_{\rm tot}\, r_{\rm Hall}\, v_{\rm Hall,\phi} $. 

To determine the size of the circumstellar disk, the angular momentum induced by the Hall effect is assumed to produce a Keplerian disk around the protostar. 
Consequently, the infalling mass is distributed into the protostar and Keplerian disk, although only the Keplerian disk is assumed to have angular momentum. 
This angular momentum is given by
\begin{equation}
\begin{split}
J_{\rm Kepler}&=\int_0^r \Sigma_{\rm disk}(r) 2 \pi r j_{\rm Kepler}\, dr \\
&=\int_0^r \Sigma_{\rm disk}(r) 2 \pi r^2 v_{\rm Kepler} \, dr \\
&=\int_0^r \Sigma_{\rm disk}(r) 2 \pi r^2 \sqrt{\frac{G M_{\rm star}}{r}} dr.  
\end{split}
\label{eq:jkepler}
\end{equation}
Therefore, using solely the total and protostellar masses, we can determine the size of the disk outer edge at which the Keplerian angular momentum corresponds to the angular momentum of the total gas that has already undergone infall. 
Hereafter, we define the size of disk's outer edge as $r_{\rm disk}$.

As described above, the disk surface density is determined based on the $Q$-parameter. 
However, since disk fragmentation may occur when $Q\lesssim 1$, it is difficult to retain a simple disk. 
Thus, we consider that $c_{\rm sd}=1$ (i.e. $Q=1$) gives the maximum surface density.
In reality, because the disk surface density is expected to be less than $c_{\rm sd}=1$, $c_{\rm sd}$ is parameterized and the value adopted in this study is provided in Table~\ref{tab:para}.
The protostellar mass, which is also employed as a parameter to determine the disk radius, is included in the same table. 
\begin{table*}
  \caption{Parameters determining the disk surface density and protostellar mass.}
  \begin{center}       
	     \begin{tabular}{|l|c|} \hline
            $c_{\rm sd}$ &  1, 0.1, 0.01  \\ \hline
            $M_{\rm star}$ [$M_\odot$] & 0.03, 0.05, 0.07, 0.1, 0.3, 0.5, 0.7 \\ \hline
          \end{tabular}
                    \label{tab:para}
        \end{center}
\end{table*}

\subsection{Keplerian Disk Induced by the Hall effect}
\label{sec:Keplerian Disk Induced by Hall effect}
\begin{figure}
\includegraphics[width=\columnwidth]{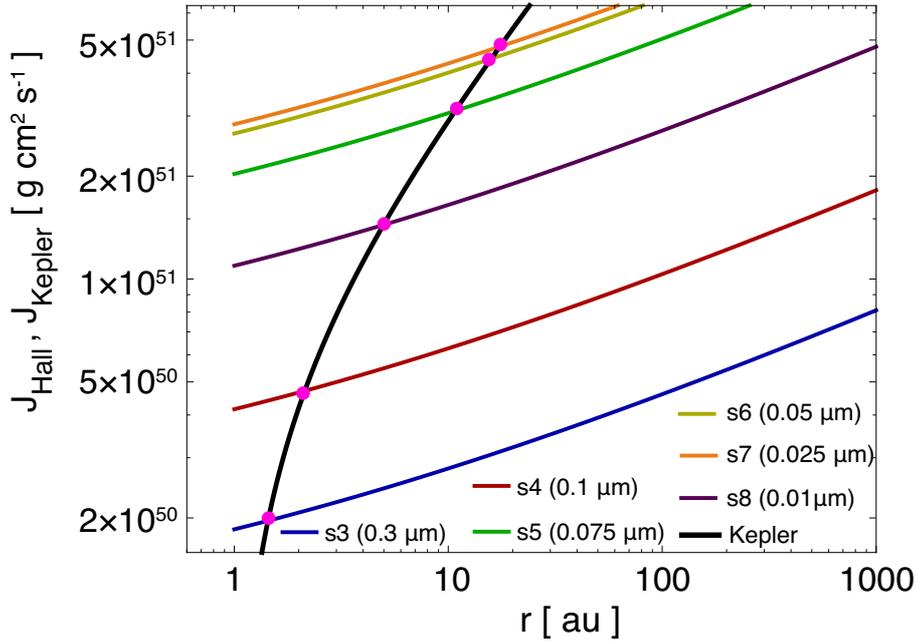}
\caption{
The angular momentum values of the Hall disks for different grain size models versus the radius, as obtained using $c_{\rm sd}=1$ and $\zeta=10^{-17}$\,s$^{-1}$, with a protostellar mass of $M_{\rm star}=0.1\msun$. 
The thick black solid line corresponds to the angular momentum of the Keplerian disk. 
The filled magenta circles indicate the intersections of the angular momenta of Hall and Keplerian disks, and correspond to the radii of the Hall disks. 
}
\label{fig;disk1701}
\end{figure}

The disk size induced by the Hall effect was determined according to the procedure:
\begin{enumerate}
\item the protostellar mass, $M_{\rm star}$, was assumed to determine the total mass, $M_{\rm tot}$ (equation (\ref{eq:mtot})),
\item the specific angular momentum, $j_{\rm Hall}= r_{\rm Hall}\, v_{\rm Hall, \phi} $, was estimated using equation~(\ref{eq:vhallmu}),
\item the total angular momentum of the accreted material was calculated as $J_{\rm Hall}=M_{\rm tot}\, j_{\rm Hall}$,
\item the total angular momentum of the Keplerian disk was estimated based on equation~(\ref{eq:jkepler}), assuming that the disk extends to a very large radius, and
\item the disk size, $r_{\rm disk}$, at which the angular momentum of the Keplerian disk equals that of the accreted material (the rotation of which is determined by the Hall effect as in steps (i) -- (ii)) was determined.
\end{enumerate}
After varying the protostellar mass, steps (i)-(v) were repeated using the different models listed in Table~\ref {tab:dust} and \ref{tab:cr}, to determine $r_{\rm disk}$ for each model. The resulting values are presented in Fig.~\ref{fig;disk1701} as filled magenta circles.
As an example of the estimation of the disk induced by the Hall effect, Fig.~\ref{fig;disk1701} plots the angular momentum values for Keplerian and Hall disks (hereafter, we refer to disks formed by the Hall effect as Hall disks). These values were obtained using $c_{\rm sd}=1$ and $\zeta=10^{-17}$\,s$^{-1}$ for different grain size models and assuming a protostellar mass of $M_{\rm star}=0.1\msun$.  
Additionally, the specific angular momentum induced by the Hall effect, $j_{\rm Hall}$, was calculated for each model (see Section \S\ref{subsec:diskmodel}).
For each line (that is, each grain size model) in this figure, the angular momentum of the Hall disk is less than that of the Keplerian disk outside the point denoted by the purple circle, meaning that a rotationally-supported disk will not form in such regions. 
Conversely, in the region (or at the radius) inside the purple circle, the angular momentum of the Hall disk is greater than that of the Keplerian disk and so a rotationally-supported disk (i.e., a Keplerian disk) can form.
Because angular momentum can be transferred inside the disk, it is difficult to precisely determine the disk radius values in this study.
Even so, a comparison of the angular momenta of the Hall disk and Keplerian disk allows a rough estimation of the disk size for a given grain size and protostellar mass. 
The data in this figure indicate that the disk size is $r_{\rm disk}=2.13$\,au in the case of model s4 ($a_{\rm single}=0.1\,{\rm \mu m}$) and 5.01\,au in the case of model s8 ($a_{\rm single}=0.01\,{\rm \mu m}$). 

\begin{figure}
\includegraphics[width=8cm]{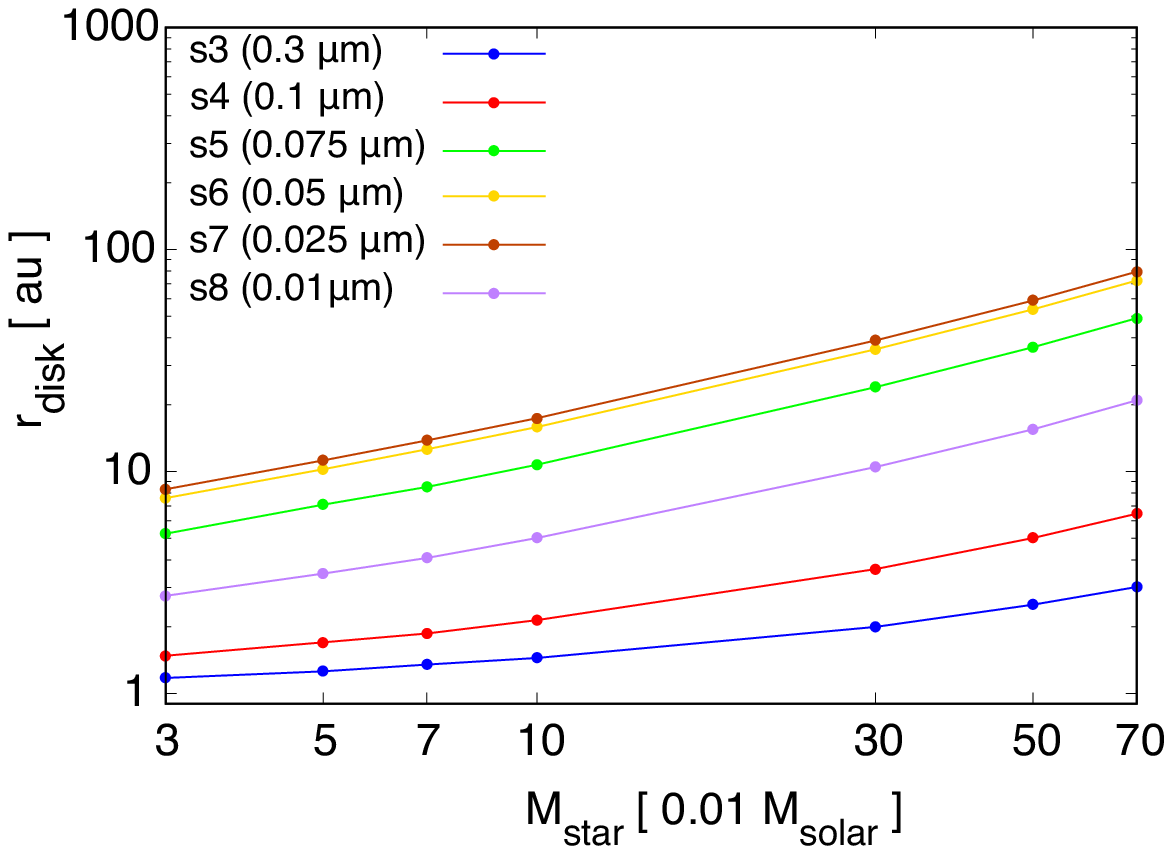}
\includegraphics[width=8cm]{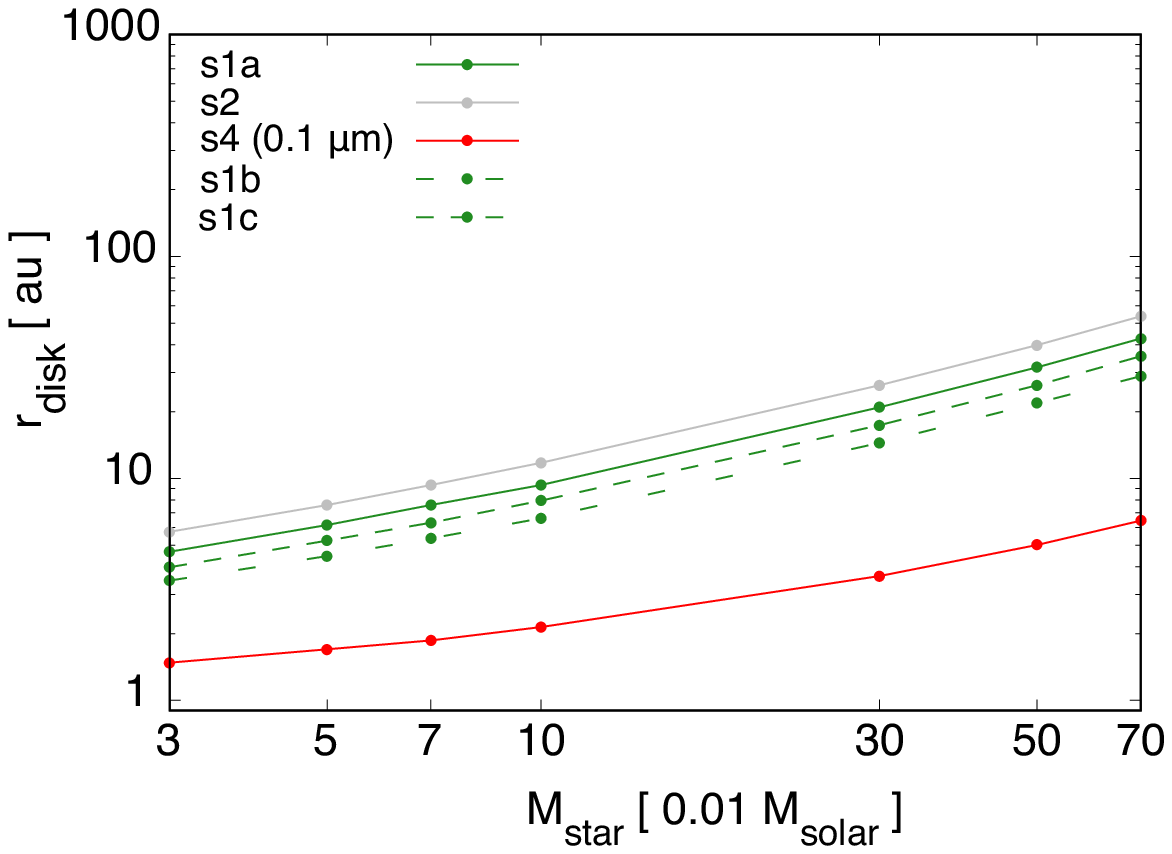}
\caption{
Disk sizes obtained using single-sized (left) and MRN (right) grain distributions as functions of the protostellar mass, as calculated with $c_{\rm sd} =1$ and $\zeta=10^{-17}$\,s$^{-1}$.
}
\label{fig:css}
\end{figure}

While varying the protostellar mass, $M_{\rm star}$, we derived the corresponding disk sizes according to the procedure described above. 
Fig.~\ref{fig:css} plots the disk sizes obtained using the single-sized (left) and MRN (right) dust grain size distribution models as functions of the protostellar mass, based on using $c_{\rm sd}=1$ and $\zeta=10^{-17}$\,s$^{-1}$.
These plots demonstrate that the size of the Keplerian disk varies considerably depending on the grain size (distribution), even if the protostellar mass is kept constant.
As an example, at $M_{\rm ps}=0.7\msun$, the disk size for $a_{\rm single}=0.025 {\rm \mu m}$ (model s7) is $r_{\rm disk}=80$\,au, while the size for $0.3 {\rm \mu m}$ (model s3) is $r_{\rm disk}=3$\,au (left panel of Fig.~\ref{fig:css}), corresponding to a variation in the disk size by a factor of 27. 
This result indicates that both the Hall effect and dust properties can affect the formation and evolution of the circumstellar disk.  

\subsection{Parameter Effects}
The effects of the various parameters on the size of the Keplerian disk induced by the Hall effect were examined by constructing a fiducial model with the surface density parameter $c_{\rm sd}=1$, protostellar mass $M_{\rm star}$=0.1$M_{\odot}$ and a cosmic ray ionization rate of $\zeta$ =$10^{-17} \rm{s^{-1}}$.  
In this subsection, we discuss the results. 

\begin{figure}
\includegraphics[width=8cm]{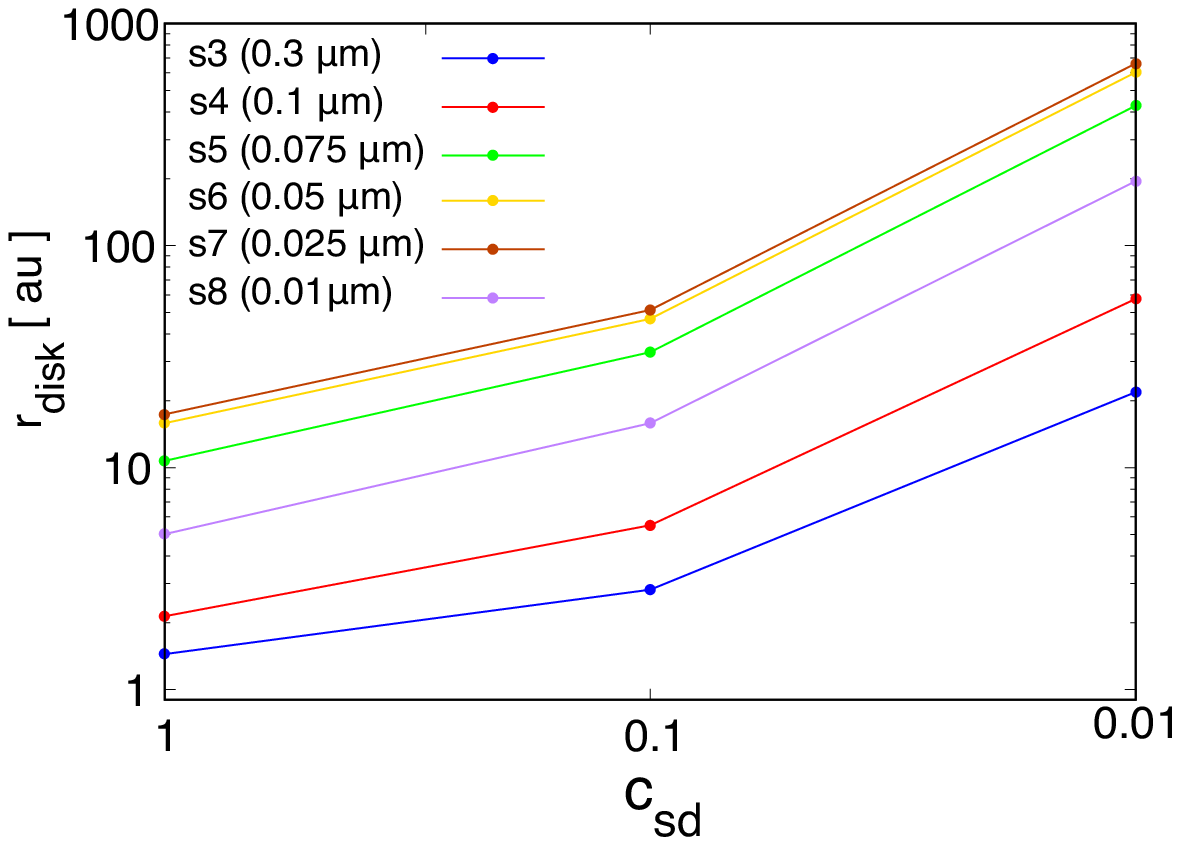}
\includegraphics[width=8cm]{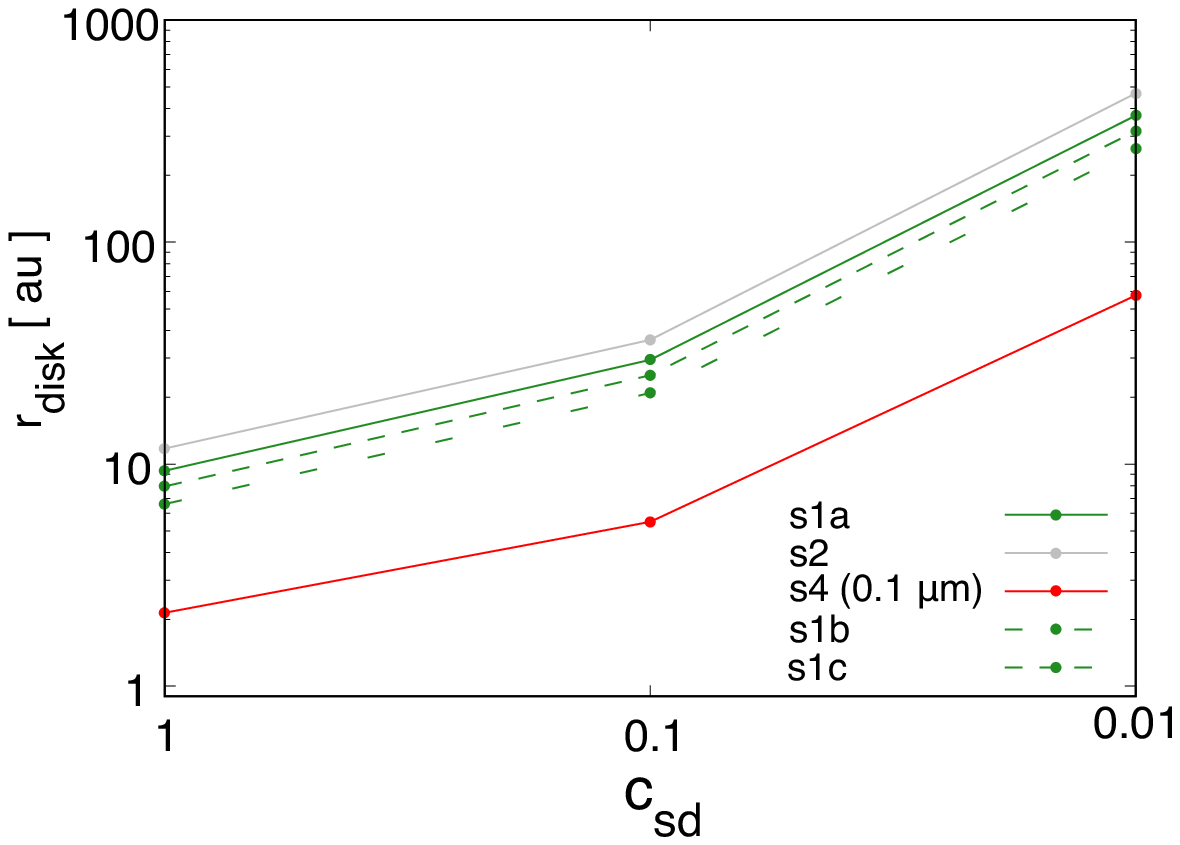}
\caption{
Disk sizes obtained from the single-sized (left) and MRN (right) dust grain distributions versus the disk surface density control parameter $c_{\rm sd}$, obtained using a prestellar mass of $M_{\rm star}=0.1\msun$ and a cosmic ray ionization rate of $\zeta=10^{-17}$\,s$^{-1}$. 
}
\label{fig:qs}
\end{figure}

Fig.~\ref{fig:qs} plots the disk sizes as functions of $c_{\rm sd}$ for each dust model listed in Tables~\ref {tab:dust}and \ref{tab:para}, obtained using a protostellar mass of $M_{\rm star}=0.1\msun$ and an ionization rate of $\zeta$ =$10^{-17} \rm{s^{-1}}$.
As described in \S\ref{subsec:diskmodel}, we constructed the disk model on the basis of $Q=1$, which corresponds to a marginally gravitationally unstable disk.
Subsequently, we calculated the disk sizes, while 
adjusting the disk surface density using the parameter $c_{\rm sd}$. As can be seen from these data, the disk size increases as the disk surface density decreases. This occurs because a less massive disk requires a large radius to have the same angular momentum as a more massive disk.
However, the disk size is not strongly correlated with the surface density or with $c_{\rm sd}$. 
As demonstrated by Fig.~\ref{fig:qs}, the disk radius increases by approximately one order of magnitude when the disk surface density is reduced by about two orders of magnitude ($c_{\rm sd}=0.01$).  

\begin{figure}
\includegraphics[width=8cm]{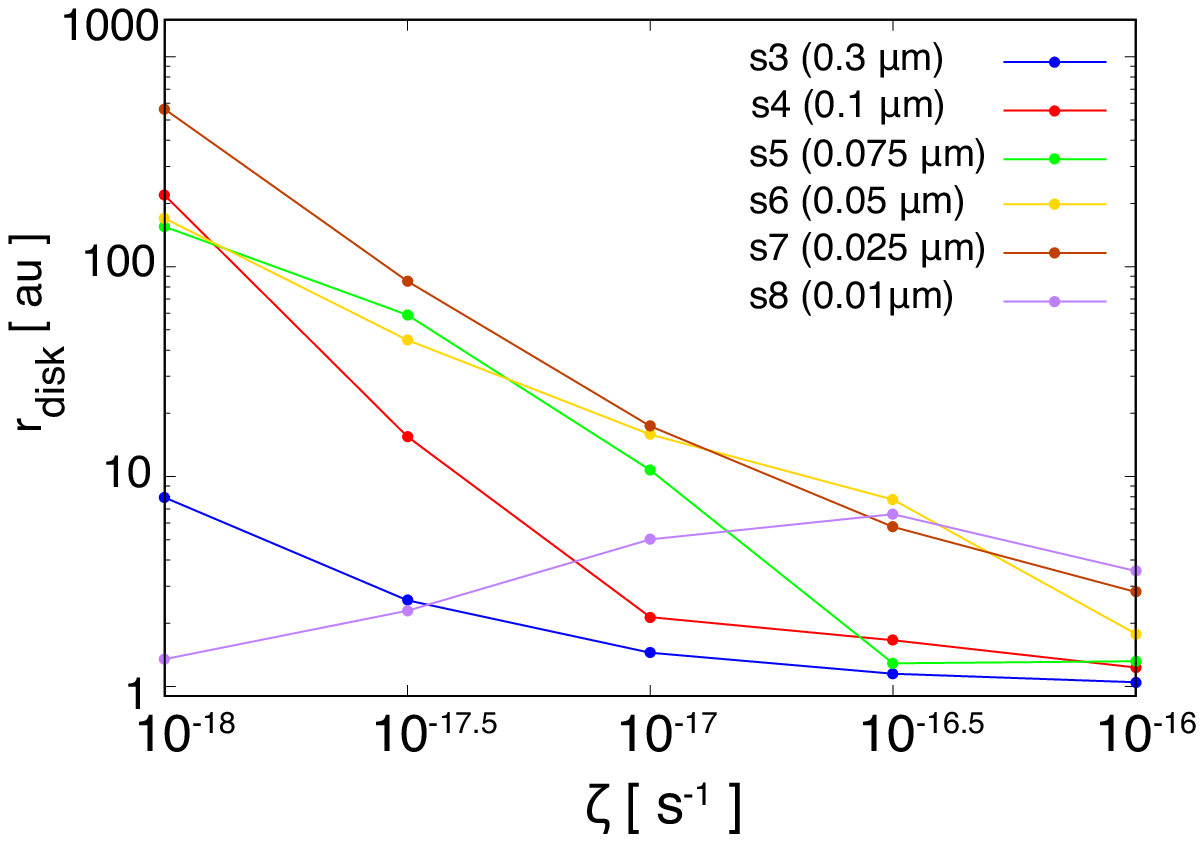}
\includegraphics[width=8cm]{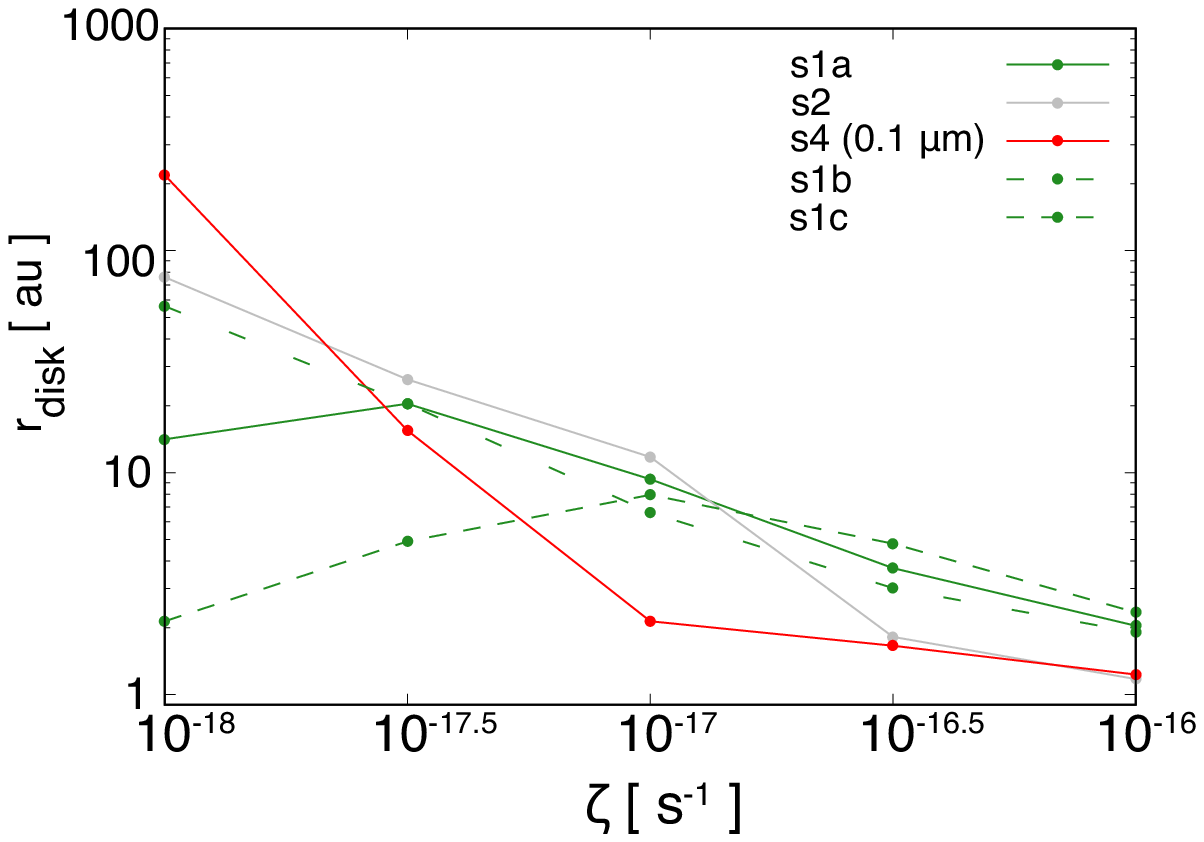}
\caption{
Disk sizes for the single-sized (left) and MRN (right) dust grain distributions as functions of the cosmic ray ionization rate, $\zeta$, obtained using $c_{\rm sd}=1$ and $M_{\rm star}=0.1\msun$.
}
\label{fig:cs}
\end{figure}

Fig.~\ref{fig:cs} summarizes the effect of the cosmic ray ionization rate $\zeta$ on the disk size for different grain size models. 
Basically, the disk size decreases as $\zeta$ increases, because the non-ideal effect (or the Hall effect) becomes ineffective in an environment in which the ionization intensity is strong. 
However, an exception is also evident such that, in the case of the small dust grain models (models s1a, s1c and s8), a disk radius peak is seen in the vicinity of $\zeta\sim 10^{-17.5}-10^{-16.5}$\,s$^{-1}$. 
These three models have greater proportions of small dust particles than the other seven models (as seen in Table\ref{tab:dust}) along with a smaller proportion of electrons, and the peak is attributed to these differences.
The calculations required to obtain the Hall coefficient are highly complex and so at present we cannot conclusively identify the cause of the peak.
However, this phenomenon will be addressed in a forthcoming paper.

\section{Caveats}
\label{sec:caveats}
As reported in \S\ref{sec:hallcoefficient}, previous simulation studies have identified the possible importance of the Hall effect during the star formation process. 
However, long-term simulations (up to the formation of a mature circumstellar disk) cannot be performed because the whistler waves propagating to the star forming cloud must be resolved using a very short time step to properly estimate the Hall effect during simulations \citep[e.g.][]{2002ApJ...570..314S}. 
Thus, in this study, we analytically estimated the impact of the Hall effect on disk evolution in star-forming clouds, and found that this effect significantly influences the star and disk formation processes depending on the dust properties and the strength of ionization sources.
However, there are some caveats associated with this investigation of the circumstellar disk induced by the Hall effect.
In this section, we discuss the validity of the assumptions used in this study.

\subsection{Other non-ideal MHD effects}
\so{
In this study, we ignored ambipolar diffusion and Ohmic dissipation in the induction equation. 
Only considering the ambipolar diffusion, \citet{2016ApJ...830L...8H} also analytically estimated the  disk radius as 
\begin{equation}
r \sim 18 \ \rm{au} \left(\frac{\eta_{\rm AD}}{0.1 \rm{s}} \right)^{\frac{2}{9}} \left(\frac{B_{\rm z}}{0.1 \rm{G}} \right)^{-\frac{4}{9}} \left(\frac{M}{0.1 M_\odot} \right)^{\frac{1}{3}},
\label{eq:Hennebelle}
\end{equation}
where $M$ is the total mass of disk and protostar and $\eta_{\rm AD}$ 
\footnote{
\so{The unit of $\eta_{\rm AD}$ in equation (\ref{eq:Hennebelle}) differs from the unit used in this study ($\rm cm^2 s^{-1}$).
The difference in resistivities between them is $c^2/4 \pi$ (for detailed definition, see \citealt{2016A&A...592A..18M}). }
} is the ambipolar diffusion coefficient.
When the protostellar mass is  $M_{\rm star} = 0.1 M_{\rm solar}$, the disk radius derived from \citet{2016ApJ...830L...8H} is comparable to that derived from this study (see Fig.\ref{fig:css}), which  implies that we cannot ignore both ambipolar diffusion and the Hall effect when considering the disk formation. 

In addition to the Hall effect and ambipolar diffusion, Ohmic dissipation also affects the disk formation (e.g. \citet{2011PASJ...63..555M}). 
Moreover, in analytical works, we ignored some mechanisms of angular momentum transfer such as protostellar outflow that significantly reduces the angular momentum of the disk. 
Thus, in order to fully investigate the disk evolution, we need to execute non-ideal MHD simulations including all the non-ideal MHD terms in future.}

\so{\subsection{Assumptions and limitations of the disk model}
In this subsection, we discuss the assumptions and limitations used in this study.
Especially, we focus on the timescale for convergence of the Hall velocity, the initial rotation, and the size growth of grains.}
\subsubsection{Convergence timescale of the Hall velocity}
\label{subsec:halltimescale}
In Section \S\ref{subsec:Hallvelocity}, we assumed that gas particles instantaneously receive the angular momentum (or the Hall velocity) at the moment at which they pass through the Hall radius (see Section \S\ref{subsec:diskmodel}). 
In the collapsing cloud, the Hall velocity results from the Lorenz force of the toroidal field component induced by the Hall effect. 
We assumed the toroidal field produced by the Hall effect at a specific, constant value.
However, it is necessary to estimate the saturation timescale in order to confirm 
that the gas particles indeed obtain the Hall velocity estimated in Section \S\ref{subsec:Hallvelocity}.

The specific angular momentum of the gas fluid induced by the Hall effect \so{is estimated from the magnetic torque exerted by the magnetic tension and} is described as
\begin{equation}
\begin{split}
j(t_{\rm Hall}) &= \left[\bm{r} \times \left(\frac{(\nabla \times \bm{B}) \times \bm{B}}{4 \pi \rho} \right) \right]_z t_{\rm Hall}, \\
&\sim \frac{r}{4 \pi \rho} \frac{B_{r,s} B_z}{H} t_{\rm Hall},
\label{eq:halltimeangularmomentum}
\end{split}
\end{equation}
where $t_{\rm Hall}$ is the timescale until a constant (or time-independent) toroidal field is induced by the magnetic tension force or Hall effect.
\so{We use the same approximation as used in the derivation of the Hall velocity (see eqs. (\ref{eq:vhallder}) and (\ref{eq:vhall3})}).
In these calculations, a Hall radius of $r_{\rm Hall}=300\,\rm{au}$ and a density of $\rho=10^{-15}\rm{g \ cm^{-3}}$ are employed (as per Section \S\ref{subsec:Hallvelocity}).
Using  $B_z$ (eq.~\ref{eq:magstrength}), $H$ (equation (\ref{eq:scaleheightH})) and  $B_{r,s}$ (equation (\ref{eq:magneticflux})) and their fiducial values,  equation~(\ref{eq:halltimeangularmomentum}) can be written as
\begin{equation}
j(t_{\rm Hall}) \sim 4.0 \times 10^{11}\, \left(\frac{t_{\rm Hall}}{1~\rm s}\right) \ \rm{cm^2 \ s^{-1}}.
\label{eq:j1}
\end{equation}
In addition, using the saturated Hall velocity (equation (\ref{eq:vhallmu})), we can write the specific angular momentum at $r_{\rm Hall}$ = 300\,au as 
\begin{equation}
j(t_{\rm sat}) = r v_\phi \sim 2.1 \times 10^{18} \ \rm{cm^2 \ s^{-1}}.
\label{eq:j2}
\end{equation}
When the steady state (or saturation) is reached, $j({t_{\rm Hall}})=j({t_{\rm sat}})$ is established.
Thus, the saturation timescale is estimated as 
\begin{equation}
t_{\rm Hall} \sim 5.1 \times 10^6 \  \rm{s}.
\end{equation}
The dynamic timescale of the collapsing cloud roughly corresponds to the freefall timescale, $t_{\rm ff}$.
The $t_{\rm ff}$ at $n_{\rm{Hall}} = 10^8\, \rm{cm^{-3}}$ is given by 
\begin{equation}
t_{\rm ff} = \sqrt{\frac{3 \pi}{32 G \rho}} = 1.4 \times 10^{11} \ \rm{s}.
\end{equation}
Thus, the saturation timescale is much shorter than the freefall timescale $t_{\rm Hall} \ll t_{\rm ff}$, suggesting that the (saturated) toroidal field is instantaneously generated under a given set of cloud parameters, and supplies the Hall velocity estimated in this study to the infalling gas particles.

\so{
\subsubsection{Initial Rotation}
In this study, we ignored the rotation  of prestellar clouds for simplicity. 
Observations show that molecular cloud cores have non-negligible rotations or angular momenta \citep[e.g.][]{1993ApJ...406..528G,2002ApJ...572..238C}.
Thus, our assumption of non-rotating cores is not very correct. 
However, we could estimate the angular momentum purely induced by the Hall effect, and found that the Hall effect significantly influences the disk formation.
This indicates that the Hall effect cannot be ignored to investigate the disk formation, independent of the initial cloud rotation.

\subsubsection{Grain growth during star formation process}
We do not include the time-dependent grain growth. 
However, in reality, grains would grow by collisions between grains or sticking molecules onto grains.
In addition, collisional destruction should be occurred in the star forming cloud.

\citet{2014MNRAS.439.3121C} executed the simulations to estimate the dust growth under very low-metallicity environments. 
They showed that the grain growth can change the star formation process under such environments. 
Our results showed that the grain sizes strongly affect the Hall effect and resulting disk size.
Thus, we require simulations including the time-dependent change of grain sizes (or grain growth) in a future study.}

\subsection{The Effect of Magnetic Field Strength}
\label{subsec:magneticfielddependence}
A value of $\beta$=100 was used to estimate the magnetic field strength in conjunction with equation~(\ref{eq:magstrength}), which is consistent with recent numerical simulations \citep[e.g.,][]{2018MNRAS.476.2063W}.
However, since only the very early phase of disk formation has been investigated by simulations, the magnetic field strength and plasma beta of the disk would be expected to vary over time.

\begin{figure}
\includegraphics[width=\columnwidth]{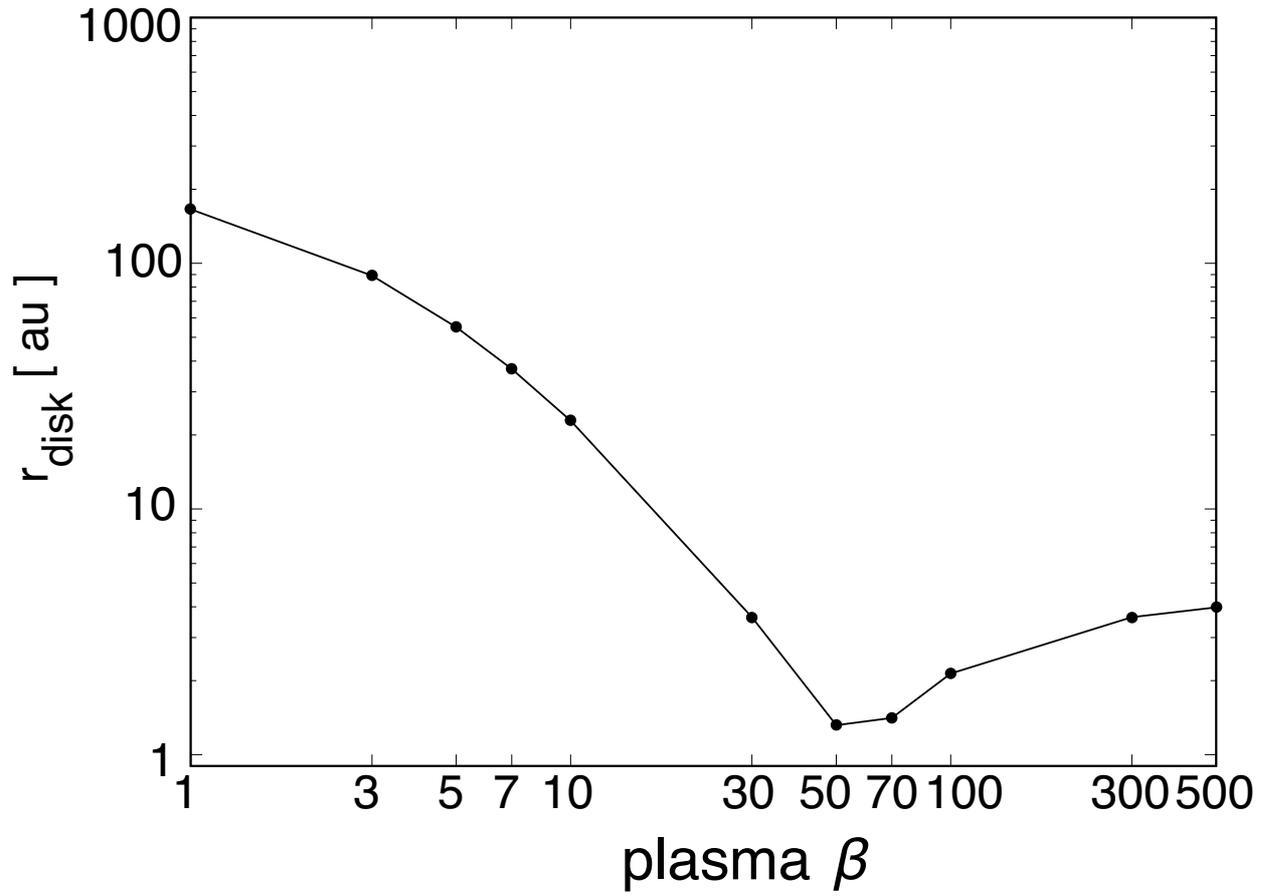}
\caption{
Calculated disk size as function of plasma $\beta$.
}
\label{fig;diskb}
\end{figure}

In this subsection, we assess the effect of the magnetic field strength (or plasma beta) on the disk size.  
Based on \S\ref{subsec:diskmodel}, we again calculated the disk size, employing $\beta$ values over the range of $1\le \beta \le 500 $. 
Fig. \ref{fig;diskb} plots the disk size values against the plasma beta, as determined using the fiducial parameters $a_{\rm single} = 0.1$\,$\mu$\,m and $\zeta = 10^{-17}\, \rm{s^{-1}}$.
This figure confirms that the disk size decreases as $\beta$ increases over the range of $1 \lesssim \beta \lesssim 50$, but increases in the case of $\beta$ values in the range of $\beta \gtrsim 50$.
At $n_{\rm{Hall}} = 10^8 \rm{cm^{-3}}$, the Hall coefficient, $\eta_H$, becomes large as the magnetic field is strengthened (see Fig. 6 in \citet{2016A&A...592A..18M}). 
In the present analysis, when $1 \lesssim \beta \lesssim 50$, variations in $\eta_H$ are primarily responsible for changes in $\beta$ . 
In contrast, when $50 \lesssim \beta \lesssim 500$, the decrease in the vertical component of the magnetic field, $B_z$, determines the disk size (see equation (\ref{eq:vhallmu})). 
However, the dependence of the magnetic field on $\eta_{\rm H}$ is not straightforward (see equations (\ref{eq:sigmav}) - (\ref{eq:betanu})), and thus we cannot provide a simple explanation of the trend seen in Fig.~\ref{fig;diskb}.
During the star formation process, although the fluctuations in the magnetic field strength are complex, our results suggest that it is important to consider the magnetic field strength when assessing the Hall effect.

It is evident that the disk size varies with the magnetic field strength, and that the magnetic field is accumulated and amplified in the disk because it is coupled with the infalling matter. 
In contrast, the magnetic field is dissipated by Ohmic dissipation and ambipolar diffusion in the disk, and the magneto-rotational instability plays a role in determining the magnetic field strength in the disk. 
These effects are beyond the scope of this study, and the determination of magnetic field strength in the disk would not be possible even using state-of-art simulations.
\red{The results of this work imply that the magnetic field strength is improtant for investigating the angular momentum induced by the Hall effect. Thus, both the magnetic field strength and Hall effect would influence the formation and evolution of circumstellar disks.}

\clearpage
\section{Summary}
\label{sec:summary}
In this study, we analytically estimated the influence of the Hall effect on the formation and evolution of circumstellar disks.
The chemical reactions in the collapsing cloud were initially determined to derive the Hall coefficient based on prior publications, but using a variety of dust grain models and ionization intensities.
The results show that the Hall coefficient strongly depends both on the assumed dust model and the ionization intensity. 

Subsequently, a simple disk model was employed to evaluate the size of the circumstellar disk solely on the basis of the Hall effect. 
Depending on the grain size and ionization intensity values selected, disk sizes of $\sim 3-100$\,au ($\sim 2-20$\,au) were obtained for a protostar mass of $0.5\msun$ ($0.1\msun$). 
It was \so{expected} that \so{at $n_{\rm Hall} = 10^8 \ {\rm cm^{-3}}$, which number density is used in the disk model},  smaller dust grains tend to form a larger disk in the grain size range of $0.025\rm{\mu}m \lesssim a_{\rm{single}} \lesssim 0.3\rm{\mu}m$, because small grains significantly reduce the fraction of charged particles and amplify the non-ideal MHD effects. 
Conversely, in the range of $0.01\rm{\mu}m \lesssim a_{\rm{single}} \lesssim 0.025\rm{\mu}m$, the disk size decreases as the grain size becomes smaller, because charged dust grains are much more abundant than charged particles. 
As such, a peak in the disk size change data appears at $a_{\rm{single}}=0.025\rm{\mu}m$ in the case of $\zeta=10^{-17} \rm{s^{-1}}$.
In addition, large disks tend to be associated with weak ionization intensities, as a result of coupling between charged species and neutral gas.

The Keplerian disks observed around Class 0 objects have a size of $\sim10-100$\,au, which is comparable to the disk size induced by the Hall effect.
Thus, the Hall effect evidently contributes to the disk evolution, depending on the star-forming environment.
It is important to note that the simple disk model in the work reported herein was not comprehensive.
In particular, the time required for the growth of dust grains was neglected. 
Nevertheless, this work demonstrates that the Hall coefficient is significantly dependent on the average grain size, and so investigations of disk evolution must take into account the growth of dust particles.

\section*{Acknowledgements}
We thank the referee for his/her useful comments.
The authors gratefully acknowledge helpful discussions with ~K. Tomida.
This work was supported by JSPS KAKENHI grants (numbers 17H02869, JP17K05387, JP17H06360 and 17KK0096).
This research used computational resources within the high-performance computing infrastructure (HPCI) system provided by the Cyberscience Center, Tohoku University, the Cybermedia Center, Osaka University, and the Earth Simulator, JAMSTEC via the HPCI System Research Project (project numbers hp170047 and hp180001).
Simulations were also performed by 2017 and 2018 Koubo Kadai on the Earth Simulator (NEC SX-ACE) at JAMSTEC. 



\bibliographystyle{mnras}
\bibliography{ms} 

\clearpage
\appendix
\section{Reaction Rates}
\label{sec:reactinrate}
All the reactions used in this study, along with the associated reaction rates, are summarized in Table~\ref{tab:rea}. 
The reaction between a charged particle i (that is, an ion) and an electron, e$^-$, is written as k$_{\rm i}$, while k$_{\rm ij}$ represents the reaction rate between charged species i and neutral species j, as in Table~\ref{tab:num}. 
Note that the HCO$^+$ reaction rate was used as the m$^+$ reaction rate, because HCO$^+$ is the most abundant molecular ion after H$_3^+$. 
In the case that a single charged species could undergo multiple reactions, these were distinguished using the symbols $a - e$. 
As an example, H$_3^+$ can participate in five different reactions, with reaction rates $k_{12a}, k_{12b}, k_{12c}, k_{12d}$ and $k_{12e}$.
As described in Section \S\ref{subsec:chemicalnetwork}, these reaction rates were acquired from the UMIST database \citep{2013A&A...550A..36M}.

\begin{table}  
	     \begin{center}
	    \caption{Reactions and reaction rates used in this study}
\renewcommand{\arraystretch}{1.4}
\begin{tabular}{cc}
            \hline
            Reaction & Rate ($\rm{cm^3\, s^{-1}}$)   \\ \hline
            $\rm{H_3^+ + e^- \to H_2 + H}$ & $k_{1a}=2.34 \times 10^{-8} (\frac{T}{300})^{-0.5}$ \\
            $\rm{H_3^+ + e^- \to H+H+H}$ & $k_{1b}=4.36 \times 10^{-8} (\frac{T}{300})^{-0.5}$ \\
            $\rm{HCO^+ + e^- \to CO + H}$ & $k_2=2.4 \times 10^{-7} (\frac{T}{300})^{-0.7}$  \\
            $\rm{Mg^+ + e^- \to Mg}$ + $h \nu$ & $k_3=2.78 \times 10^{-12} (\frac{T}{300})^{-0.7}$  \\
            $\rm{He^+ + e^- \to He}$ + $h \nu$ & $k_4=5.36 \times 10^{-12} (\frac{T}{300})^{-0.5}$  \\
            $\rm{C^+ + e^- \to C}$ + $h \nu$ & $k_5=2.36 \times 10^{-12} (\frac{T}{300})^{-0.3} \rm{exp}(\frac{-17.6}{T})$  \\
            $\rm{H^+ + e^- \to H}$ + $h \nu$ & $k_6=3.50 \times 10^{-12} (\frac{T}{300})^{-0.8}$  \\ 
            $\rm{H_3^+ + O \to H_2O^+ + H}$ & $k_{12a}=3.42 \times 10^{-10} (\frac{T}{300})^{-0.2}  \rm{xO}$ \\
            $\rm{H_3^+ + O \to OH^+ + H_2}$ & $k_{12b}=7.98 \times 10^{-10} (\frac{T}{300})^{-0.2}  \rm{xO}$ \\
            $\rm{H_3^+ + CO \to HCO^+ + H_2}$ & $k_{12c}=1.36 \times 10^{-9} (\frac{T}{300})^{-0.1} \rm{exp}(\frac{3.40}{T}) \rm{xCO}$ \\
            $\rm{H_3^+ + CO \to HOC^+ + H_2}$ & $k_{12d}=8.49 \times 10^{-10} (\frac{T}{300})^{0.1} \rm{exp}(\frac{-5.20}{T}) \rm{xCO}$ \\
            $\rm{H_3^+ + O_2 \to O_2H^+ + H_2}$ & $k_{12e}=9.30 \times 10^{-10}  \rm{exp}(\frac{-100.00}{T}) \rm{xO_2}$ \\
            $\rm{H_3^+ + Mg \to Mg^+ + H + H_2}$ & $k_{13}=1.0 \times 10^{-9} \rm{xM}$  \\
            $\rm{HCO^+ + Mg \to Mg^+ + HCO}$ & $k_{23}=2.9 \times 10^{-9} \rm{xM}$  \\
            $\rm{He^+ + O_2 \to O^+ + O + He}$ & $k_{42}=1.1 \times 10^{-9} \rm{xO_2}$  \\
            $\rm{He^+ + CO \to C^+ + O + He}$ & $k_{45}=1.6 \times 10^{-9} \rm{xCO}$  \\
            $\rm{He^+ + H_2 \to H^+ + H + He}$ & $k_{46}=3.7 \times 10^{-14} \rm{exp}(\frac{-35.0}{T}) \rm{xH_2}$  \\
            $\rm{C^+ + H_2 \to CH_2^+}$ + $ h \nu$ & $k_{52a}=4.0 \times 10^{-16} \left(\frac{T}{300} \right)^{-0.2} \rm{xH_2}$ \\
            $\rm{C^+ + O_2 \to CO^+ + O}$  & $k_{52b}=3.42 \times 10^{-10} \rm{xO_2}$ \\
             $\rm{C^+ + O_2 \to CO + O^+}$  & $k_{52c}=4.54 \times 10^{-10} \rm{xO_2}$ \\
            $\rm{C^+ + Mg \to Mg^+ + C}$ & $k_{53}=1.1 \times 10^{-9} \rm{xM}$  \\
            $\rm{H^+ + O \to O^+ + H}$ & $k_{62a}=6.86 \times 10^{-10} (\frac{T}{300})^{0.3} \rm{exp} (\frac{-224.30}{T}) \rm{xO}$  \\
            $\rm{H^+ + O_2 \to O_2^+ + H}$ & $k_{62b}=2.00 \times 10^{-9} \rm{xO_2}$  \\
            $\rm{H^+ + Mg \to Mg^+ + H}$ & $k_{63}=1.1 \times 10^{-9} \rm{xM}$  \\
            $\rm{He^+ + H_2 \to H_2^+ + He}$ & $k_{41}=7.2 \times 10^{-15} \rm{xH_2}$  \\
            \hline
          \end{tabular}
          \label{tab:rea}
          \end{center}
\renewcommand{\arraystretch}{1.0}
\end{table}

\clearpage
\section{Dust Reactions}
\label{secdust}
In this appendix, we address collisions between dust grains and charged particles and the absorption of charged particles onto the surfaces of dust grains. 
Three types of dust grains are considered: positively charged, negatively charged and neutral. 
The reactions of these grains with ions and electrons are also considered, such that six reactions are required, as described below.
According to \citet{1987ApJ...320..803D}, the reaction rates can be summarized as itemized below.
\begin{enumerate}
\renewcommand{\labelenumi}{[\arabic{enumi}]}
\item The collisional rate between charged grains and ions ($i$ = 1 -- 6), written as
\begin{equation}
k_{id} (s,z)= \left \{
\begin{array}{l}
\sqrt{\frac{8 k_B T}{\pi m_i}} \sigma_s \ \rm{exp} \left(-\frac{\it{z}}{(1+|\it{z}|^{-1/2})\tau_s} \right) \ \left\{1+ \left(\frac{1}{4\tau_s+3\it{z}} \right)^{1/2} \right\}^2 \qquad (\it{z} \ \rm{>0}) \\
\sqrt{\frac{8 k_B T}{\pi m_i}} \sigma_s \ \left(1-\frac{z}{\tau_s} \right) \ \left\{1+ \left(\frac{2}{\tau_s-2z} \right)^{1/2} \right\} \qquad (z<0).
\end{array}
\right.
\end{equation} 

\item The collisional rate between charged grains and electrons ($i$ = 0), written as
\begin{equation}
k_{id} (s,z)= \left \{
\begin{array}{l}
\sqrt{\frac{8 k_B T}{\pi m_e}} \sigma_s \ \left( 1+\frac{z}{\tau_s} \right) \left\{ 1+\left(\frac{2}{\tau_s+2z} \right)^{1/2} \right\} \qquad (z>0) \\
\sqrt{\frac{8 k_B T}{\pi m_e}} \sigma_s \ \rm{exp} \left(\frac{\it{z}}{(1+|\it{z}|^{-1/2})\tau_s} \right) \ \left\{1+ \left(\frac{1}{4\tau_s-3\it{z}} \right)^{1/2} \right\}^2 \qquad (\it{z} \ \rm{<0}).
\end{array}
\right.
\end{equation} 

\item The collisional rate between neutral grains, electrons and ions ($i$ = 0 -- 6), written as 
\begin{equation}
k_{nd} (s,0) =\sqrt{\frac{8 k_B T}{\pi m_e}} \sigma_s \left\{1+ \left(\frac{\pi}{2 \tau_s} \right)^{1/2} \right\}.
\end{equation}
\end{enumerate}
Here, $z$, $m_e$, $\sigma_{s}$ and $\tau_{s}$ indicate the charge on the dust particles, the electron mass, the collision cross-section and the reduced temperature, respectively.
The reduced temperature, $\tau_{s}$, is described as 
\begin{equation}
\tau_{s}=\frac{a_s k_B T}{e^2},
\end{equation}
and the collision cross-section, $\sigma_{s}$, is 
\begin{equation}
\sigma_s=\pi {a_{s}}^2,
\end{equation}
where the suffix $s$ indicates the dust grains in the $s$-th bin (see Section \S\ref{sec:dustsize}).

\clearpage
\section{Reaction Equations}
\label{sec:reactionequations}
The reaction equations used in this study were as follows. 
The term $x_i$ ($i$=0 -- 6) refers to the fractional abundance of each charged species (for each index, see Table~\ref{tab:num}) and $x_{\rm d}(s,z)$ refers to the fractional abundance of dust grains, where $s$ is the bin number and $z$ is the charge of dust grains.
The reaction rates are summarized in Appendix \S\ref{sec:reactinrate} and the cosmic ray rates, $\zeta$, can be found in Section \S\ref{subsec:chemicalnetwork}. 
\label{sec:reactioneq}
\begin{eqnarray}
\begin{aligned}
\frac{dx_0}{dt} &= (0.97+0.84+0.03) \zeta -  n_{\rm H} (k_1 x_1 +k_2 x_2 +k_3 x_3 +k_4 x_4  \\
&+k_5 x_5 +k_6 x_6 ) x_0 - n_{\rm H} \biggl( \displaystyle \sum_{s=1}^{s_{\rm max}} \Bigl(\displaystyle \sum_{z=z_{\rm min}}^{-1} k_{ed}(s,z) x_{\rm d}(s,z)   \\ 
&+k_{nd}(s)x_{\rm d}(s,0) +\displaystyle \sum_{z=1}^{z_{\rm max}} k_{ed}(s,z)x_{\rm d}(s,z) \Bigr) \biggr) x_0.
\end{aligned}
\end{eqnarray}

\begin{eqnarray}
\begin{aligned}
\frac{dx_1}{dt} &= 0.97\zeta - n_{\rm H} \bigl( (k_{1a}+k_{1b}) x_0   \\
 &+ (k_{12a}+k_{12b}+k_{12c}+k_{12d}+k_{12e}) x_2 - k_{41} x_4 \bigr) x_1 - \delta_{1,\rm{d}}.
\end{aligned}
\end{eqnarray}

\begin{eqnarray}
\begin{aligned}
\frac{dx_2}{dt}  &= - n_{\rm H} \Bigl( k_2 x_0 - (k_{12a}+k_{12b}+k_{12c}+k_{12d}+k_{12e}) x_1 \\
&+  k_{23} x_3 - k_{42} x_4 - (k_{52a}+k_{52b}+k_{52c}) x_5 - (k_{62a}+k_{62b}) x_6 \Bigr) x_2 \\
&- \delta_{2,\rm{d}}.
\end{aligned}
\end{eqnarray}

\begin{eqnarray}
\begin{aligned}
\frac{dx_3}{dt} = - n_{\rm H} \left( k_3 x_0 - k_{13} x_1 - k_{23} x_2 - k_{53} x_5 - k_{63} x_6  \right) x_3 - \delta_{3,\rm{d}}.
\end{aligned}
\end{eqnarray}

\begin{eqnarray}
\begin{aligned}
\frac{dx_4}{dt} = 0.84 \zeta - n_{\rm H} \left( k_4 x_0 + k_{42} x_2 + k_{45} x_5 + k_{46} x_6 + k_{41} x_1  \right) x_4 - \delta_{4,\rm{d}}.
\end{aligned}
\end{eqnarray}

\begin{eqnarray}
\begin{aligned}
\frac{dx_5}{dt} = - n_{\rm H} \left( k_5 x_0 - k_{45} x_4 + (k_{52a}+k_{52b}+k_{52c}) x_2 + k_{53} x_3 \right) x_5 - \delta_{5,\rm{d}}.
\end{aligned}
\end{eqnarray}

\begin{eqnarray}
\begin{aligned}
\frac{dx_6}{dt} =0.03 \zeta - n_{\rm H} \left( k_6 x_0 - k_{46} x_4 + (k_{62a}+k_{62b}) x_2 + k_{63} x_3 \right) x_6 - \delta_{6,\rm{d}}.
\end{aligned}
\end{eqnarray}

\begin{eqnarray}
\begin{aligned}
\frac{dx_{\rm d}(z)}{dt} &= n_{\rm H} \displaystyle \sum_{s=1}^{s_{\rm{max}}} \Bigl( -k_{ed} (s,z) x_{\rm d}(s,z) x_0 - \displaystyle \sum_{i=1}^6 k_{id} (s,z) x_{\rm d}(s,z) x_i  \\
&+ k_{ed}(s,z+1) x_{\rm d}(s,z+1) x_0 + k_{id}(s,z-1) x_{\rm d}(s,z-1) x_i \Bigr),
\end{aligned}
\end{eqnarray}

In the above, $\delta_{i,{\rm d}}$ indicates the reaction following a collision between a charged particle {\it i} and a dust grain, and is given by  
\begin{eqnarray}
\begin{aligned}
\delta_{i,{\rm d}} &= n_{\rm H} \biggl\{ \displaystyle \sum_{s=1}^{s_{\rm{max}}} \Bigl(\displaystyle \sum_{z=z_{min}}^{-1} k_{i-d}(s,z) x_{\rm d}(s,z) +k_{nd}(s)x_d(s,0) \\
&+\displaystyle \sum_{z=1}^{z_{max}} x_{i+d}(s,z)x_{\rm d}(s,z) \Bigr) \biggr\} x_i.
\end{aligned}
\end{eqnarray}

\clearpage
\so{\section{Equilibrium timescale vs. freefall timescale}
\label{sec;quilibriumtime}
In this appendix, we compare the elapsed time required to obtain an equilibrium state with the freefall timescale at a given density.
Figure~\ref{fig:reactiontime} shows that chemical abundances of charged species considered in this study reach an equilibrium state within the freefall timescale. 

In all density, the equilibrium is achieved before the freefall time. 
Thus we confirm that the elapsed time necessary for reaching an equilibrium state is much shorter than the freefall timescale over the density range of $10^4\cm < n_{\rm H} < 10^{14}\cm$.
The validity of the chemical equilibrium is also seen in Fig.13 of \citet{2016A&A...592A..18M}.

\begin{figure*}
\begin{tabular}{cc}
\begin{minipage}{0.5\hsize}
\includegraphics[width=\columnwidth]{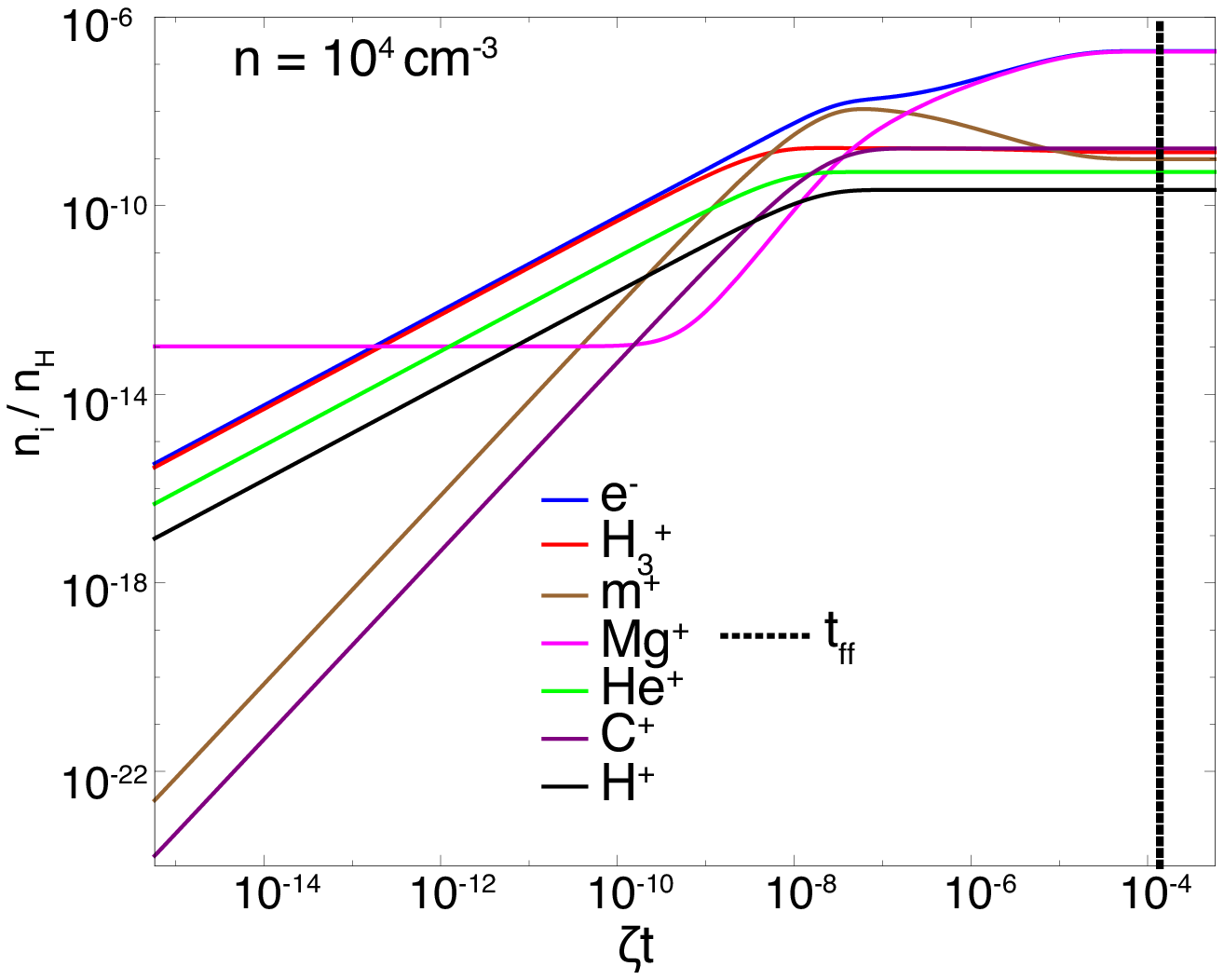}
\end{minipage}
\hspace{-5pt}
\begin{minipage}{0.5\hsize}
\includegraphics[width=\columnwidth]{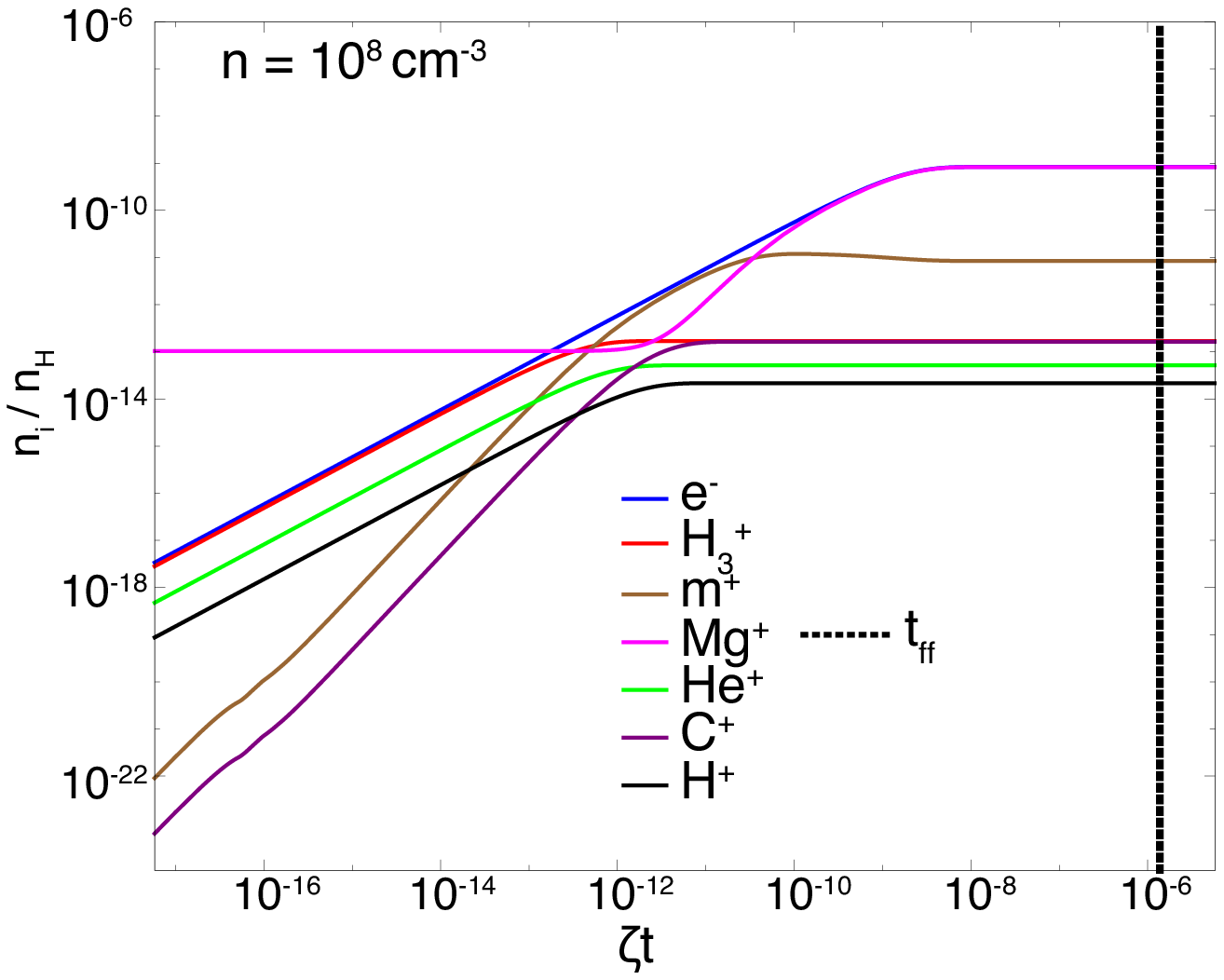}
\end{minipage}
\hspace{-5pt}
\\
\begin{minipage}{0.5\hsize}
\includegraphics[width=\columnwidth]{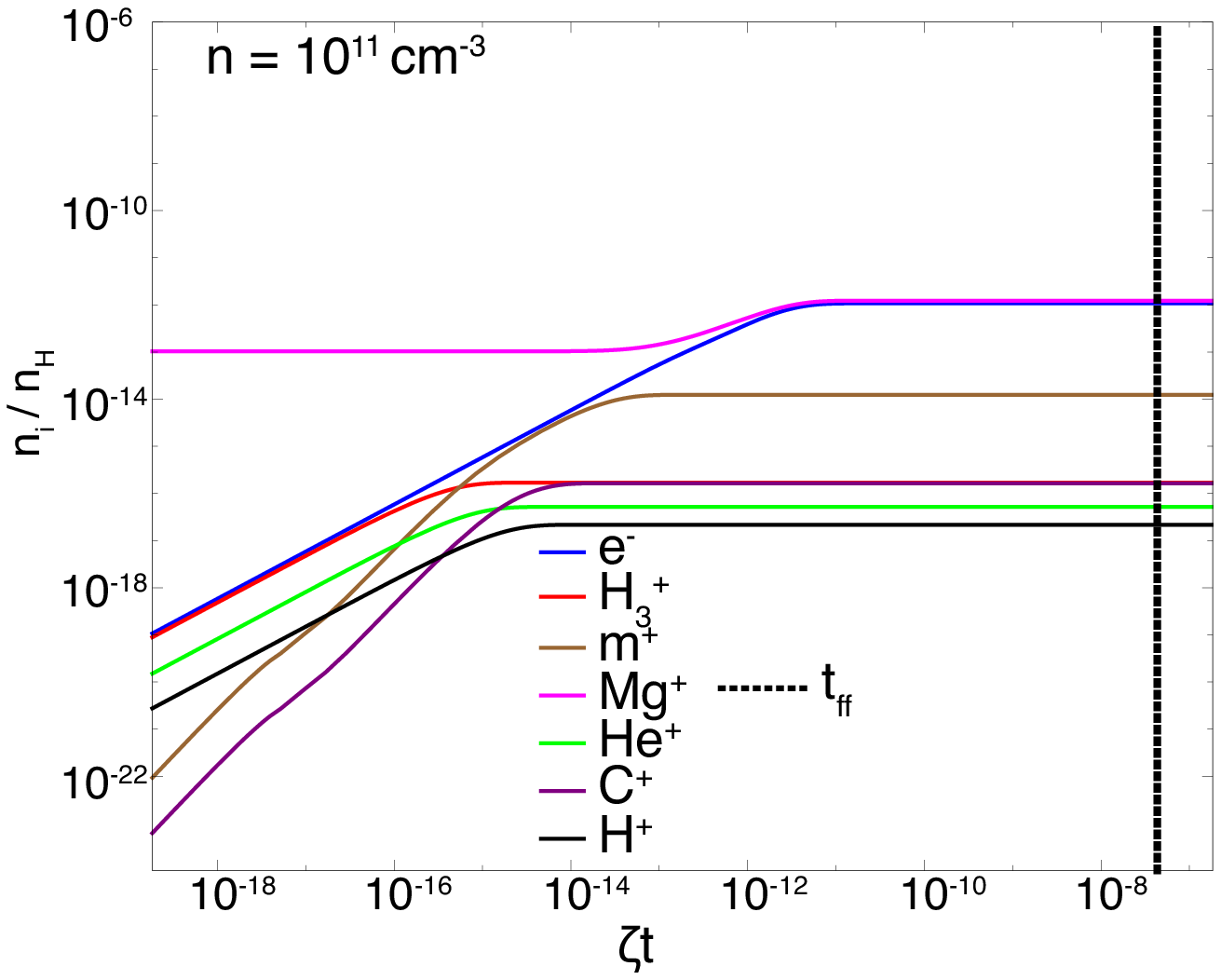}
\end{minipage}
\hspace{-5pt}
\begin{minipage}{0.5\hsize}
\includegraphics[width=\columnwidth]{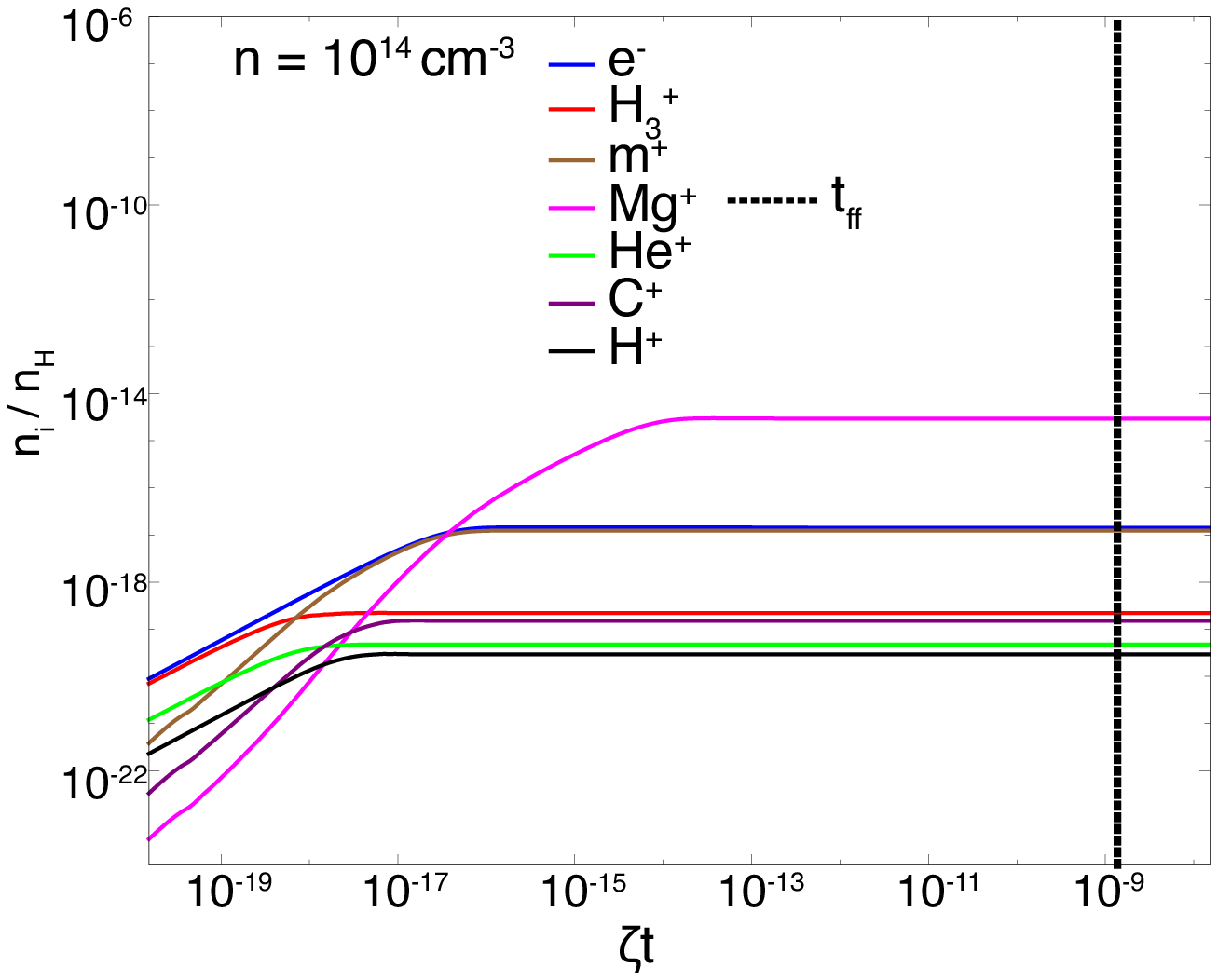}
\end{minipage}
\hspace{-5pt}
\end{tabular}\caption{Chemical abundances of charged species for different gas number density $10^4, 10^8, 10^{11}, 10^{14} \rm{cm^{-3}}$ as a function of $\zeta t$ ($\zeta = 10^{-17} \rm{s^{-1}}$).
The freefall timescale is given by $t_{\rm ff} = \sqrt{\frac{3 \pi}{32 G \rho}}$.}
\label{fig:reactiontime}
\end{figure*}}

\bsp	
\label{lastpage}
\end{document}